\documentclass{article}

\usepackage{arxiv}

\usepackage[utf8]{inputenc} 
\usepackage[T1]{fontenc}    
\usepackage{hyperref}       
\hypersetup{
  colorlinks   = true, 
  urlcolor     = ForestGreen, 
  linkcolor    = BrickRed, 
  citecolor   = blue 
}
\usepackage{url}            
\usepackage{booktabs}       
\usepackage{amsfonts}       
\usepackage{nicefrac}       
\usepackage{microtype}      
\usepackage{comment}

\usepackage{subfigure} 

\usepackage{framed,multirow}

\usepackage{amsmath}
\usepackage{algorithm}
\usepackage{algpseudocode}
\usepackage{graphicx}

\usepackage{amssymb}
\usepackage{latexsym}
\usepackage{bm}
\usepackage{amssymb}
\usepackage{array}
\usepackage{multirow}
\usepackage[english]{babel}
\usepackage[numbers]{natbib}
\usepackage{footnote}
\usepackage{tabularx}
\makesavenoteenv{tabular}
\makesavenoteenv{table}

\usepackage{cleveref}
\crefname{equation}{Eq.}{Eqs.}
\crefname{table}{Table}{Tables}
\crefname{figure}{Fig.}{Figs.}
\crefname{section}{Section}{Sections}
\crefname{subsection}{Section}{Secs.}
\Crefname{figure}{Figs.}{Figs.}
\Crefname{Algorithm}{Algorihtm}{Algorihtm}

\usepackage[dvipsnames]{xcolor}
\definecolor{newcolor}{rgb}{.8,.349,.1}

\title{Hamiltonian MCMC Methods for Estimating Rare Events Probabilities in High-Dimensional Problems}

\author{
 Konstantinos G.~Papakonstantinou \\
Department of Civil and Environmental Engineering\\
The Pennsylvania State University\\
University Park, PA 16802 \\
\texttt{kpapakon@psu.edu} \\
   \And
    Hamed~Nikbakht\\
Department of Civil and Environmental Engineering\\
The Pennsylvania State University\\
University Park, PA 16802 \\
\texttt{hun35@psu.edu} \\
   \And
    Elsayed~Eshra\\
Department of Civil and Environmental Engineering\\
The Pennsylvania State University\\
University Park, PA 16802 \\
\texttt{eme5375@psu.edu} \\
}

\begin{document}
\maketitle

\begin{abstract}
Accurate and efficient estimation of rare events probabilities is of significant importance, since often the occurrences of such events have widespread impacts.  The focus in this work is on precisely quantifying these probabilities, often encountered in reliability analysis of complex engineering systems, based on an introduced framework termed \textit{Approximate Sampling Target with Post-processing Adjustment} (ASTPA), which herein is integrated with and supported by gradient-based Hamiltonian Markov Chain Monte Carlo (HMCMC) methods. The developed techniques in this paper are applicable from low- to high-dimensional stochastic spaces, and the basic idea is to construct a relevant target distribution by weighting the original random variable space through a one-dimensional output likelihood model, using the limit-state function. To sample from this target distribution, we exploit HMCMC algorithms, a family of MCMC methods that adopts physical system dynamics, rather than solely using a proposal probability distribution, to generate distant sequential samples, and we develop a new \textit{Quasi-Newton mass preconditioned} HMCMC scheme (QNp-HMCMC), which is particularly efficient and suitable for high-dimensional spaces. To eventually compute the rare event probability, an original post-sampling step is devised using an \textit{inverse importance sampling} procedure based on the already obtained samples. The statistical properties of the estimator are analyzed as well, and the performance of the proposed methodology is examined in detail and compared against Subset Simulation in a series of challenging low- and high-dimensional problems.
\end{abstract}


\section{Introduction}
In this work, we develop a framework for estimation of rare events probabilities, a commonly encountered important problem in several engineering and scientific applications, often observed in the form of probability of failure ($P_{F}$) estimation or, alternatively, reliability estimation. In many practical applications, failure probabilities are fortunately very low, from $10^{-4}$ to even $10^{-9}$ and lower, and calculating such small probabilities presents many numerical and mathematical challenges, particularly in cases with high dimensional random spaces and/or expensive computational models, that practically limit the afforded number of model calls. The number of model calls is thus of great importance in these problems and one of the critical parameters that limits or prohibits use of several available techniques in the literature.\par

The reliability estimation problem has a long history in the engineering community  
\citep{melchers2018structural,ditlevsen1996structural,au2014engineering,nikolaidisengineering,lemaire2009structural}. One of the significant early successes was the discovery of the so called First Order Reliability Method (FORM) \citep{rackwitz2001reliability,breitung201540}, long investigated by Der Kiureghian, Ditlevsen and co-workers \citep{der1986structural,ditlevsen1986methods}, and many others, e.g., Shinozuka \citep{shinozuka1983basic}, providing also several enhancements, including second order effects (SORM) by Breitung \citep{breitung1984asymptotic}.
In FORM/SORM methods, the search for the most probable failure point (MPP) is usually performed by gradient-based optimization methods \citep{liu1991optimization,haukaas2006strategies}. 
Although these asymptotic approximation methods are usually computationally inexpensive, they have several limitations and may involve considerable errors, particularly in high-dimensional problems or in problems with highly nonlinear limit-state functions \citep{rackwitz2001reliability,valdebenito2010role}. 
As such, various sampling-based methods have also been suggested by the reliability community, e.g., Schuëller and Pradlwarter \citep{schueller2007benchmark}, to tackle the problem in its utmost generality, with crude Monte Carlo approaches being prohibitive for this type of problems due to their excessive computational demands.
Only some of many notable contributions can be seen in \citep{papaioannou2015mcmc,zuev2011modified,zuev2012bayesian,au2016rare,wang2019hamiltonian}, describing and studying the state-of-the art-Subset Simulation and its enhancements, originally presented in \citep{au2001estimation}, and in \citep{au1999new,au2003important,papaioannou2016sequential} utilizing importance sampling schemes, often also combined with the cross-entropy method \citep{rubinstein2004cross,kurtz2013cross,geyer2019cross,yang2017cross,papaioannou2019improved}. Alternative approaches include directional and line Sampling \citep{koutsourelakis2004reliability,pradlwarter2007application,bjerager1988probability,nie2000directional}, the PHI2 method for time-variant reliability \citep{andrieu2004phi2}, and asymptotic sampling strategies \cite{bucher2009asymptotic,naess2009system}, among others. The problem of estimating rare event probabilities has also attracted a great deal of attention in other relevant communities and in mathematical literature, with several suggested methods sharing many similarities with FORM/SORM approaches, e.g.,\citep{tong2022large}, and Subset Simulation, such as in approaches involving Sequential Monte Carlo samplers for rare events \citep{johansen2005sequential,cerou2012sequential,del2006sequential}, interacting particle system methodologies \citep{del2005genealogical,cerou2006genetic}, multilevel splitting methods \citep{guyader2011simulation,walter2015moving} and forward flux sampling \citep{allen2009forward}, to name but a few. \par     

In this paper, we are presenting a new solution approach to the problem by combining gradient-based approaches, already familiar to the engineering community and often available in computational tools, with Markov Chain Monte Carlo (MCMC) sampling methods in the form of Hamiltonian MCMC.
\newcommand{\cell}[1]{\begin{tabular}{@{}l@{}}#1\end{tabular}}
MCMC methods \citep{brooks2011handbook} are plausibly the most broadly
accepted ones to generate samples from target distributions, in cases where direct sampling is not possible. Despite notable successes, many MCMC methods scale poorly with the number of dimensions and can become inefficient. For complicated multivariate models, classic methods such as random-walk Metropolis-Hastings \citep{metropolis1953equation} and Gibbs sampling \citep{geman1984stochastic} may require an unacceptably long time and number of samples to adequately explore the target distribution. Originally developed by Duane et al. \citep{duane1987hybrid} and to a large extent understood and popularized through the works of Neal \citep{neal2012bayesian,neal2011mcmc}, Hamiltonian Markov Chain Monte Carlo (HMCMC), usually called Hamiltonian Monte Carlo (HMC) in the literature, produces Markov chain samples based on Hamiltonian dynamics principles, is characterized by much better scalability \citep{neal2011mcmc,beskos2013optimal} and faster mixing rates, is capable of generating samples with much weaker auto-correlation, even in complex high-dimensional random spaces \citep{betancourt2017conceptual}, and has enjoyed broad-spectrum successes in most general settings \cite{carpenter2017stan}. Balanced against these features and achievements is of course the need for multiple gradient evaluations in each HMCMC iteration, making the method more computationally intensive per iteration than other algorithms, such as random-walk Metropolis-Hastings and Metropolis-adjusted Langevin \citep{roberts2001optimal,roberts1998optimal}, for example. Girolami and Calderhead introduced a Riemannian Manifold Hamiltonian Monte Carlo (RMHMC) approach in \citep{girolami2011riemann,chen2022riemannian} that has demonstrated significant successes in many challenging problems but requires computing higher-order derivatives of the target distribution. Overall, the two impediments to using Hamiltonian MCMC methods are the required gradients, since analytical formulas are not always available and numerical techniques are computationally costly, particularly in high dimensions, and the heedful tuning of the involved parameters \citep{neal2011mcmc}. The first issue can in certain cases be solved by automatic differentiation (e.g. \citep{carpenter2017stan,griewank1989automatic}) and stochastic gradient approaches \citep{chen2014stochastic}, while for the second a fully automated state-of-the-art HMCMC algorithm has been developed by Hoffman and Gelman, known as the No-U-Turn Sampler (NUTS \citep{hoffman2014no}). NUTS introduces, among others, an expensive tree building procedure, in order to trace when the Hamiltonian trajectory turns back on itself. Many of these approaches are not however relevant and/or applicable to the analyzed problem in this work, since, in general, many rare event and reliability problems involve complex, computationally expensive models, complicating and/or precluding use of automatic differentiation and data-based stochastic gradient techniques, as well as methodologies that require a considerably high number of model calls per problem, such as NUTS. 
\par 

A new computationally efficient sampling framework for estimation of rare events probabilities is thus presented in this work, having exceptional performance in quantifying low failure probabilities for any type of reliability problems described in both low and high dimensional stochastic spaces. The introduced methodology is termed \textit{Approximate Sampling Target with Post-processing Adjustment} (ASTPA) and comprises a sampling and a post-processing phase. The sampling target in ASTPA is constructed by appropriately combining the multi-dimensional random variable space with a cumulative distribution function that utilizes the limit-state function. Having acquired the samples, an adjustment step is then applied, in order to account for the fact that the samples are drawn from an approximate target distribution, and to thus correctly quantify the rare event probability. An original \textit{inverse importance sampling} procedure is devised for this adjustment step, taking its name from the fact that the samples are already available. Although the ASTPA framework is general and can be combined with any appropriate Monte Carlo sampling method, it becomes substantially efficient when directly supported by gradient-based Hamiltonian Markov Chain Monte Carlo (HMCMC) samplers. To address the scalability issues a typical HMCMC sampler may manifest in high-dimensional spaces, a new \textit{Quasi-Newton mass preconditioned HMCMC} approach is also developed. This new sampling scheme follows an approximate Newton direction and estimates the pertinent Hessian in its burn-in stage, only based on gradient information and the BFGS approximation, and eventually utilizes the computed Hessian as a preconditioned mass matrix in the main non-adaptive sampling stage. An approximate analytical expression for the uncertainty of the computed estimation is also derived, showcasing significant accuracy with numerical results, and all involved user-defined parameters of ASTPA are thoroughly analyzed and general default values are suggested. Finally, to fully examine the capabilities of the proposed methodology, its performance is demonstrated and compared against Subset Simulation in a series of challenging low- and high-dimensional problems. 

\section{Failure Probability Estimation}\label{secPF}
The failure probability \textit{P\textsubscript{F}} for a system, that is the probability of a defined unacceptable system performance, can be expressed as a $d$-fold integral, as:
\begin{equation} \label{eq:1}
\textit{P\textsubscript{F}}= \mathop{\mathbb{E}}[I_{F} (\boldsymbol{\theta})] = \int_{g(\boldsymbol{\theta})\leq 0} I_{F} (\boldsymbol{\theta}) \pi_{\Theta}(\boldsymbol{\theta}) d\boldsymbol{\theta} 
\end{equation} 
where \boldsymbol{$\theta$} is the random vector $[\theta_{1},..., \theta_{d}]$$^{T}$, $F \subset \mathbb{R}^{d}$ is the failure event, g(\textbf{\boldsymbol{$\theta$}}) is the limit-state function that can include one or several distinct failure modes and defines the system failure by g(\textbf{\boldsymbol{$\theta$}})$\leq$ 0, \textit{I(.)} denotes the indicator function with \textit{$I_{F}$} (\boldsymbol{$\theta$}) = 1 if \boldsymbol{$\theta$} $\in$ g(\textbf{\boldsymbol{$\theta$}})$\leq$ 0 and $I_{F}$(\boldsymbol{$\theta$}) = 0 otherwise, $\mathop{\mathbb{E}}$ is the expectation operator, and $\pi_{\Theta}$ is the joint probability density function (PDF) for $\boldsymbol{\Theta}$. As is common practice for problems of this type, in this work the joint PDF of $\boldsymbol{\Theta}$ is the standard normal one, due to its rotational symmetry and exponential probability decay. In most cases this is not restrictive, since it is uncomplicated to transform the original random variables \textbf{X} to $\boldsymbol{\Theta}$, e.g. \citep{hohenbichler1981non}. When this is not the case, however, and the probabilistic characterization of  \textbf{X} can be defined in terms of marginal distributions and correlations, the Nataf distribution (equivalent to Gaussian copula) is commonly used to model the joint PDF, and the mapping to the standard normal space can be then accomplished \citep{der1986structural,lebrun2009innovating}.\par

The focus in this work is to analyze the described integration in \cref{eq:1} under very general settings, including the following challenging
sampling context: \textbf{\textit{(i)}} Computation of \cref{eq:1} can only be done in approximate ways; \textbf{\textit{(ii)}} the relationship between \boldsymbol{$\theta$} and $I_{F}$ is not explicitly known and for any \boldsymbol{$\theta$} we can merely check whether it belongs to the failure set or not, i.e. calculate the value $I_{F}$(\boldsymbol{$\theta$}); \textbf{\textit{(iii)}} the computational effort for evaluating $I_{F}$(\boldsymbol{$\theta$}) for each value of \boldsymbol{$\theta$} is assumed to be quite significant, so that it is essential to minimize the number of such function evaluations (model calls); \textbf{\textit{(iv)}} the probability of failure \textit{P\textsubscript{F}} is assumed to be very small, e.g. in order of $\textit{P\textsubscript{F}} \sim 10^{-4} - 10^{-9}$; \textbf{\textit{(v)}} the parameter space $\mathbb{R}^{d}$ is assumed to be high-dimensional, in the order of $10^{2}$ and more, for example. Under these general settings, several sampling methods, including direct Monte Carlo approaches and NUTS \citep{hoffman2014no}, become highly inefficient and fail to address the problem effectively. Subset Simulation (SuS) \citep{au2001estimation} has however proven successful and robust in dealing with problems of this type and is shown to outperform other relevant methods in numerous papers, e.g. \citep{papaioannou2016sequential,pradlwarter2007application}. SuS relies on a modified component-wise Metropolis MCMC method that can successfully work in high dimensions and does not require a burn-in sampling stage. A notable  adaptive conditional sampling (aCS) methodology within the SuS framework is also introduced by Papaioannou et al. in \citep{papaioannou2015mcmc}, providing important advantages and enhanced performance in several cases. Relevant SuS variants are thus utilized in this work, for  validation and comparison purposes with our presented methodology that completely deviates from SuS and is efficiently supported by 
a direct Hamiltonian MCMC sampling approach \citep{nikbakht2019HMCMC}. 

\section{Hamiltonian Markov Chain Monte Carlo}\label{section2}
\subsection{Standard HMCMC with leapfrog integrator}\label{sec3.1}
Based on the aforementioned discussion in the previous sections, Hamiltonian dynamics can be used to produce distant Markov chain samples, thereby avoiding the slow exploration of the state space that results from the diffusive behavior of simple random-walk proposals. This Hamiltonian approach was firstly introduced to molecular simulations by Alder and Wainwright in \citep{alder1959studies}, in which the motion of the molecules was deterministic. Duane et al. in \citep{duane1987hybrid} united the MCMC and molecular dynamics approaches. Given $d$-dimensional variables of interest $\boldsymbol{\theta}$ with (unnormalized) density $\pi_{\Theta}$(.), the Hamiltonian Monte Carlo method introduces $d$-dimensional auxiliary momentum variables $\textbf{z}$ and samples from the joint
distribution characterized by:
\setlength{\abovedisplayskip}{3pt}
\setlength{\belowdisplayskip}{3pt}
\begin{align} \label{eq:8}
\pi(\boldsymbol{\theta},\textbf{z}) \propto  \pi_{\Theta}(\boldsymbol{\theta})\ \pi_{Z \arrowvert\Theta}(\textbf{z} \arrowvert \boldsymbol{\theta})
\end{align} 
where $\pi_{Z\arrowvert \Theta}(.\arrowvert \boldsymbol{\theta})$ is proposed to be a symmetric distribution. With $\pi_{\Theta}(\boldsymbol{\theta})$ and $\pi_{Z\arrowvert \Theta}(\textbf{z}\arrowvert \boldsymbol{\theta})$ being uniquely described up to normalizing constants, the functions $U(\boldsymbol{\theta}) = -\log\pi_{\Theta}(\boldsymbol{\theta})$ and $K(\boldsymbol{\theta},\textbf{z})=-\log\pi_{Z\arrowvert \Theta}(\textbf{z}\arrowvert \boldsymbol{\theta})$ are introduced as the potential energy and kinetic energy, owing to the concept of the canonical distribution \citep{neal2011mcmc} and the physical laws which motivate the Hamiltonian Markov Chain Monte Carlo algorithm. The total energy $H(\boldsymbol{\theta},\textbf{z})=U(\boldsymbol{\theta})+K(\boldsymbol{\theta},\textbf{z})$ is often termed the \textit{Hamiltonian} $H$. The kinetic energy function is unconstrained and can be formed in various ways according to the implementation. In most typical cases, the momentum is sampled by a zero-mean normal distribution \citep{neal2011mcmc,betancourt2017conceptual}, and accordingly the kinetic energy can be written as: $K(\boldsymbol{\theta},\textbf{z}) = -\log\pi_{Z\arrowvert \Theta}(\textbf{z}\arrowvert \boldsymbol{\theta})=-\log\pi_{Z}(\textbf{z}) = \frac{1}{2}\textbf{z}^{T} \textbf{M}^{-1} \textbf{z}$, where $\textbf{M}$ is a symmetric, positive-definite inverse covariance matrix, termed mass matrix. \par
HMCMC generates a Metropolis proposal on the joint state-space $(\boldsymbol{\theta},\textbf{z})$ by sampling the momentum and simulating trajectories of Hamiltonian dynamics in which the time evolution of the state $(\boldsymbol{\theta},\textbf{z})$ is governed by Hamilton's equations, expressed typically by: 
\setlength{\abovedisplayskip}{6pt}
\setlength{\belowdisplayskip}{6pt}
\begin{equation}
\frac{d\boldsymbol{\theta}}{dt} = \frac{\partial H}{\partial \textbf{z}} =\frac{\partial K}{\partial \textbf{z}}= \textbf{M}^{-1} \textbf{z} ,\ \ \  \ 
\frac{d \textbf{z}}{dt} = -\frac{\partial H}{\partial \boldsymbol{\theta}} = -\frac{\partial U}{\partial \boldsymbol{\theta}}= \nabla_{\theta} \mathcal{L}(\boldsymbol{\theta}) \label{eq:10}
\end{equation}  
where $\mathcal{L}(\boldsymbol{\theta})$ denotes here the log-density of the target distribution. Hamiltonian dynamics prove to be an effective proposal generation mechanism because the distribution 
$\pi(\boldsymbol{\theta},\textbf{z})$ is invariant under the dynamics of \cref{eq:10}. These dynamics enable a proposal, triggered by an approximate solution of \cref{eq:10}, to be distant from the current state, yet with high probability acceptance. The solution to \cref{eq:10} is analytically intractable in general and thus the Hamiltonian equations need to be numerically solved by discretizing time using some small step size, $\varepsilon$. A symplectic integrator that can be used for the numerical solution is the \textit{leapfrog} one and works as follows:
\setlength{\abovedisplayskip}{0pt}
\begin{equation} 
\textbf{z}\textsubscript{t+$\varepsilon$/2} = \textbf{z}\textsubscript{t} - (\dfrac{\varepsilon}{2})\dfrac{\partial U}{\partial \boldsymbol{\theta}} (\boldsymbol{\theta}\textsubscript{t}) \label{eq:11} ,\ \ \  \
\boldsymbol{\theta}\textsubscript{t+$\varepsilon$} = \boldsymbol{\theta}\textsubscript{t} + \varepsilon \dfrac{\partial K}{\partial \textbf{z}} (\textbf{z}\textsubscript{t+$\varepsilon$/2}) ,\ \ \  \
\textbf{z}\textsubscript{t+$\varepsilon$} = \textbf{z}\textsubscript{t+$\varepsilon$/2} - (\dfrac{\varepsilon}{2})\dfrac{\partial U}{\partial \boldsymbol{\theta}} (\boldsymbol{\theta}\textsubscript{t+$\varepsilon$}) 
\end{equation}
The main advantage of using the leapfrog integrator is its simplicity, that is volume-preserving, and that it is reversible, due to its symmetry, by simply negating $\textbf{z}$, in order to generate a valid Metropolis proposal. See Neal \citep{neal2011mcmc} and Betancourt \citep{betancourt2017conceptual} for more details on energy-conservation, reversibility and volume-preserving integrators and their connections to HMCMC. It is noted that in the above leapfrog integration algorithm, the computationally expensive part is the one model call per step to acquire the $\dfrac{\partial U}{\partial \boldsymbol{\theta}}$ term. With $\tau$ the trajectory or else path length, taking $\textit{L}=\tau/\varepsilon$ leapfrog steps approximates the evolution $(\boldsymbol{\theta}(0),\textbf{z}(0)) \longrightarrow (\boldsymbol{\theta}(\tau),\textbf{z}(\tau))$, providing the exact solution in the limit $\varepsilon \longrightarrow 0$.\par 

\begin{algorithm}[t!]
\caption{Hamiltonian Markov Chain Monte Carlo}\label{alg:stdHMCMC}
\begin{algorithmic}[1]
\Procedure{HMCMC}{$\boldsymbol{\theta}^{0}$, $\varepsilon$, \textit{L}, $\mathcal{L}(\boldsymbol{\theta})$, \textit{$N_{Iter}$}}
\For{\texttt{$m=1$ $to$ $N_{Iter}$}}
\State $\textbf{z}^{0}$$\sim$$\textbf{N}(\textbf{0},\textbf{I})$\Comment{momentum sampling from standard normal distribution}
\State $\boldsymbol{\theta}^{m}$ $\gets$ $\boldsymbol{\theta}^{m-1}$, $\tilde{\boldsymbol{\theta}}$ $\gets$ $\boldsymbol{\theta}^{m-1}$, $\tilde{\textbf{z}}$ $\gets$ $\textbf{z}^{0}$
\For{\texttt{$i=1$ $to$ $L$}}
\State $\tilde{\boldsymbol{\theta}}$, $\tilde{\textbf{z}}$ $\gets$ Leapfrog($\tilde{\boldsymbol{\theta}}$, $\tilde{\textbf{z}}$, $\varepsilon$) \Comment{leapfrog integration}
\EndFor\label{HMCMCfor2}
\State $with$ $probability$:\\ 
       \hspace{1cm} 
       $\alpha$ = min$\bigg\{$1,$\dfrac{\exp(\mathcal{L}(\tilde{\boldsymbol{\theta}})-\dfrac{1}{2} \tilde{\textbf{z}}^{T}\tilde{\textbf{z}})}{\exp(\mathcal{L}(\boldsymbol{\theta}^{m-1})-\dfrac{1}{2}{\textbf{z}^{0}}^{T}\textbf{z}^{0})}$
       $\bigg\}$ \Comment{Metropolis step}\\
       \hspace{1cm} $\boldsymbol{\theta}^{m}$ $\gets$ $\tilde{\boldsymbol{\theta}}$, $\textbf{z}^{m}$ $\gets$ -$\tilde{\textbf{z}}$
\EndFor\label{HMCMCfor}
\EndProcedure
\end{algorithmic}
\end{algorithm}
 
Typically, a simple Gaussian momentum is used for the Hamiltonian, $\pi_{Z \arrowvert\Theta}(\textbf{z} \arrowvert \boldsymbol{\theta}) = \pi_{Z}(\textbf{z}) = \textbf{N}(\textbf{0},\textbf{M})$ (or $\textbf{z} \sim \textbf{N}(\textbf{0},\textbf{M})$) and the mass matrix $\textbf{M}$ is often set to the identity matrix, \textbf{I}. A generic procedure for drawing \textit{$N_{Iter}$} samples via HMCMC is described in \Cref{alg:stdHMCMC}, where again $\mathcal{L}(\boldsymbol{\theta})$ is the log-density of the target distribution of interest, $\boldsymbol{\theta}^{0}$ are initial values, and \textit{L} is the number of leapfrog steps, as explained before. For each HMCMC step, the momentum is first resampled and then the \textit{L} leapfrog updates are performed, as seen in \cref{eq:11}, before a typical accept/reject MCMC Metropolis step takes place.
\subsection{HMCMC parameters}\label{sec3.2}
The HMCMC performance and efficiency is well known to rely on selecting suitable values for the $\varepsilon$ and \textit{L} parameters. For a fixed trajectory length $\tau$, the stepsize $\varepsilon$ balances the trade-off between accuracy and computational cost. In this work, we select the stepsize $\varepsilon$ in such a way so that the corresponding average acceptance rate is approximately 65$\%$, as values between 60$\%$ and 80$\%$ are typically assumed optimal \citep{neal2011mcmc,beskos2013optimal,hoffman2014no}. The \textit{dual averaging} algorithm of Hoffman and Gelman \citep{hoffman2014no} is adopted to perform this task, used here only in the burn-in phase, to tune $\varepsilon$. To determine the number of leapfrog steps, $L$, we estimate an appropriate to use trajectory length $\tau$ based on a few simulation runs, so as to have a sufficiently high so called normalized \textit{Expected Square Jumping Distance} (\textit{ESJD}), $\tau^{-1/2}\mathop{\mathbb{E}}$\(\lVert\theta^{(m+1)} (\tau) - \theta^{(m)} (\tau)\rVert\)$^{2}$, as introduced in \citep{wang2013adaptive}, and then we randomly perturb each trajectory length $\tau^{(m)}$ in the range $[0.9\tau,1.1\tau]$ to further avoid periodicity ($m$ denotes the $m$-th iteration of HMCMC). In all our numerical experiments herein, we determine $L$ and $\tau$ in this manner, as we have found it to work well in practice. The role of these parameters and techniques for determining them have been quite extensively studied in the literature and for more details we refer to \citep{neal2011mcmc,beskos2013optimal,hoffman2014no}.
\section{Quasi-Newton mass preconditioned HMCMC (QNp-HMCMC)}\label{section3}
In complex high-dimensional problems, the performance of the typical HMCMC sampler, presented as \Cref{alg:stdHMCMC}, may deteriorate and a prohibitive number of model calls could be required. A variety of methods have been proposed in the literature to address this issue. Among others, a Riemannian Manifold Hamiltonian Monte Carlo (RMHMC) has been suggested in \citep{girolami2011riemann} that takes advantage of the manifold structure of the variable space, at the cost of calculating second- and third-order derivatives of distributions and using a generalized leapfrog scheme, requiring additional model calls per leapfrog step. Although possible in some cases regarding second-order derivatives, e.g. \citep{andriotis2018nonlinear}, in the majority of cases higher-order derivatives are not provided by computational models, such as finite element models. In addition, the computational cost still increases importantly and extra model calls per leapfrog step are usually restrictive for computationally expensive models. \par
In this work, we instead address the high-dimensionality performance issue in a different, Newton-type context, without needing additional model calls per leapfrog step and by only using the Hessian information of the target distribution, either the exact one, when relevant information is provided freely by the computational model, or, even more general, an approximate one that does not increase the computational cost. An approximate Hessian can be given in a systematic manner based on already available gradient information, similar to Quasi-Newton methods used in nonlinear programming \citep{nocedal2006numerical}. The well-known BFGS approximation \citep{nocedal2006numerical} is thus utilized for our Quasi-Newton type Hamiltonian MCMC approach and all numerical examples in this work are analyzed accordingly, based solely on this most general approximate Hessian case. Let $\boldsymbol{\theta}\in\mathbb{R}^{d}$, consistent with the previous sections. Given the $k$-$th$ estimate $\textbf{W}_{k}$, where $\textbf{W}_{k}$ is an approximation to the inverse Hessian at $\boldsymbol{\theta}_{k}$, the BFGS update $\textbf{W}_{k+1}$ can be expressed as:
\begin{align}
\textbf{W}_{k+1} = (\textbf{I}-\dfrac{\textbf{\textit{s}}_{k} \textbf{\textit{y}}_{k}^{T}}{\textbf{\textit{y}}_{k}^{T} \textbf{\textit{s}}_{k}})\textbf{W}_{k}(\textbf{I}-\dfrac{\textbf{\textit{y}}_{k} \textbf{\textit{s}}_{k}^{T}}{\textbf{\textit{y}}_{k}^{T} \textbf{\textit{s}}_{k}})+\dfrac{\textbf{\textit{s}}_{k} \textbf{\textit{s}}_{k}^{T}}{ \textbf{\textit{y}}_{k}^{T}  \textbf{\textit{s}}_{k}} \label{eq:14}
\end{align}
where \textbf{I} is the identity matrix, $\textbf{\textit{s}}_{k}=\boldsymbol{\theta}_{k+1}-\boldsymbol{\theta}_{k}$, and $\textbf{\textit{y}}_{k}= -\nabla \mathcal{L}(\boldsymbol{\theta}_{k+1})+\nabla \mathcal{L}(\boldsymbol{\theta}_{k})$ where $\mathcal{L}:\mathbb{R}^{d} \longrightarrow\mathbb{R}$ denotes the log density of the target distribution, as before. There is a long history of efficient BFGS updates for very large systems and several numerical techniques can be used, including sparse and limited-memory approaches. \par
Two relevant studies in the literature on Quasi-Newton extensions and connections to MCMC algorithms can be found in \citep{zhang2011quasi,fu2016quasi}. Our developed method, however, has fundamental differences that are summarized in that we are focusing on Hamiltonian methods, we are using two integrated coupled phases, an adaptive and a non-adaptive, and finally we consistently incorporate the Quasi-Newton outcomes in both stages of momentum sampling and simulation of Hamiltonian dynamics. In more detail, in the adaptive burn-in phase of the algorithm we are still sampling the momentum from an identity mass matrix, $\textbf{M}=\textbf{I}$, but the ODEs of \cref{eq:10} now become:    
\begin{equation}
\dot{\boldsymbol{\theta}}=\textbf{W}\textbf{M}^{-1}\textbf{z},\ \ \  \ 
\dot{\textbf{z}}=\textbf{W}\nabla_{\theta} \mathcal{L}(\boldsymbol{\theta}) \label{eqQN}
\end{equation}
which is equivalent to the implicit linear transformation $\boldsymbol{\theta}'=\textbf{W}\boldsymbol{\theta}$, and $\textbf{W}$ is given by \cref{eq:14}. Accordingly, the leapfrog integrator is then reformulated as:
\begin{equation} 
\textbf{z}\textsubscript{t+$\varepsilon$/2} = \textbf{z}\textsubscript{t} + (\dfrac{\varepsilon}{2}) \textbf{W}\nabla_{\boldsymbol{\theta}}\mathcal{L}(\boldsymbol{\theta}\textsubscript{t}) ,\ \ \  \
\boldsymbol{\theta}\textsubscript{t+$\varepsilon$} = \boldsymbol{\theta}\textsubscript{t} +\varepsilon\textbf{W}\textbf{z}\textsubscript{t+$\varepsilon$/2} ,\ \ \  \
\textbf{z}\textsubscript{t+$\varepsilon$} = \textbf{z}\textsubscript{t+$\varepsilon$/2} + (\dfrac{\varepsilon}{2})\textbf{W}\nabla_{\boldsymbol{\theta}}\mathcal{L}(\boldsymbol{\theta}\textsubscript{t+$\varepsilon$}) 
\end{equation} 
Hence, this new dynamic scheme efficiently and compatibly adjusts both the $\textbf{z}$ and $\boldsymbol{\theta}$ evolutions based on the local structure of the target distribution, and also features a Quasi-Newton direction for the momentum variables, allowing large jumps across the state space. The final estimation of the approximated inverse Hessian matrix, $\textbf{W}$, in the adaptive burn-in phase is then used in the subsequent non-adaptive phase of the algorithm as a preconditioned mass (inverse covariance) matrix, $\textbf{M}=\textbf{W}^{-1}$, used to sample the Gaussian momentum $\textbf{z} \sim \textbf{N}(\textbf{0},\textbf{M})$. As such, typical Hamiltonian dynamics are now used, albeit with this properly constructed mass matrix that takes into account the scale and correlations of the position variables, leading to significant efficiency gains, particularly in high-dimensional problems. The BFGS procedure in \cref{eq:14} normally provides a symmetric, positive-definite $\textbf{W}$ matrix in an optimization context. However, in our case we are using BFGS under different settings that may not satisfy the curvature condition $\textbf{\textit{s}}_{k}^{T} \textbf{\textit{y}}_{k} > 0$, resulting in occasional deviations from positive-definiteness. Several standard techniques can be then implemented to ensure positive-definiteness, such as a damped BFGS updating \citep{nocedal2006numerical} or the simple addition $\textbf{W}_{new} = \textbf{W}_{old} + \delta \textbf{I}$, where $\delta\geq 0$ is some appropriate number. A straightforward method to determine $\delta$ is to choose it larger than the absolute value of the minimum eigenvalue of $\textbf{W}_{old}$. Another technique involves utilizing a semidefinite programming approach to identify an optimized diagonal matrix to add to $\textbf{W}_{old}$. Alternatively, $\textbf{W}$ can be updated only when the curvature condition is satisfied, which directly guarantees positive definiteness. To further ensure the stability of the sampler, a positive threshold can be introduced to the curvature condition instead of zero, e.g., $\textbf{\textit{s}}_{k}^{T} \textbf{\textit{y}}_{k} > 10$. This latter approach has been used and has worked well in this work. Since the final estimation of $\textbf{W}$ in the adaptive burn-in phase is then extensively utilized in the subsequent non-adaptive phase, we suggest use of a directly provided positive-definite matrix $\textbf{W}$ at this step. This can simply be accomplished by adding one more burn-in iteration step at the end of the burn-in phase, until an appropriate sample, directly supported by a positive definite $\textbf{W}$ matrix is identified.\par

Our derived \textit{Quasi-Newton mass preconditioned Hamiltonian Markov Chain Monte Carlo} (QNp-HMCMC) method is concisely summarized and presented in \Cref{QNHMCMC}. Overall, QNp-HMCMC is a practical, efficient approach that only requires already available gradient information and provides important insight about the geometry of the target distribution, eventually improving computational performance and enabling faster mixing.

\section{Approximate Sampling Target with Post-processing Adjustment (ASTPA)}\label{section4}
\begin{figure}[t!]
 \centering
  \begin{tabular}{c ccccc}
   \includegraphics[trim=0cm 0cm 0cm 3.5cm,width=.3\textwidth,keepaspectratio]{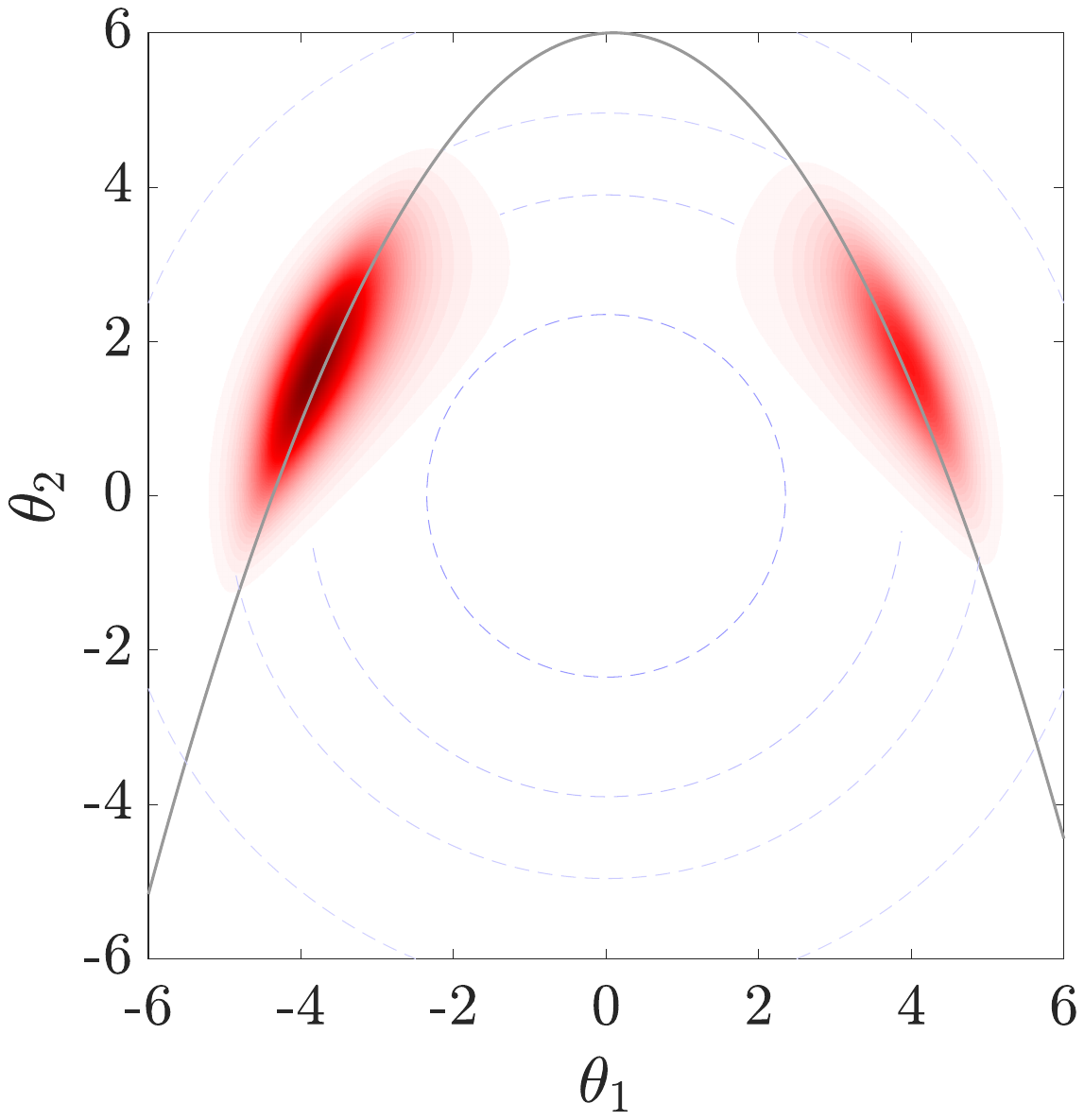}&
   \includegraphics[trim=0cm 0cm 0cm 3.5cm,width=.3\textwidth,keepaspectratio]{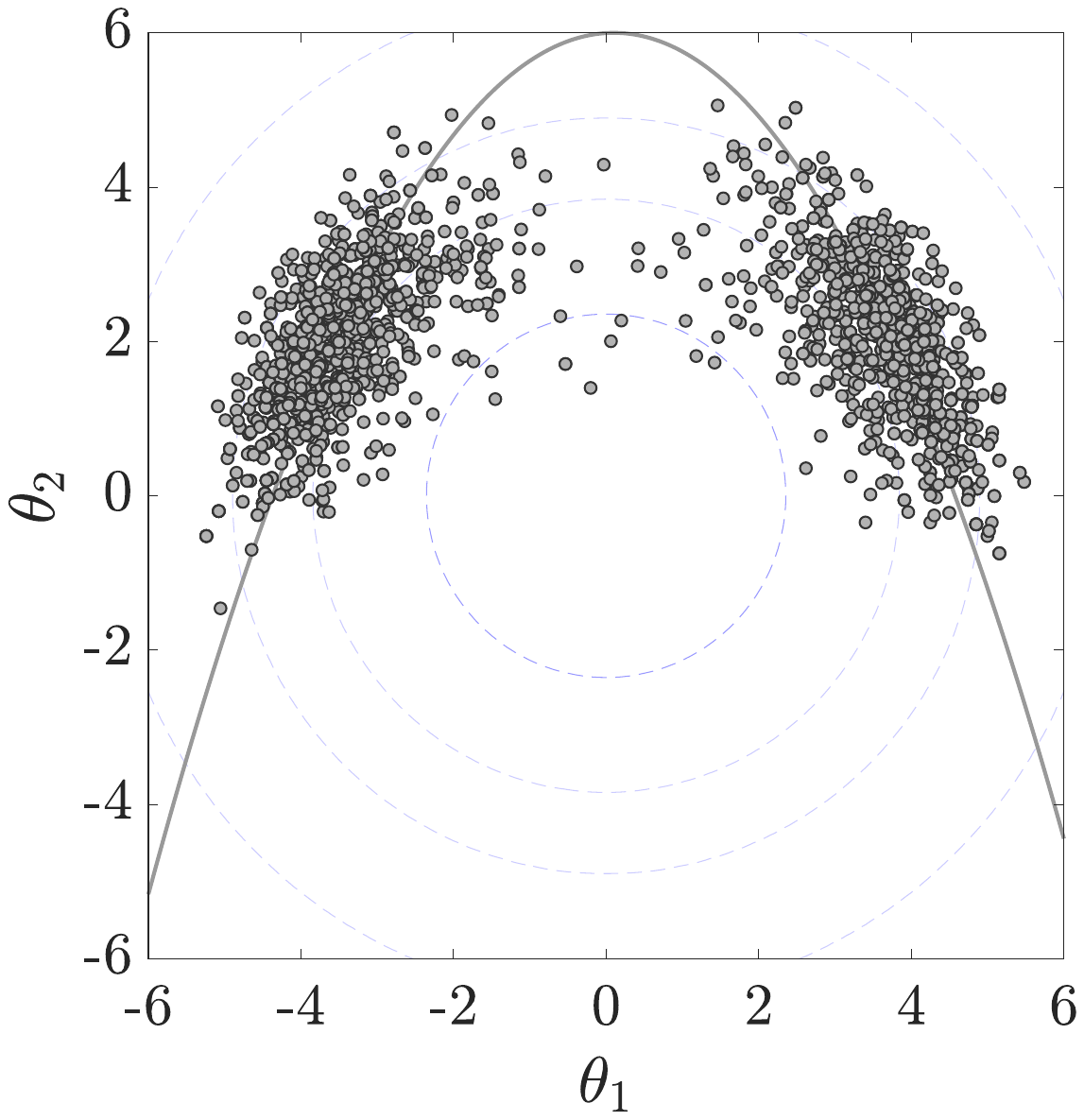}&
    \includegraphics[trim=0cm 0cm 0cm 3.5cm,width=.3\textwidth,keepaspectratio]{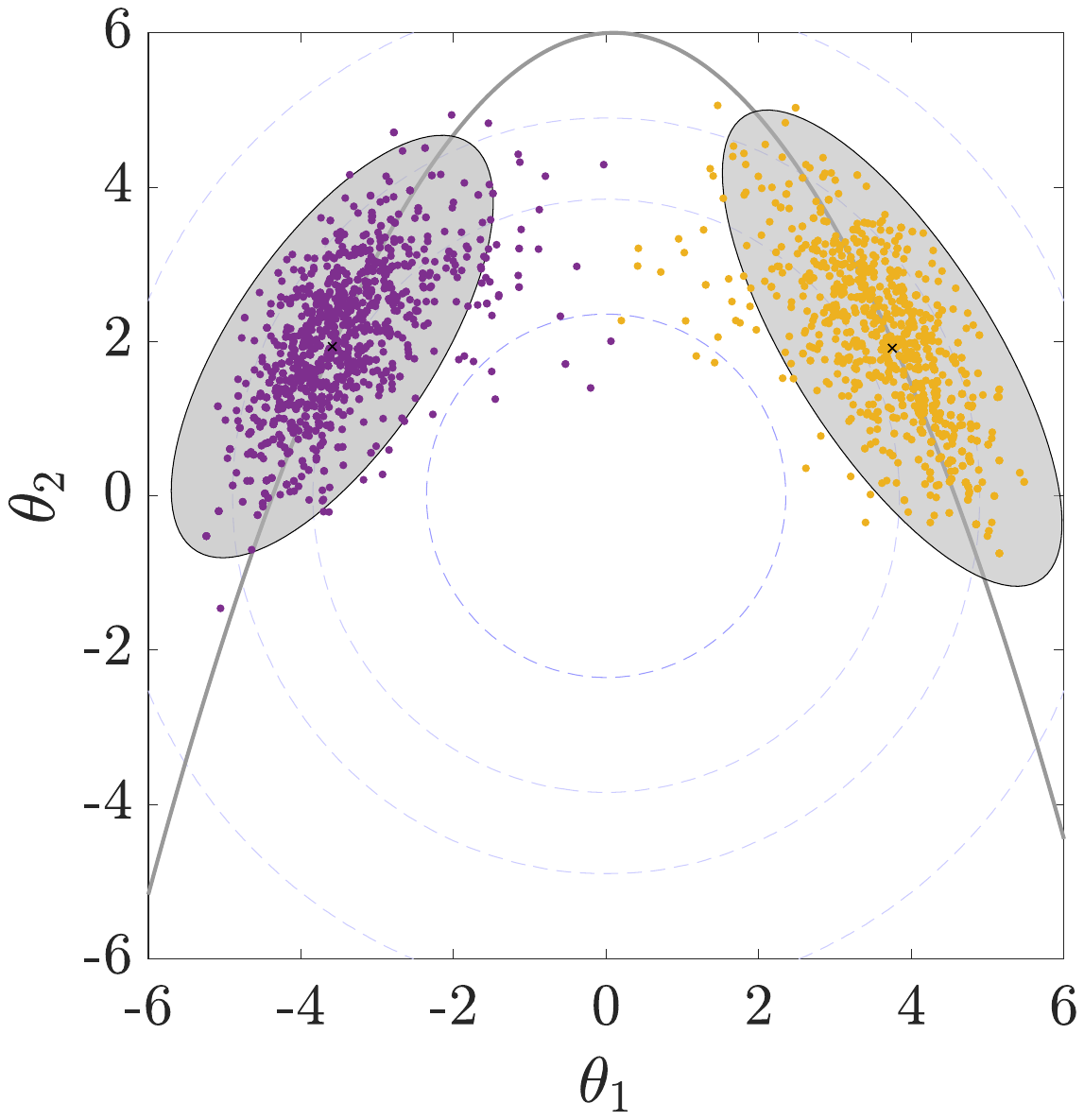}
    \vspace*{-0.3in}
\\\hline\noalign{\smallskip}
 \end{tabular}
 \caption{An analytical target distribution, simulated target distribution samples based on our QNp-HMCMC method, and a fitted Gaussian Mixture Model describing the simulated samples, from left to right plot, respectively.}
 \label{fig2}
\end{figure}

In order for an appropriate number of samples to discover and enter the relevant regions, contributing to the rare event probability estimation, a suitable approximate target distribution is constructed in this work, as analyzed in \Cref{sec4.1}, then sampled by Hamiltonian MCMC methods that can effectively reach regions of interest. For their initial stage, our HMCMC samplers have an adaptive annealed phase. This adaptive phase will be thoroughly explained later in \Cref{burnin}. To estimate the pertinent probability, \cref{eq:1} needs to be then adjusted accordingly, since the samples are sampled from our constructed target distribution and not $\pi_{\Theta}(\boldsymbol{\theta})$. An original post-sampling step is devised at this stage, using an inverse importance sampling procedure, described in \Cref{IIS}, i.e. first sample, then choose the importance sampling density automatically, based on the samples. 
\cref{fig2} concisely portrays the overall approach by using a bimodal target distribution with $P_{F}\sim 3.95\times10^{-5}$. The gray curves represent the parabolic limit-state function $g(\boldsymbol{\theta})$ of this problem, with the failure domain being outside $g(\boldsymbol{\theta})$. The left figure displays the constructed target distribution, which in this simple 2D case can be visualized. The middle figure shows drawn samples from the target distribution by our suggested QNp-HMCMC algorithm, described in \Cref{section3}, and the right figure demonstrates the inverse importance sampling step. All these different steps will be discussed in detail in the following sections.

\subsection{Target distribution formulation}\label{sec4.1}
The basic idea is to construct an approximate sampling target distribution that places higher importance to the failure regions, to efficiently guide the samples to these domains of interest, and then the probability of failure can be quantified using an inverse importance sampling procedure. \cref{eq:1} can be computed by directly sampling $I_{F}$ $(\boldsymbol{\theta})$ $\pi_{\Theta}(\boldsymbol{\theta})$. However, this direct approach cannot be practically and effectively used in most general cases since the support domain of the non-smooth indicator function $I_{F} (\boldsymbol{\theta})$ is only the failure regions (\boldsymbol{$\theta$} $\in$ g(\textbf{\boldsymbol{$\theta$}})$\leq$ 0), making it challenging of locating and adequately sampling the failure domains, especially in cases of high-dimensional and multi-modal spaces. In this work, the indicator function is hence approximated by a one-dimensional output likelihood function, that is based on the limit-state expression $g(\boldsymbol{\theta})$, supporting the entire domain $\Theta$. This likelihood function, $\ell_{g_{\boldsymbol\theta}}$, is expressed as a \textit{logistic} cumulative distribution function, $F_{cdf}$, with, mean, $\mu_{g}$, and a dispersion factor $\sigma$, as:

\begin{algorithm}[t!]
\caption{Quasi-Newton mass preconditioned Hamiltonian Markov Chain Monte Carlo (QNp-HMCMC)}\label{QNHMCMC}
\begin{algorithmic}[1]
\Procedure{QNp-HMCMC}{$\boldsymbol{\theta}^{0}$, $\varepsilon$, \textit{L}, $\mathcal{L}(\boldsymbol{\theta})$, \textit{BurnIn}, \textit{$N_{Iter}$}}\\
 \hspace{0.5cm}\textbf{W} = \textbf{I}
\For{\texttt{$m=1$ $to$ $N_{Iter}$}}
\If {$m$ $\leq$ $BurnIn$}
\State $\textbf{z}^{0}$$\sim$$\textbf{N}(\textbf{0},\textbf{M})$\Comment{where $\textbf{M}=\textbf{I}$}
\State $\boldsymbol{\theta}^{m}$ $\gets$ $\boldsymbol{\theta}^{m-1}$, $\tilde{\boldsymbol{\theta}}$ $\gets$ $\boldsymbol{\theta}^{m-1}$, $\tilde{\textbf{z}}$ $\gets$ $\textbf{z}^{0}$, $\textbf{B}$ $\gets$ $\textbf{W}$
\For{\texttt{$i=1$ $to$ $L$}}
\State $\tilde{\boldsymbol{\theta}}$, $\tilde{\textbf{z}}$ $\gets$ Leapfrog-BurnIn($\tilde{\boldsymbol{\theta}}$, $\tilde{\textbf{z}}$, $\varepsilon$, $\textbf{B}$)\\
\hspace{2cm} Update \textbf{W} using \cref{eq:14}
\EndFor
\State $with$ $probability$:\\ 
       \hspace{1.5cm}$\alpha$ = min$\bigg\{$1,$\dfrac{\exp(\mathcal{L}(\tilde{\boldsymbol{\theta}})-\dfrac{1}{2} \tilde{\textbf{z}}^{T}\tilde{\textbf{z}})}{\exp(\mathcal{L}(\boldsymbol{\theta}^{m-1})-\dfrac{1}{2}{\textbf{z}^{0}}^{T}\textbf{z}^{0})}$
       $\bigg\}$\\
       \hspace{1.5cm} $\boldsymbol{\theta}^{m}$ $\gets$ $\tilde{\boldsymbol{\theta}}$, $\textbf{z}^{m}$ $\gets$ -$\tilde{\textbf{z}}$\Comment{If proposal rejected: $\textbf{W}$ $\gets$ $\textbf{B}$}
\Else \Comment{If $m$ $>$ $BurnIn$}      
\State $\textbf{z}^{0}$$\sim$$\textbf{N}(\textbf{0},\textbf{M})$\Comment{where $\textbf{M}=\boldsymbol{\Sigma}^{-1}=\textbf{W}^{-1}$}
\State $\boldsymbol{\theta}^{m}$ $\gets$ $\boldsymbol{\theta}^{m-1}$, $\tilde{\boldsymbol{\theta}}$ $\gets$ $\boldsymbol{\theta}^{m-1}$, $\tilde{\textbf{z}}$ $\gets$ $\textbf{z}^{0}$
\For{\texttt{$i=1$ $to$ $L$}}
\State $\tilde{\boldsymbol{\theta}}$, $\tilde{\textbf{z}}$ $\gets$ Leapfrog($\tilde{\boldsymbol{\theta}}$, $\tilde{\textbf{z}}$, $\varepsilon$, $\textbf{M}$)
\EndFor
\State $with$ $probability$:\\ 

       \hspace{1.5cm}$\alpha$ = min$\bigg\{$1,$\dfrac{\exp(\mathcal{L}(\tilde{\boldsymbol{\theta}})-\dfrac{1}{2} \tilde{\textbf{z}}^{T}\ \textbf{M}^{-1}\tilde{\textbf{z}})}{\exp(\mathcal{L}(\boldsymbol{\theta}^{m-1})-\dfrac{1}{2}{\textbf{z}^{0}}^{T}\ \textbf{M}^{-1}\textbf{z}^{0})}$
       $\bigg\}$\\
       \hspace{1.5cm} $\boldsymbol{\theta}^{m}$ $\gets$ $\tilde{\boldsymbol{\theta}}$, $\textbf{z}^{m}$ $\gets$ -$\tilde{\textbf{z}}$
\EndIf  
\EndFor
\EndProcedure
\\
\\
\\
\\
\\
\\
\Function {Leapfrog-BurnIn}{$\boldsymbol{\theta}, \textbf{z}, \varepsilon, \textbf{B}$}
\State $\tilde{\textbf{z}} \gets \textbf{z}+(\varepsilon/2)\textbf{B}\nabla_{\boldsymbol{\theta}}\mathcal{L}(\boldsymbol{\theta})$
\State $\tilde{\boldsymbol{\theta}} \gets \boldsymbol{\theta}+\varepsilon\textbf{B}\tilde{\textbf{z}}$
\State $\tilde{\textbf{z}} \gets \tilde{\textbf{z}}+(\varepsilon/2)\textbf{B}\nabla_{\boldsymbol{\theta}}\mathcal{L}(\tilde{\boldsymbol{\theta}})$\\
\Return $\tilde{\boldsymbol{\theta}}$, $\tilde{\textbf{z}}$ 
\EndFunction
\\
\\
\Function {Leapfrog}{$\boldsymbol{\theta}, \textbf{z}, \varepsilon, \textbf{M}$}
\State $\tilde{\textbf{z}} \gets \textbf{z}+(\varepsilon/2)\nabla_{\boldsymbol{\theta}}\mathcal{L}(\boldsymbol{\theta})$
\State $\tilde{\boldsymbol{\theta}} \gets \boldsymbol{\theta}+\varepsilon\textbf{M}^{-1}\tilde{\textbf{z}}$
\State $\tilde{\textbf{z}} \gets \tilde{\textbf{z}}+(\varepsilon/2)\nabla_{\boldsymbol{\theta}}\mathcal{L}(\tilde{\boldsymbol{\theta}})$\\
\Return $\tilde{\boldsymbol{\theta}}$, $\tilde{\textbf{z}}$ 
\EndFunction
\end{algorithmic}
\end{algorithm}

\begin{equation}
\begin{aligned}
\ell_{g_{\boldsymbol\theta}} =  \, F_{cdf}\bigg  (&\dfrac{-g(\boldsymbol\theta)}{g_{c}}\bigg\arrowvert\ \mu_{g} , \ \sigma\bigg ) =  \, \dfrac{1}{\Bigg (1 + e^{\big (\dfrac{(\frac{g(\boldsymbol\theta)}{g_{c}})+\mu_{g}}{(\frac{\sqrt{3}}{\pi})\sigma} \big )} \Bigg )} 
\label{eqq8}
\end{aligned}
\end{equation}

where both $\mu_{g}$, $\sigma$, parameters are described in detail in \Cref{sec4.2}, $g(\boldsymbol\theta)$ is the limit-state function, and $g_{c}$ is a scaling constant. A similar approach to approximate the indicator function can be seen in \citep{papaioannou2016sequential,papaioannou2019improved}, whereas it was used in a completely different context, in order to quantify the probability of failure using a sequential importance sampling method. The non-normalized target PDF is then defined as:
\begin{align}
\tilde{h}(\boldsymbol{\theta}) = \ell_{g_{\boldsymbol\theta}} \, \pi_{\Theta}(\boldsymbol{\theta})= \, F_{cdf}\bigg (\dfrac{-g(\boldsymbol\theta)}{g_{c}}\bigg\arrowvert\ \mu_{g} , \ \sigma\bigg )\, \pi_{\Theta}(\boldsymbol{\theta}) \label{Tar}
\end{align}
and combining \cref{eqq8,Tar}, the $\tilde{h}(\boldsymbol{\theta})$ is finally expanded as:
\begin{align}
\tilde{h}(\boldsymbol{\theta}) = \Bigg (1 + e^{\big (\dfrac{(\frac{g(\boldsymbol\theta)}{g_{c}})+\mu_{g}}{(\frac{\sqrt{3}}{\pi})\sigma} \big )} \Bigg )^{-1} \pi_{\Theta}(\boldsymbol{\theta}) \label{finalTar}
\end{align}

where $\pi_{\Theta}(.)$ denotes the multivariate standard normal distribution PDF in this work, $\pi_{\Theta}(\boldsymbol{\theta}) = \textbf{N}(\textbf{0},\textbf{I})$, describing the multidimensional variable space $\boldsymbol{\Theta}$. This approximate target distribution $\tilde{h}(\boldsymbol{\theta})$ is smooth and supports both the safe and failure domains, emphasizing on the failure one, and can be sampled efficiently, particularly by the gradient-based HMCMC samplers which can take informed
large jumps across the state space, exploring all regions of interest. The $\mathcal{L}(\boldsymbol{\theta})$ in \cref{section2,section3}, is thus the logarithmic form of the $\tilde{h}(\boldsymbol{\theta})$ distribution, $\mathcal{L}(\boldsymbol{\theta})=\ln(\tilde{h}(\boldsymbol{\theta}))$.

 The reason for the scaling $g(\boldsymbol\theta)/g_{c}$ is mainly to properly and generally adjust the influence of the limit-state surface on the whole standard normal space $\boldsymbol{\Theta}$, and to standardize parameter values, aiming at a universal algorithm and settings. \cref{eqq88} defines recommended values for the scaling constant $g_{c}$ and the suggested $q$ range $[3 \, \,5]$ was found to work well in practice. Nonetheless, it is worth mentioning that these values may need further investigation when the proposed framework is directly used in non-Gaussian spaces.

\begin{equation}
\begin{aligned}
&g_{c}=
\begin{cases}
\frac{g(\textbf{0})}{q}, q \in [3 \, \,5]& \text{if}  \, \bigg ( \big (g(\textbf{0})>7 \big ) \bigcup  \big (0<g(\textbf{0})< 2 \big ) \bigg ) \\
1,    & \text{otherwise}
\end{cases}\label{eqq88}
\end{aligned}
\end{equation}

\begin{figure}[t!]
 \centering
  \begin{tabular}{ccccc}
   \hspace*{-.15in}
   \includegraphics[trim=0cm 0cm 0cm 3.5cm,width=.255\textwidth,keepaspectratio]{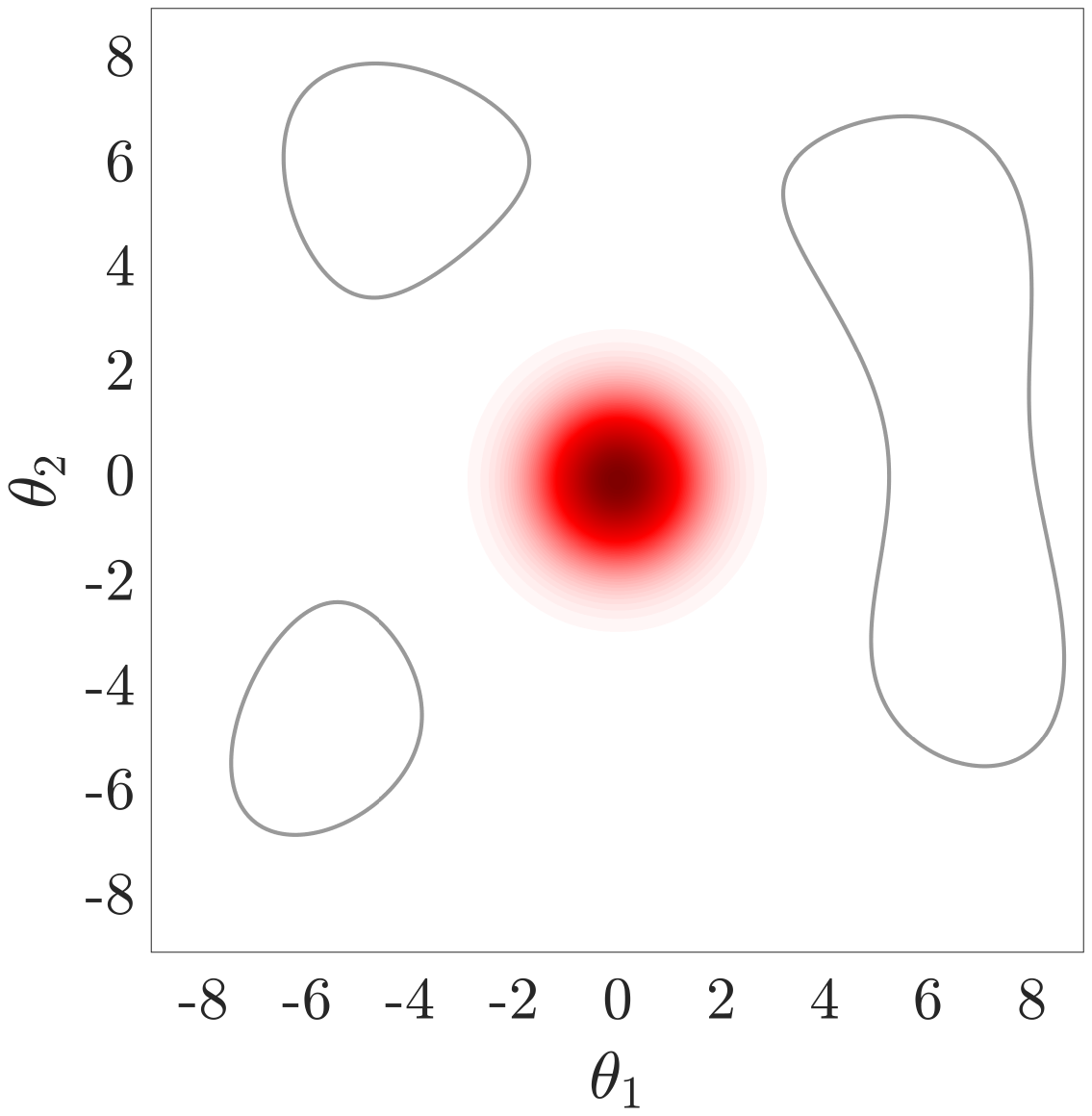}&
   \hspace*{-0.25in}
   \includegraphics[trim=0cm 0cm 0cm 3.5cm,width=.255\textwidth,keepaspectratio]{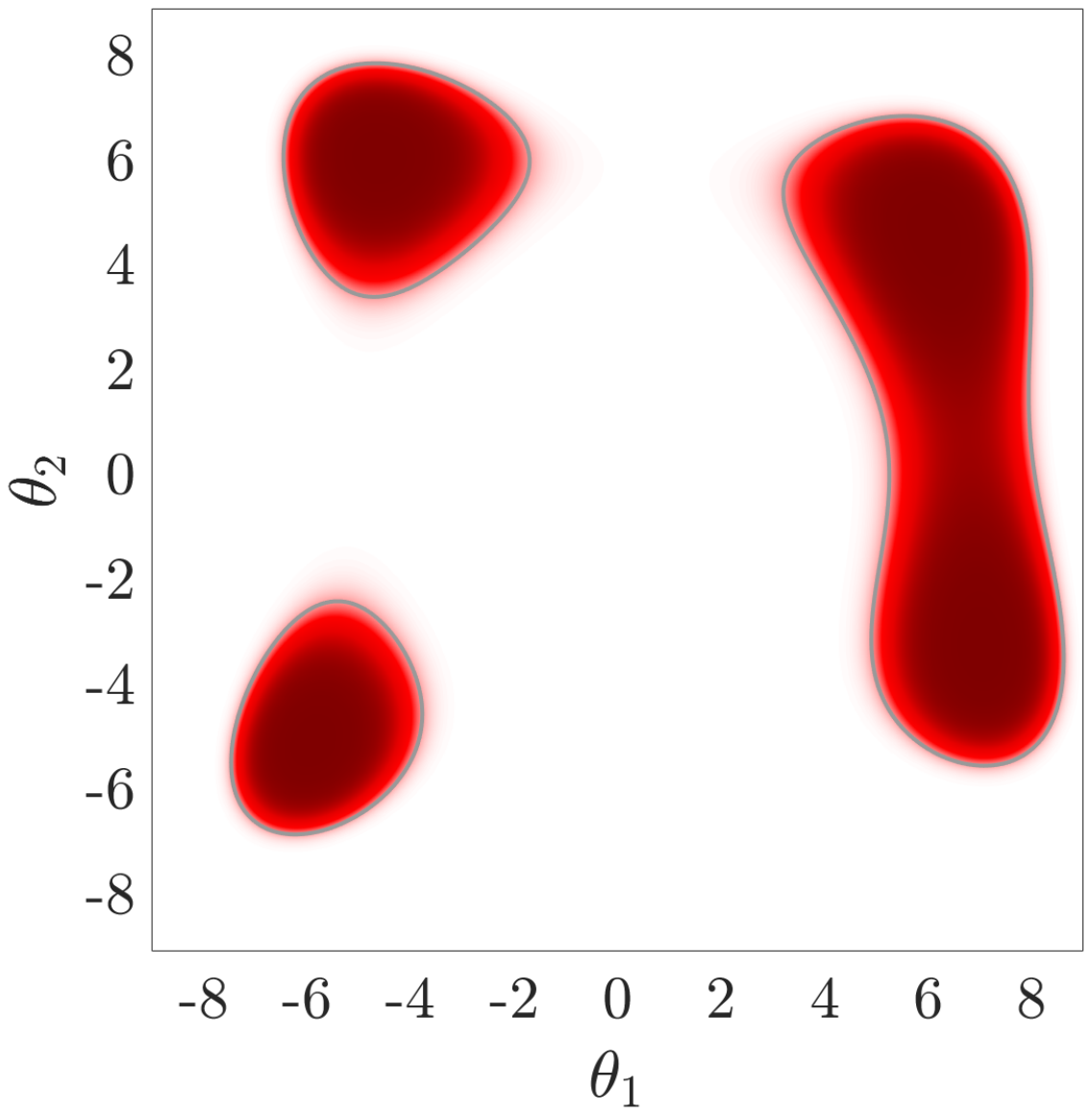}&
   \hspace*{-0.25in}
   \includegraphics[trim=0cm 0cm 0cm 3.5cm,width=.255\textwidth,keepaspectratio]{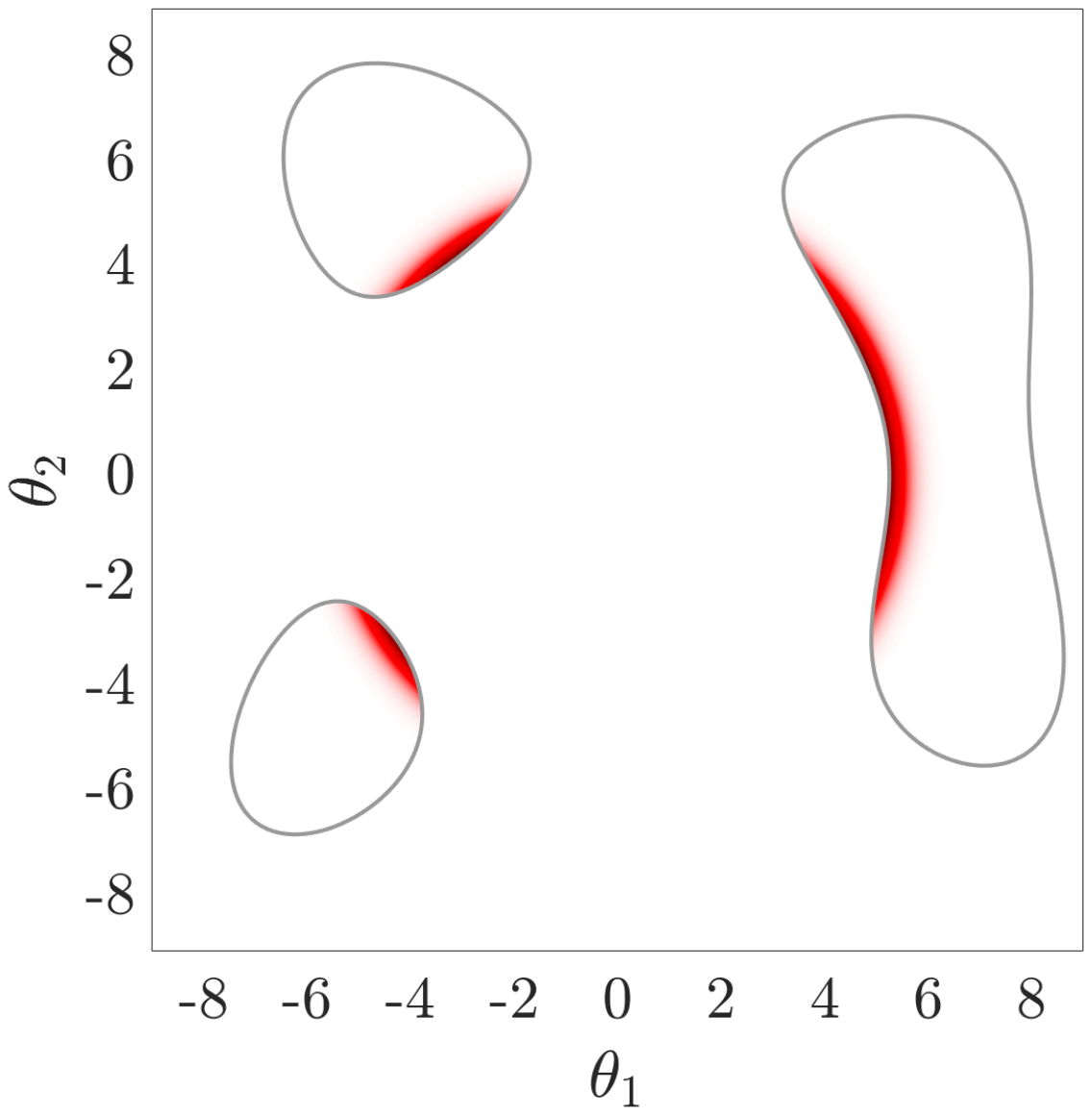}&
   \hspace*{-0.25in}
    \includegraphics[trim=0cm 0cm 0cm 3.5cm,width=.255\textwidth,keepaspectratio]{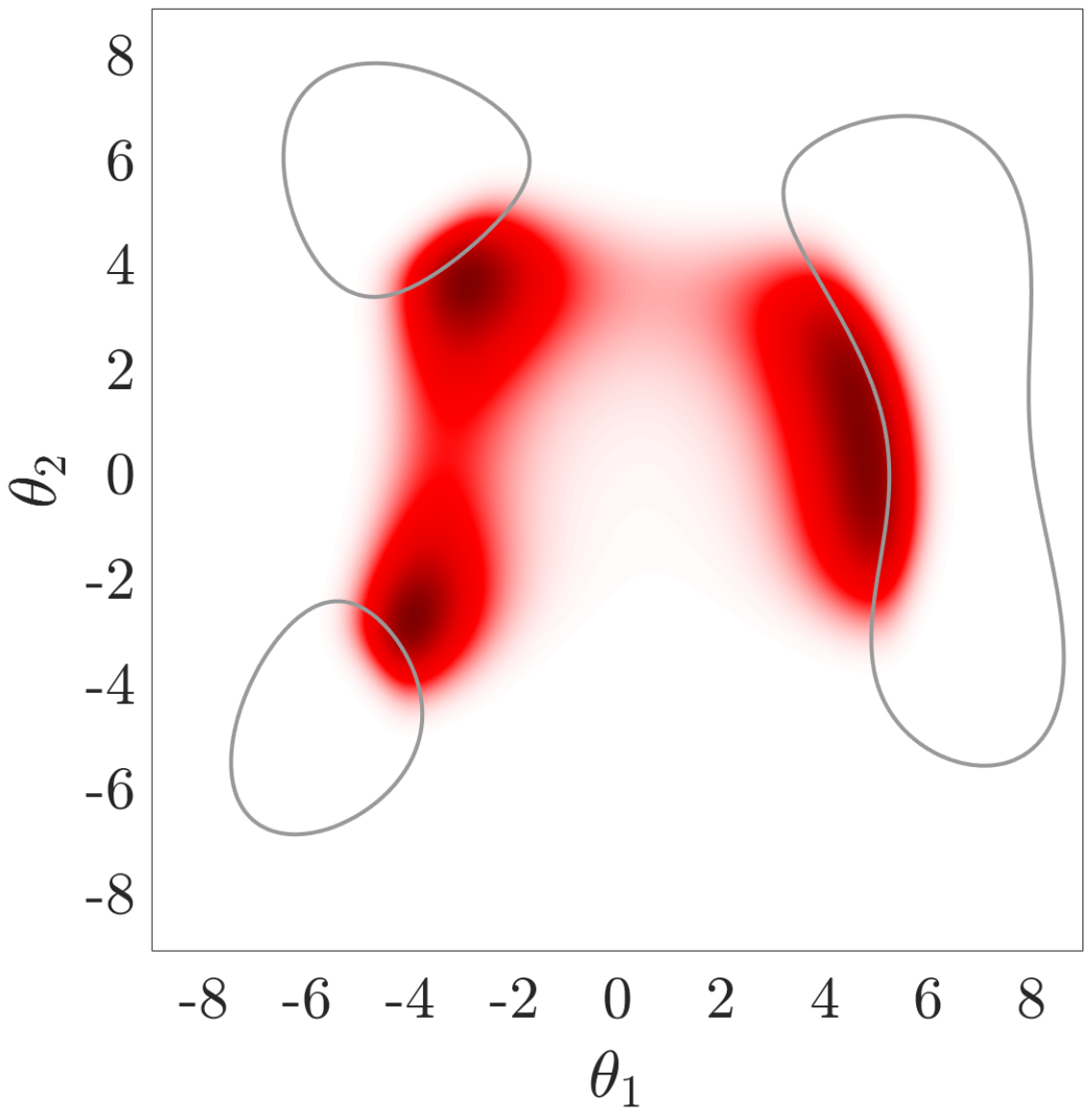}
    \vspace*{-0.25in}\\
    \hspace*{-0.25in}
    $\pi_{\Theta}(\boldsymbol{\theta})$&
    \hspace*{-0.25in}
    $\ell_{g_{\boldsymbol\theta}}; {g_{c}=\dfrac{g(\textbf{0})}{4}} $&
    \hspace*{-0.25in}
    $\tilde{h}(\boldsymbol{\theta}); {g_{c}=\text{1}}$&
    \hspace*{-0.25in}
    $\tilde{h}(\boldsymbol{\theta}); {g_{c}=\dfrac{g(\textbf{0})}{4}}$\\
    \vspace*{-0.1in}\\
    \hspace*{-0.25in}
    (a)&
    \hspace*{-0.25in}
    (b)&
    \hspace*{-0.25in}
   (c)&
    \hspace*{-0.25in}
    (d)\\
   \vspace*{-0.1in}
\\\hline\noalign{\smallskip}
 \end{tabular}
  \captionsetup{labelfont={color=Black}}
 \caption{Demonstrating the concept of constructing a target distribution (\cref{Tar} or \cref{finalTar}) as a product of a 1-D likelihood function and a multi-dimensional variable space; From left to right plot, it represents the variable space $\pi_{\Theta}(\boldsymbol{\theta})$, the likelihood function $\ell_{g_{\boldsymbol\theta}}$, the non-normalized target distribution $\tilde{h}(\boldsymbol{\theta})$ computed without the proper value of $g_{c}$, and the correct non-normalized target distribution $\tilde{h}(\boldsymbol{\theta})$, respectively.}
 \label{figggg2}
\end{figure}

\cref{figggg2} illustrates the described approach in constructing the target distribution $\tilde{h}(\boldsymbol{\theta})$ in \cref{Tar} and \cref{finalTar} under the ASTPA framework, emphasizing also on the effect of the scaling constant $g_{c}$. The grey island-shaped curve is the well-known multi-modal Himmelblau's function (particularly popular in mathematical optimization), seen also in \cref{exHimmelblau}, and it characterizes our limit-state function $g(\boldsymbol{\theta})$, with the failure domain being inside of the curves and $P_{F}\sim 2.77\times10^{-7}$ in this case. \cref{figggg2}(a) illustrates the bivariate normal distribution $\pi_{\Theta}(\boldsymbol{\theta})$ which defines the ($\theta_{1}$, $\theta_{2}$) space. The limit state function here provides $g(\textbf{0})$ $>$ $7$, so the scaling constant $g_{c}$ can be defined as $g(\textbf{0})$ $/$ $4$, following \cref{eqq88}. \cref{figggg2}(b) portrays our prescribed 1-D likelihood function $\ell_{g_{\boldsymbol\theta}}$, described by a logistic cumulative distribution function, as shown in \cref{eqq8}. \cref{figggg2}(c) shows the non-normalized target PDF $\tilde{h}(\boldsymbol{\theta})$ determined as a product of $\pi_{\Theta}(\boldsymbol{\theta})$ and $\ell_{g_{\boldsymbol\theta}}$ without considering however the scaling constant $g_{c}$ here. \cref{figggg2}(d) instead represents the non-normalized target PDF $\tilde{h}(\boldsymbol{\theta})$ constructed with the suggested value of $g_{c}$. These two figures can thus clearly explain the role of the scaling constant $g_{c}$ in attracting samples to the regions of interest.

\subsubsection{Impact of mean (\texorpdfstring{$\mu_{g}$}{mug}) and dispersion factor (\texorpdfstring{$\sigma$}{sigma}) }\label{sec4.2}
\begin{figure}[t!]
 \centering
  \begin{tabular}{c ccccc ccccc cc cc cc cc}
\\\hline\noalign{\smallskip}
   \hspace*{-0.2in}
   \includegraphics[width=.245\textwidth,keepaspectratio]{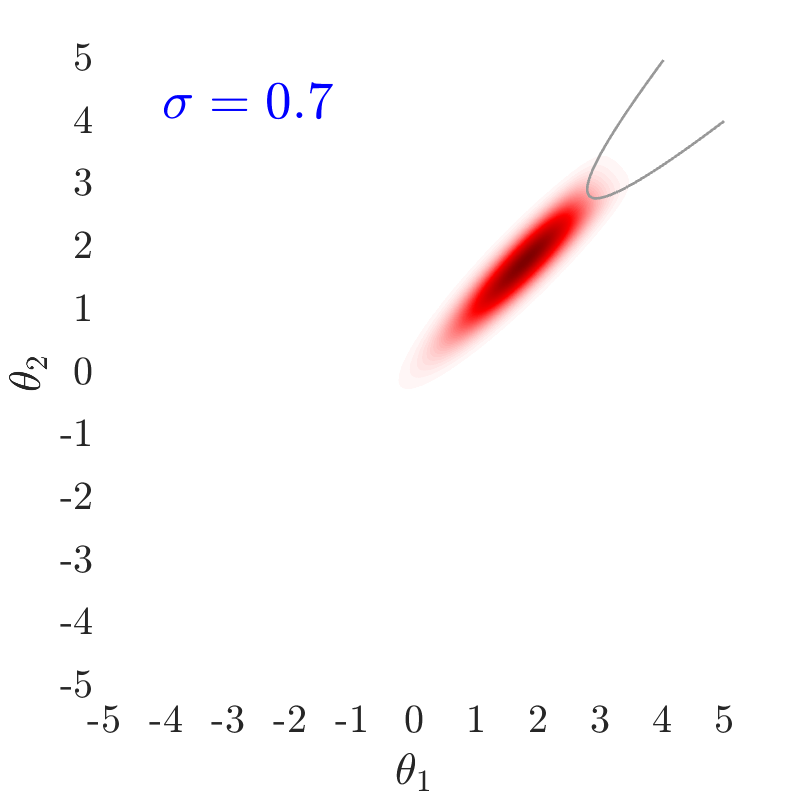}& \hspace*{-0.2in} \includegraphics[width=.245\textwidth,keepaspectratio]{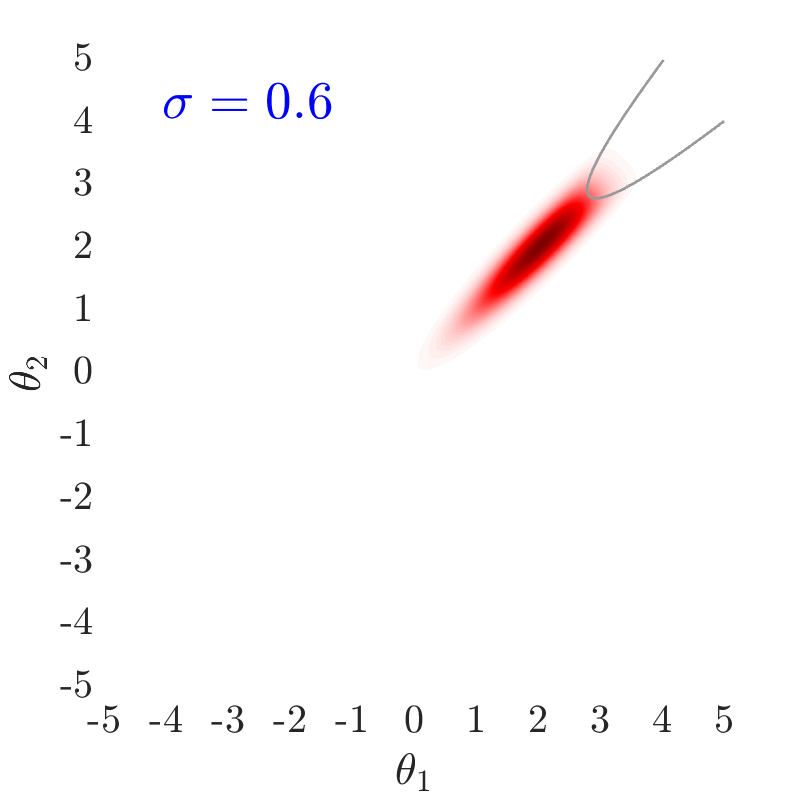}& \hspace*{-0.2in}
   \includegraphics[width=.245\textwidth,keepaspectratio]{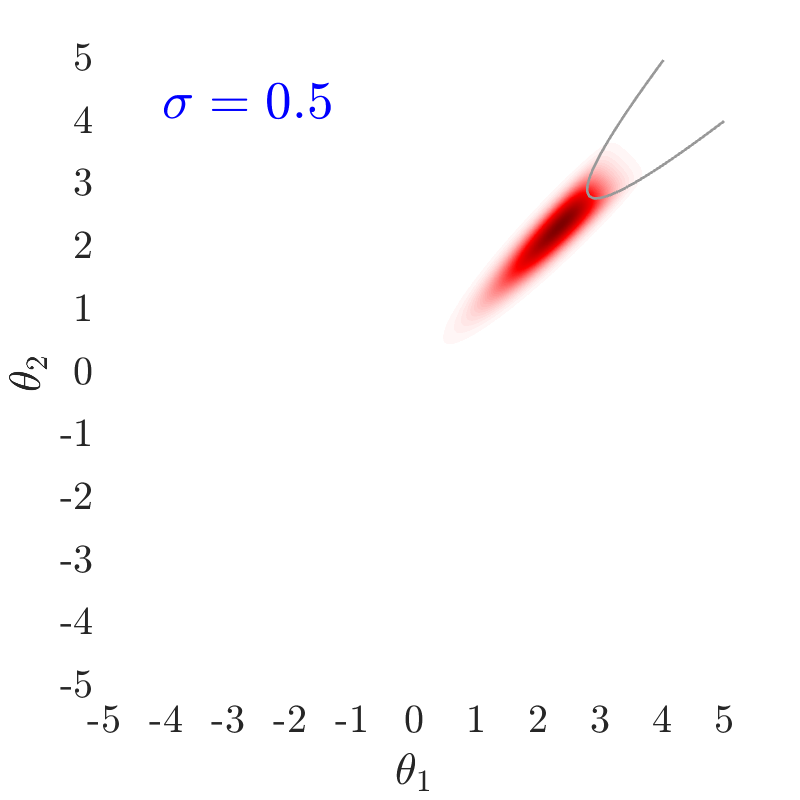}& \hspace*{-0.2in}
   \includegraphics[width=.245\textwidth,keepaspectratio]{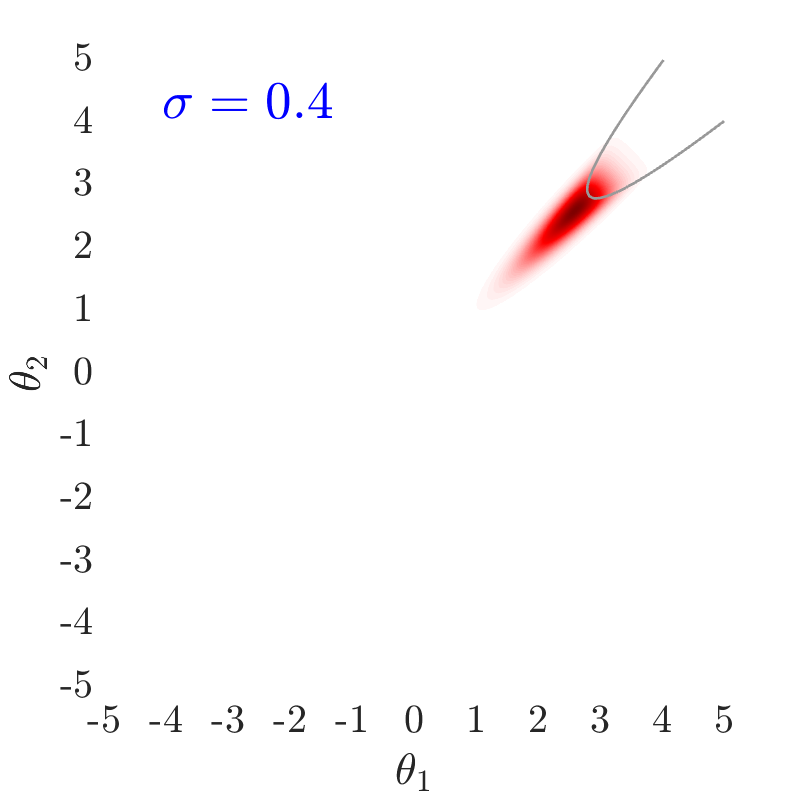}& \hspace*{-0.2in}
\\ \hspace*{-0.2in}
   \includegraphics[width=.245\textwidth,keepaspectratio]{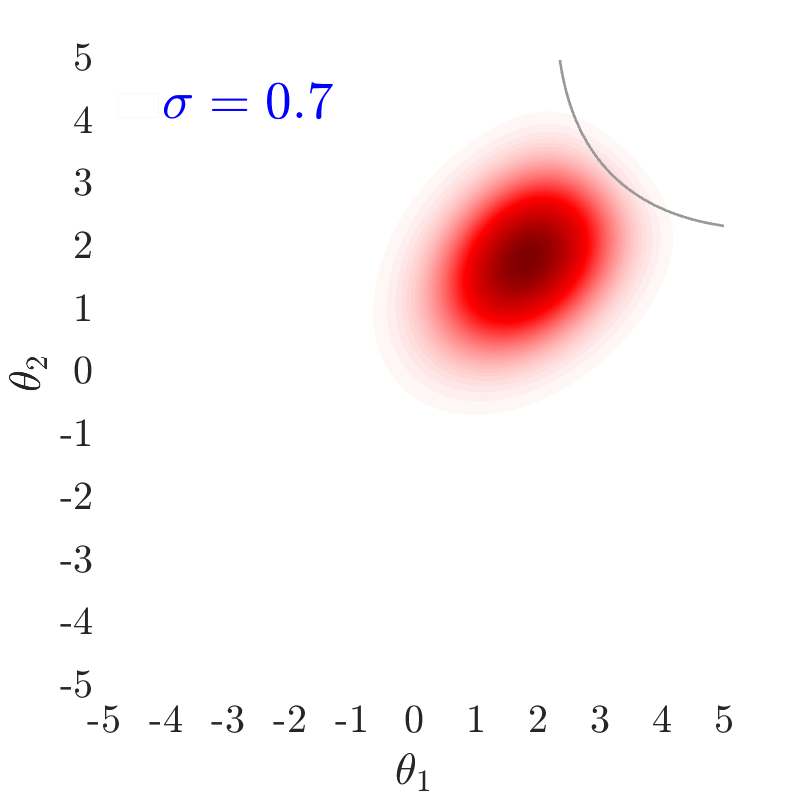}& \hspace*{-0.2in} \includegraphics[width=.245\textwidth,keepaspectratio]{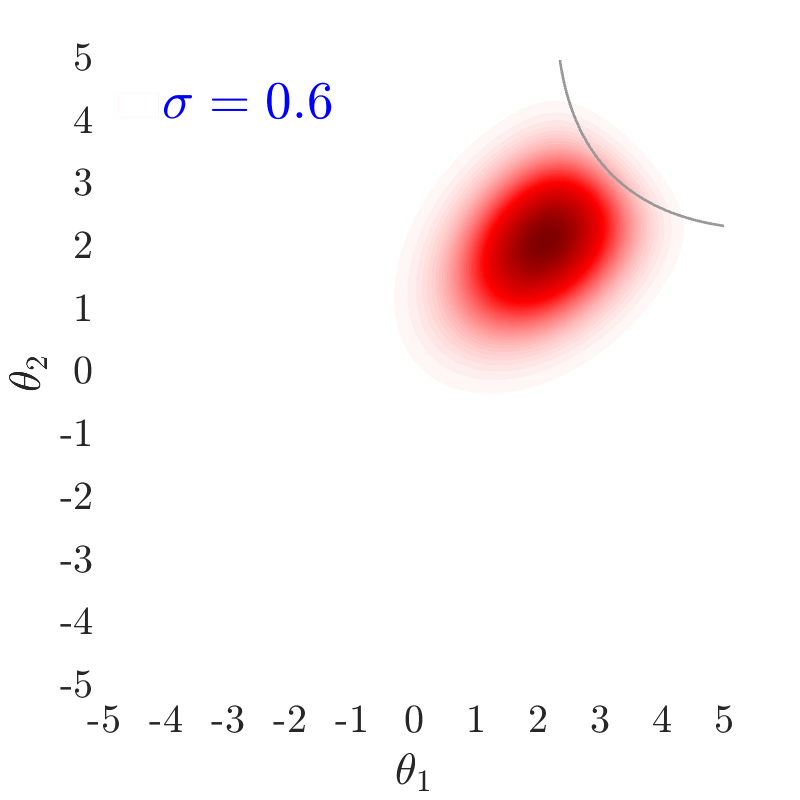}& \hspace*{-0.2in}
   \includegraphics[width=.245\textwidth,keepaspectratio]{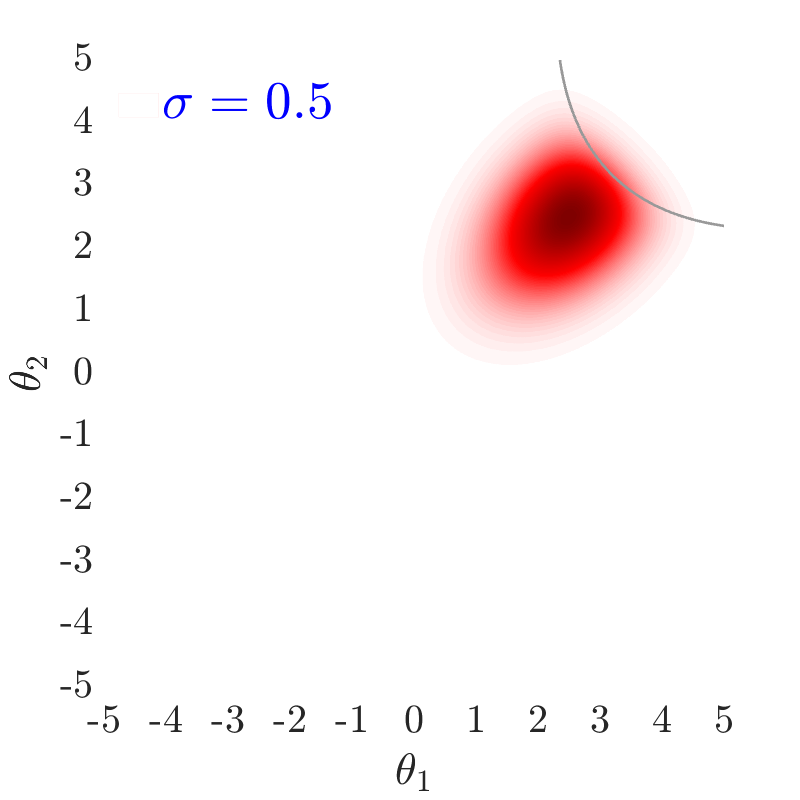}& \hspace*{-0.2in}
   \includegraphics[width=.245\textwidth,keepaspectratio]{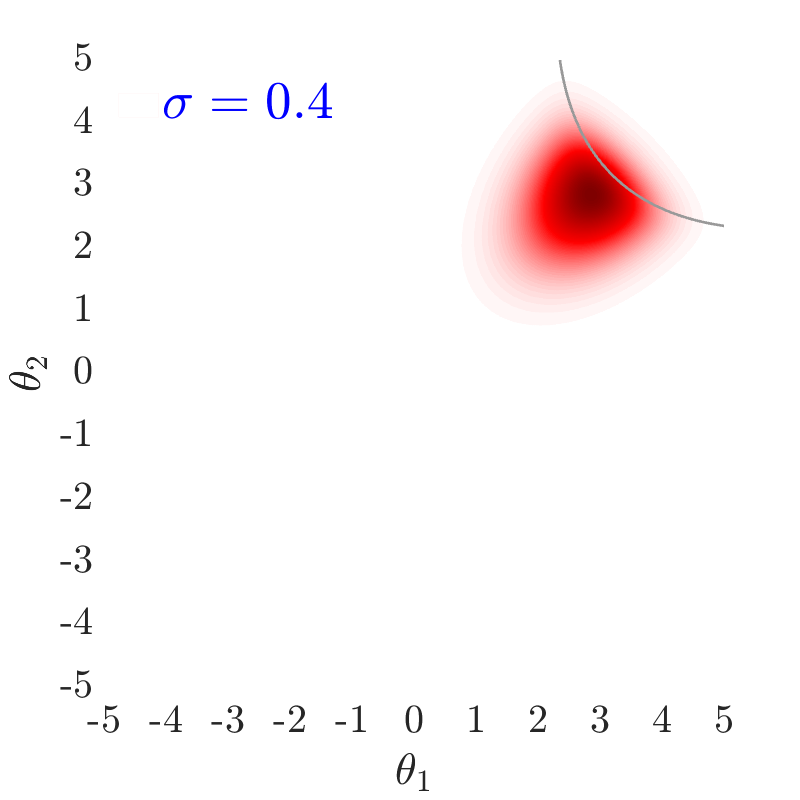}& \hspace*{-0.2in}

\smallskip\\\hline\noalign{\smallskip}
 \end{tabular}
 \caption{Effect of the likelihood dispersion factor, $\sigma$, on the target distribution for two different unimodal limit-state functions, for the $p50$ percentile parameter of $\mu_{g}$.}
 \label{figg:4}
\end{figure}

The use of $\sigma$ in this work, largely follows ideas presented in \citep{papaioannou2016sequential}, albeit at a completely different context. The value of $\sigma$ also has an important effect on the convergence and the computational efficiency of our HMCMC sampling methods. In \cref{figg:4} the effect of $\sigma$ on the target distribution is displayed based on two 2D examples on the $\boldsymbol{\Theta}$ space, with a small failure probability ($\sim10^{-6}$) and a unimodal failure region. A higher value of $\sigma$ can reduce the number of model calls as the constructed target becomes more dispersed compared to the case of lower $\sigma$ values, thus being comparatively easier to be located and sampled, but an insufficient number of samples may then reach inside the failure regions. In contrast, reducing $\sigma$ concentrates the target distribution within the regions of interest, making it, however, a more challenging sampling target, often requiring a higher number of model calls to be located and sampled. Choosing the value of the likelihood dispersion factor, $\sigma$, is therefore a trade-off. The suggested value of $\sigma$ for our methodology is in the range $[0.1 \, \,0.8]$. Fine tuning higher decimal values of $\sigma$ in that range is not usually necessary. It is generally recommended to use higher $\sigma$ values $(0.5-0.8)$ in multi-modal cases, enabling longer state jumps, even between modes, and lower values, $[0.1 \, \,0.6]$, for cases with very small failure probability, e.g. around $P_{F} \leq 10^{-6}$ and lower, and/or high-dimensional problems, in order to attract more samples into the failure domain.\par

\begin{figure}[t!]
 \centering
  \begin{tabular}{c ccccc ccccc cc cc cc cc}
\\\hline\noalign{\smallskip}
   \hspace*{-0.2in}
   \includegraphics[width=.245\textwidth,keepaspectratio]{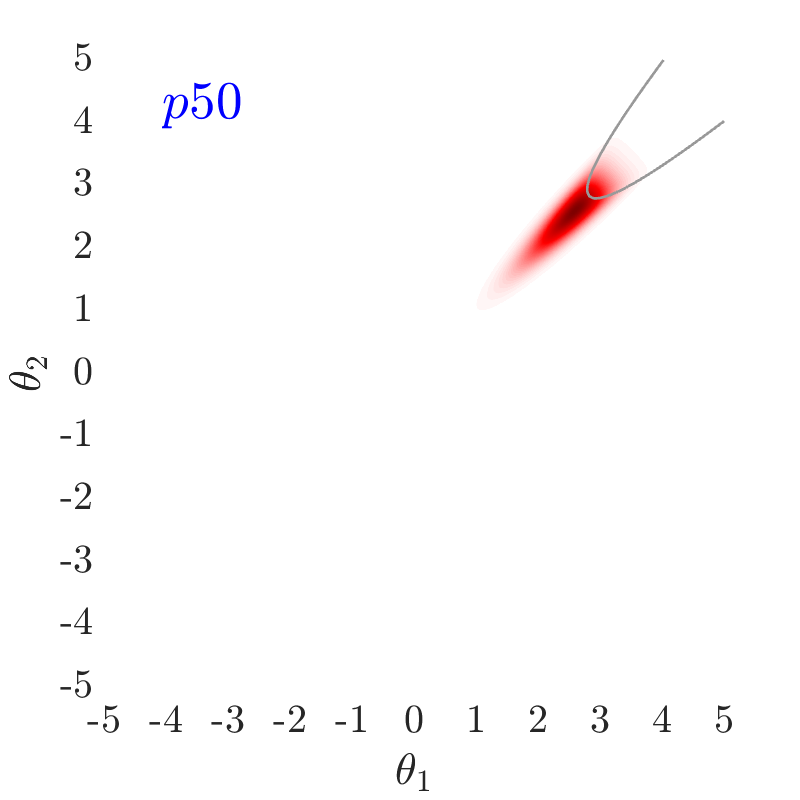}& \hspace*{-0.2in} \includegraphics[width=.245\textwidth,keepaspectratio]{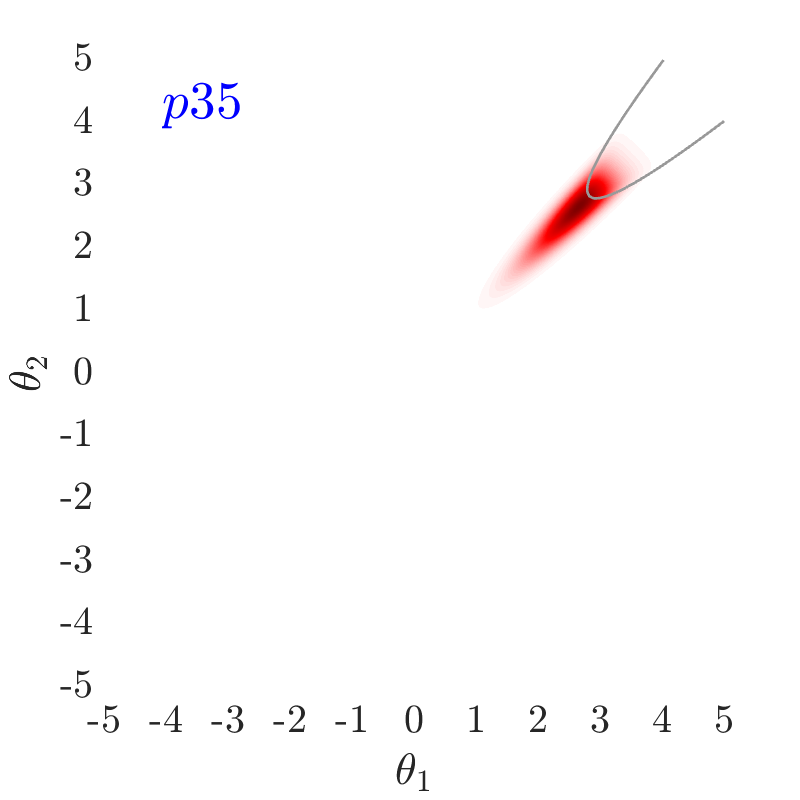}& \hspace*{-0.2in}
   \includegraphics[width=.245\textwidth,keepaspectratio]{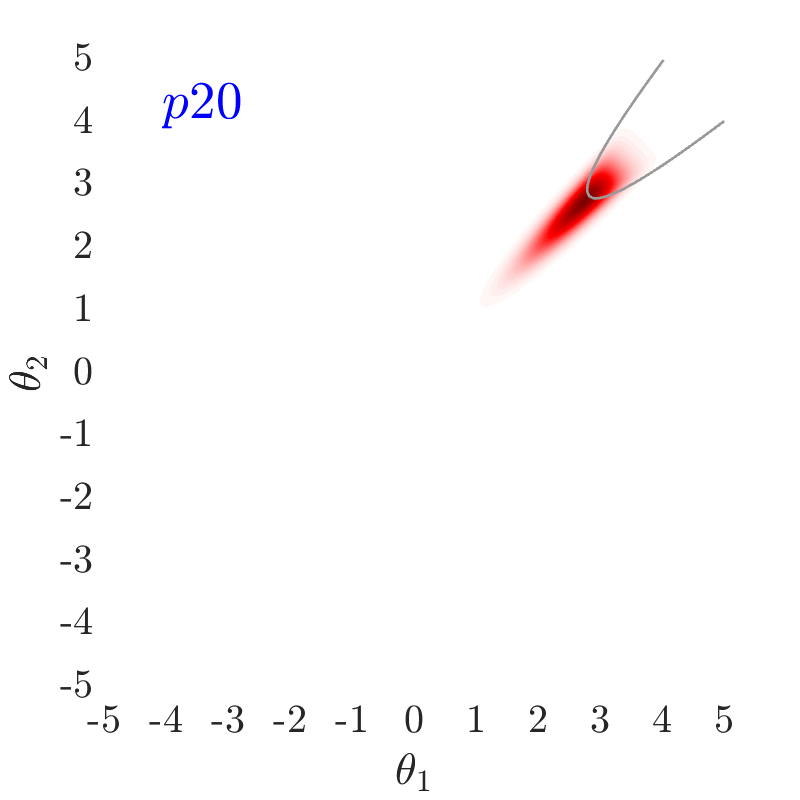}& \hspace*{-0.2in}
   \includegraphics[width=.245\textwidth,keepaspectratio]{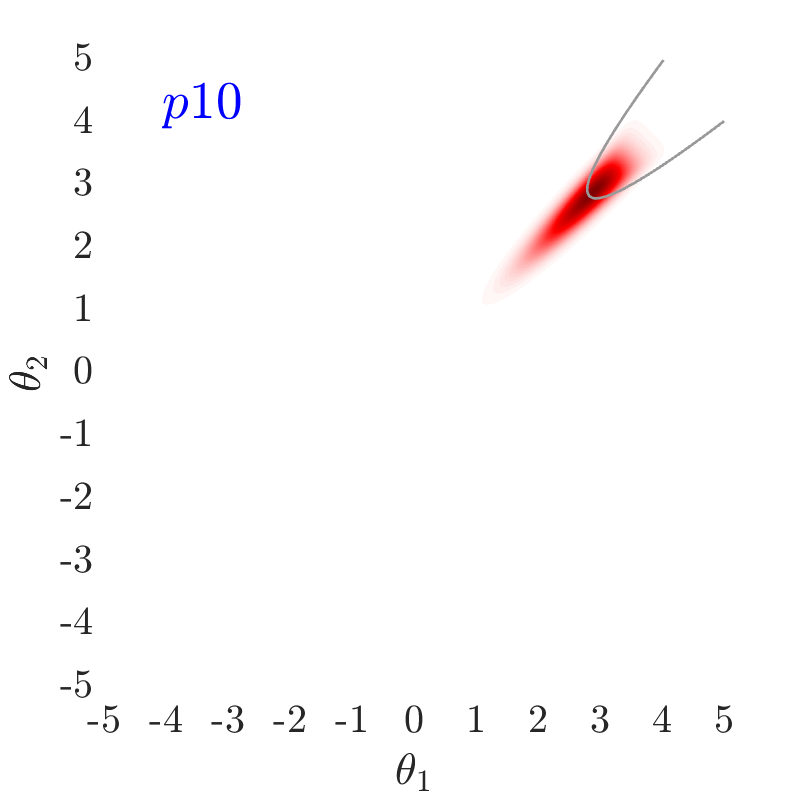}& \hspace*{-0.2in}
\\ \hspace*{-0.2in}
   \includegraphics[width=.245\textwidth,keepaspectratio]{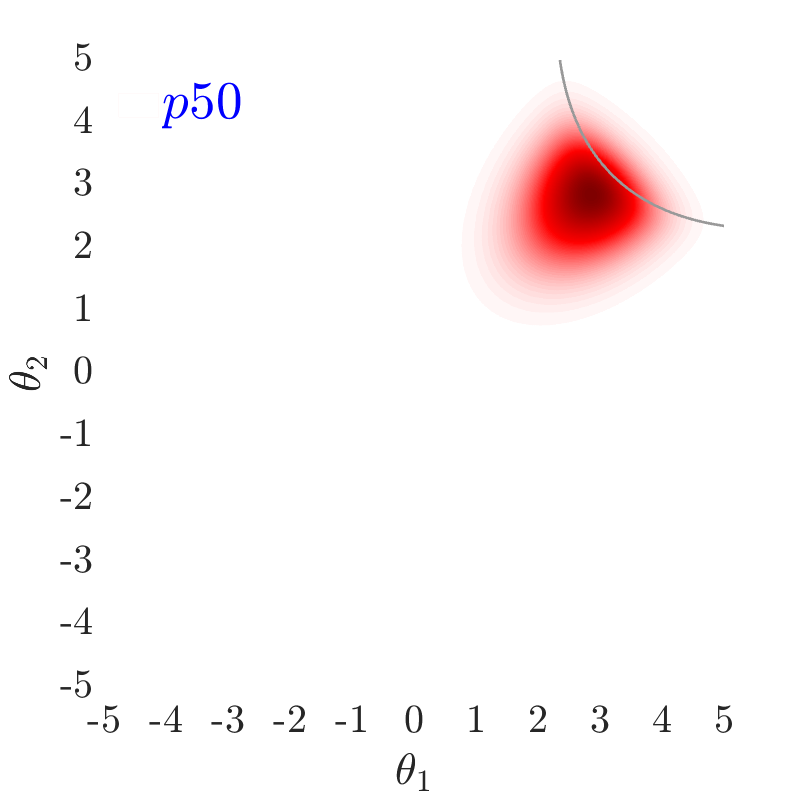}& \hspace*{-0.2in} \includegraphics[width=.245\textwidth,keepaspectratio]{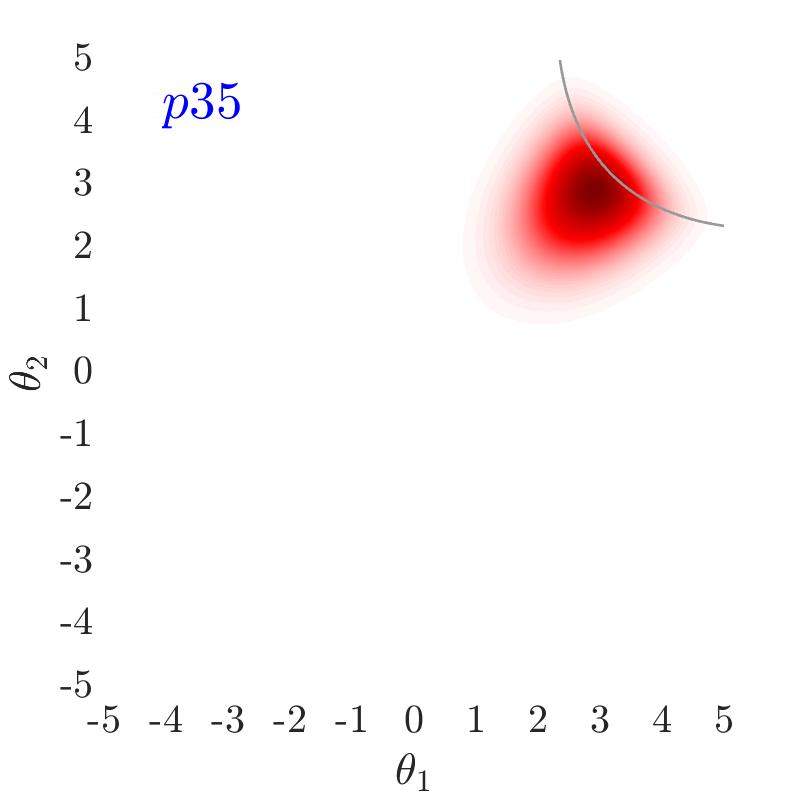}& \hspace*{-0.2in}
   \includegraphics[width=.245\textwidth,keepaspectratio]{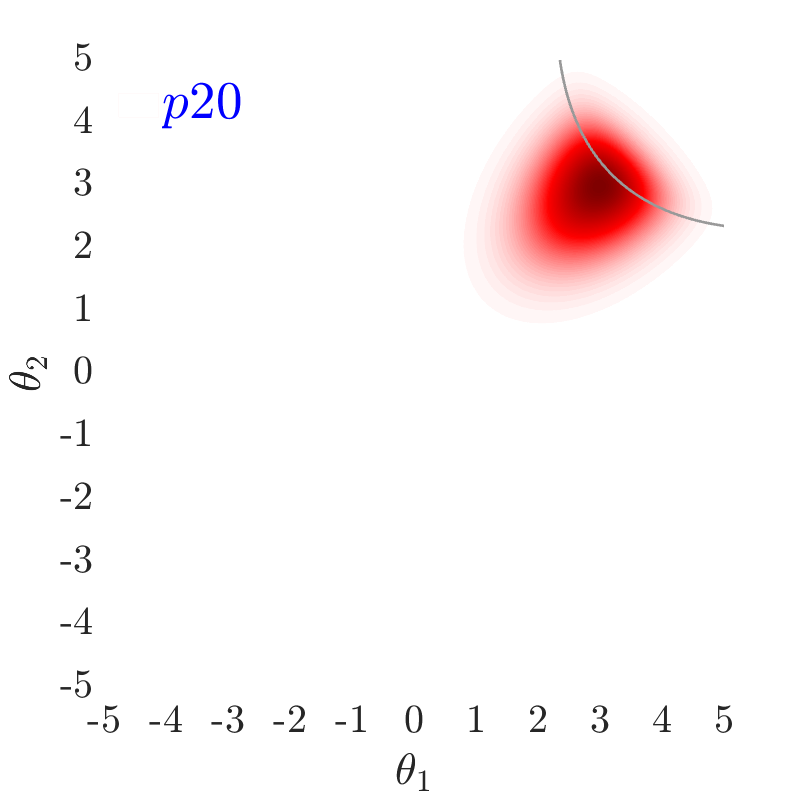}& \hspace*{-0.2in}
   \includegraphics[width=.245\textwidth,keepaspectratio]{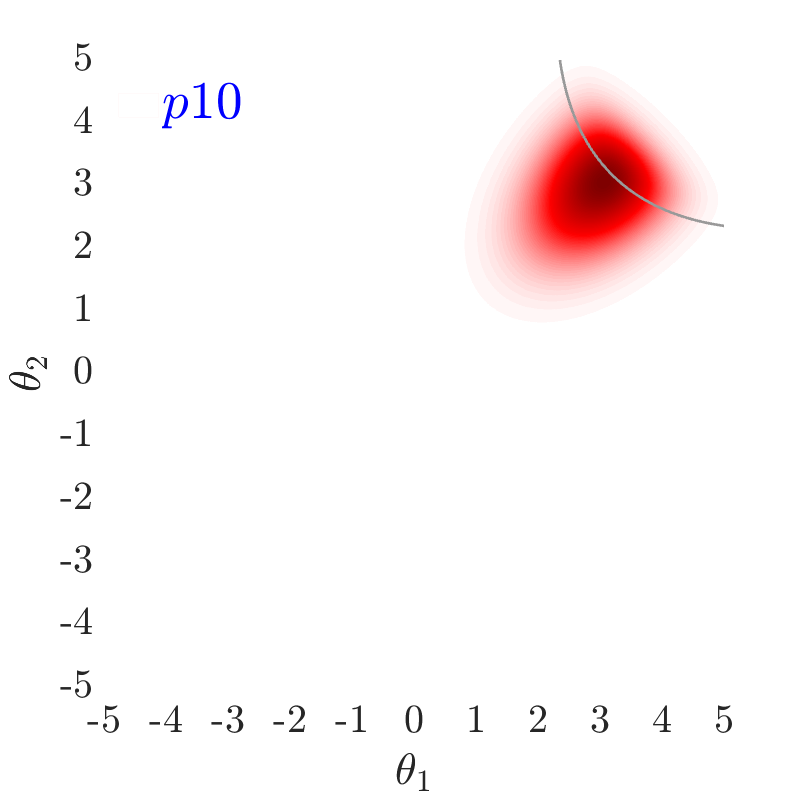}& \hspace*{-0.2in}

\smallskip\\\hline\noalign{\smallskip}
 \end{tabular}
 \caption{Effect of $\mu_{g}$ of the likelihood function on the target distribution, expressed through its percentile parameter $p$, for two different unimodal limit-state functions, and for $\sigma = 0.4$.}
 \label{fig:4}
\end{figure}

\begin{figure}[t!]
 \centering
  \begin{tabular}{c ccccc ccccc cc cc cc cc}
   \\\hline\noalign{\smallskip}
   \hspace*{-0.2in}
   \includegraphics[width=.245\textwidth,keepaspectratio]{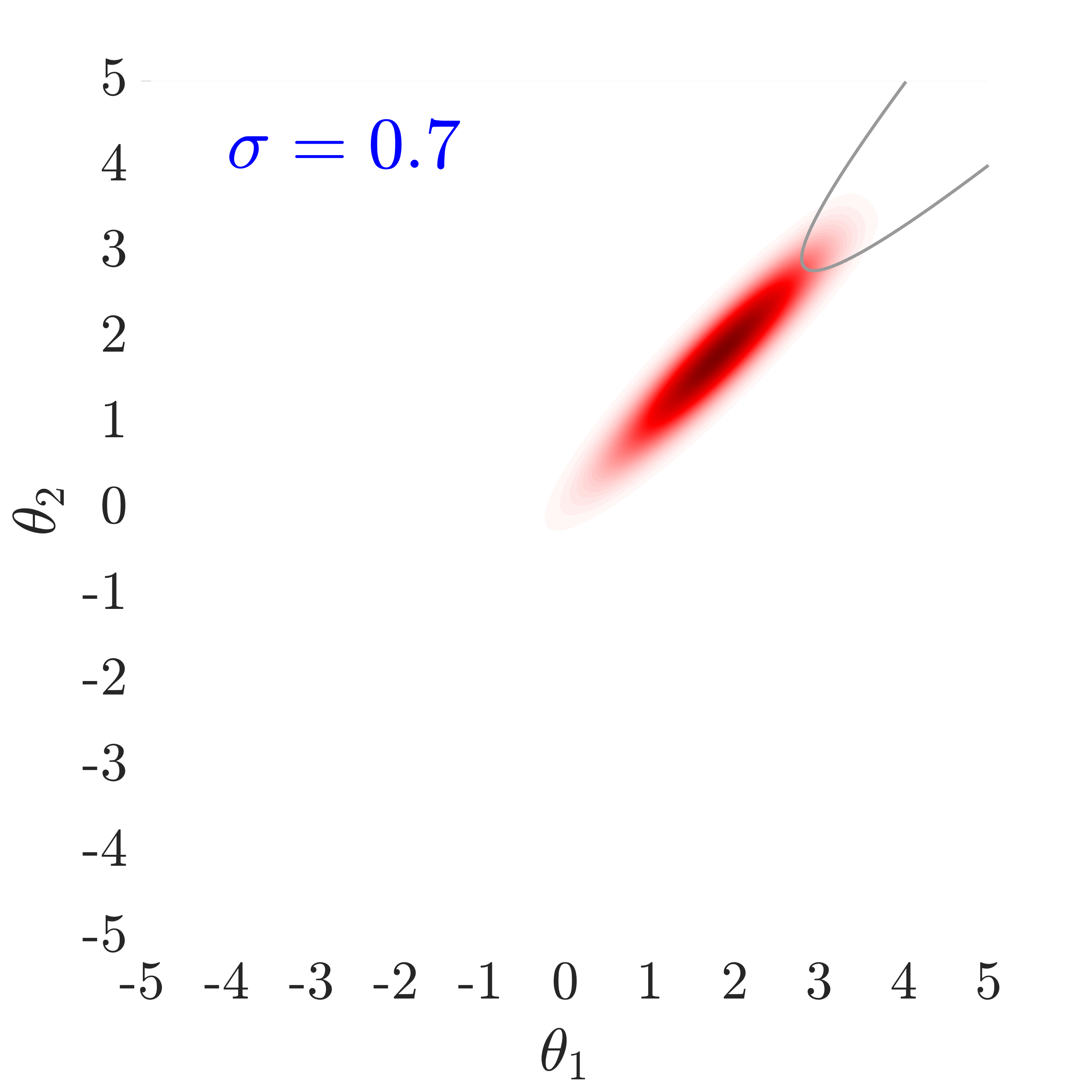}& \hspace*{-0.2in} \includegraphics[width=.245\textwidth,keepaspectratio]{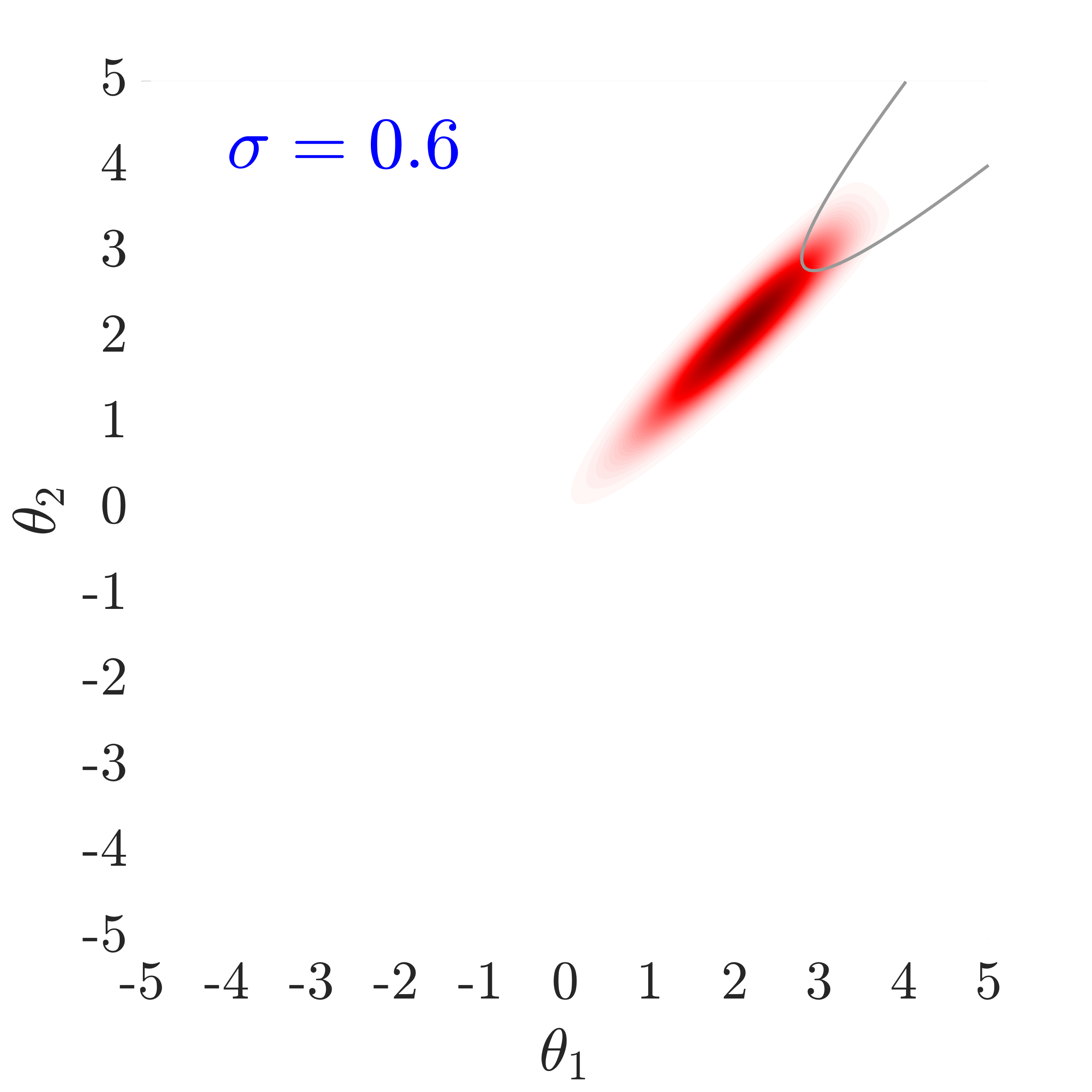}& \hspace*{-0.2in}
   \includegraphics[width=.245\textwidth,keepaspectratio]{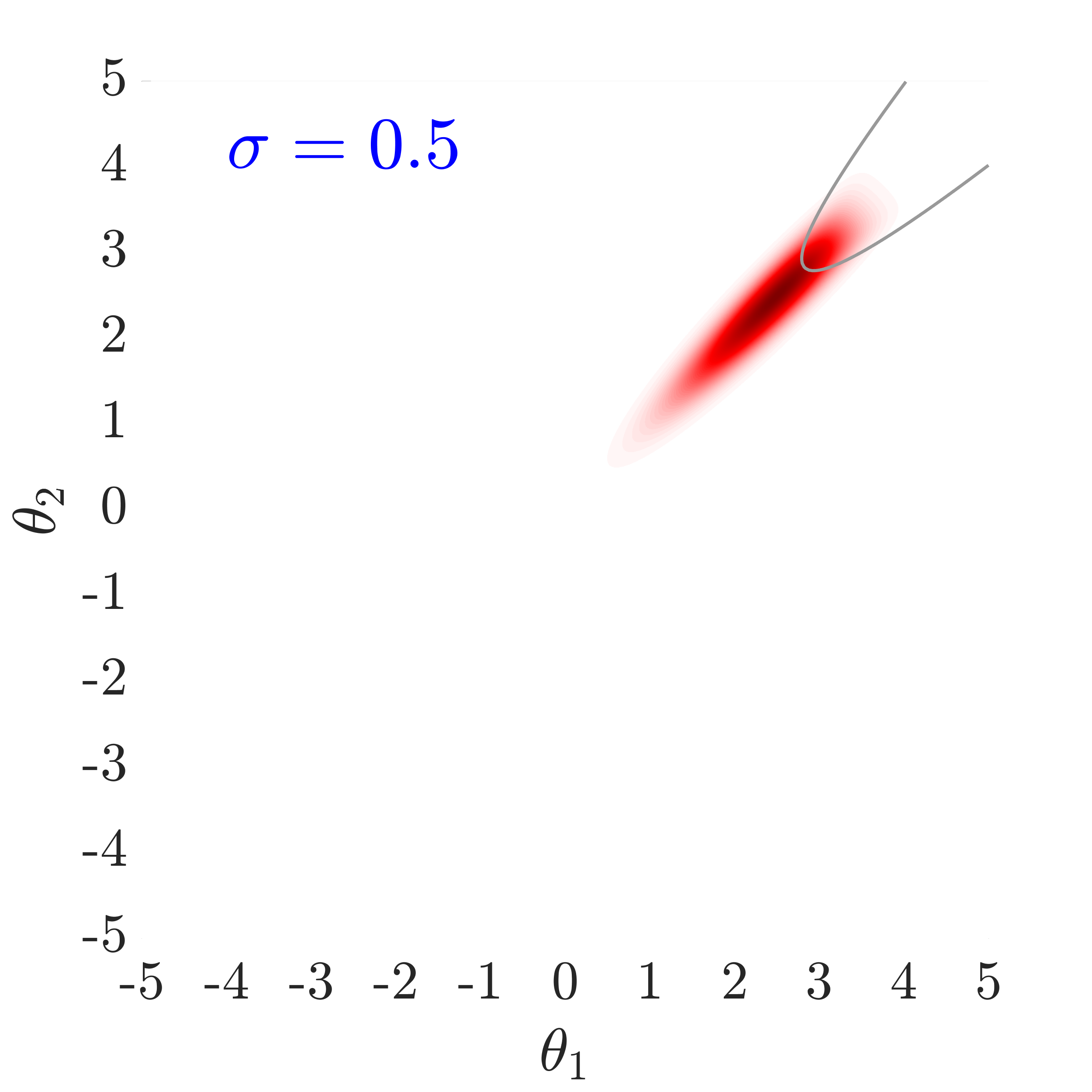}& \hspace*{-0.2in}
   \includegraphics[width=.245\textwidth,keepaspectratio]{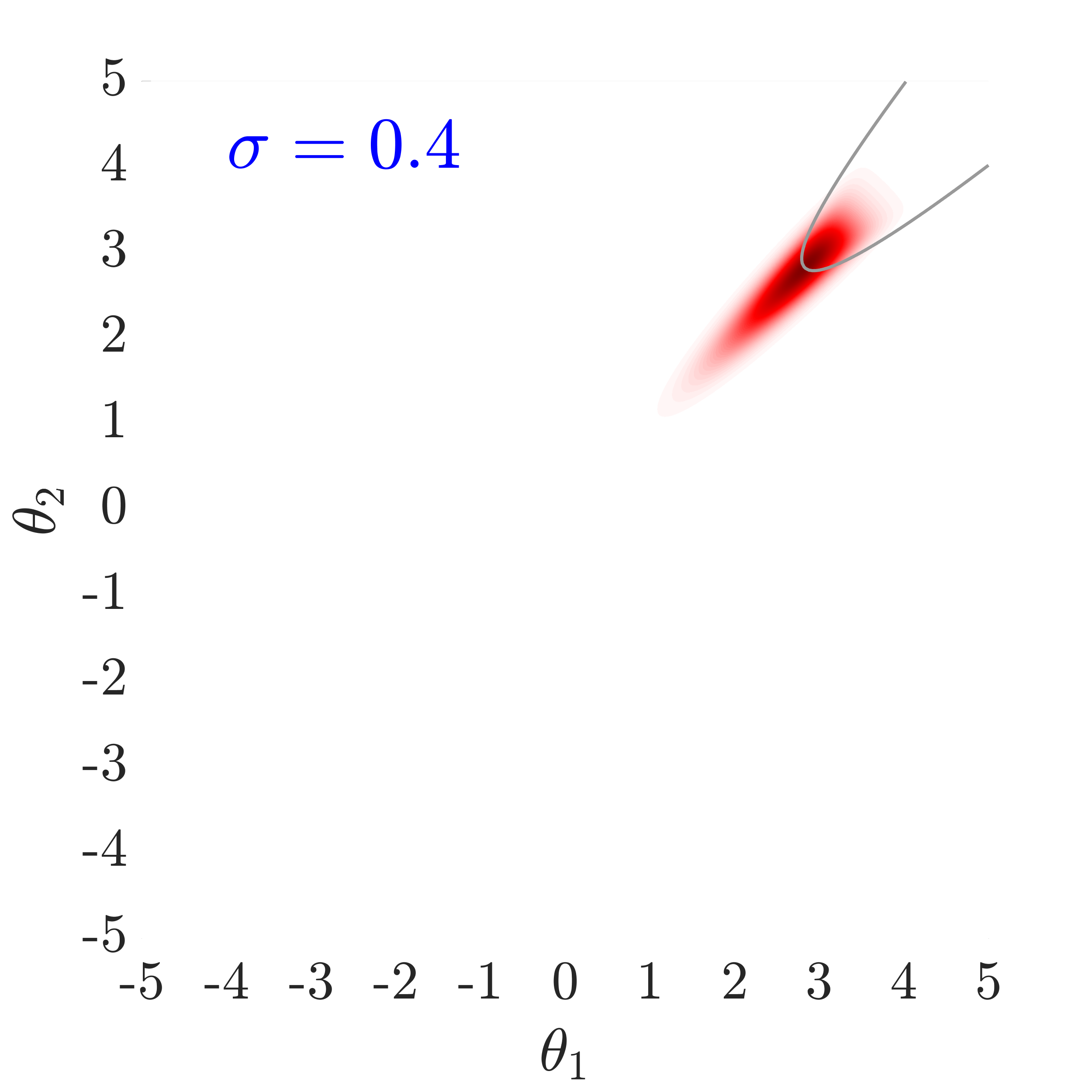}& \hspace*{-0.2in}
  \\ \hspace*{-0.2in}
   \includegraphics[width=.245\textwidth,keepaspectratio]{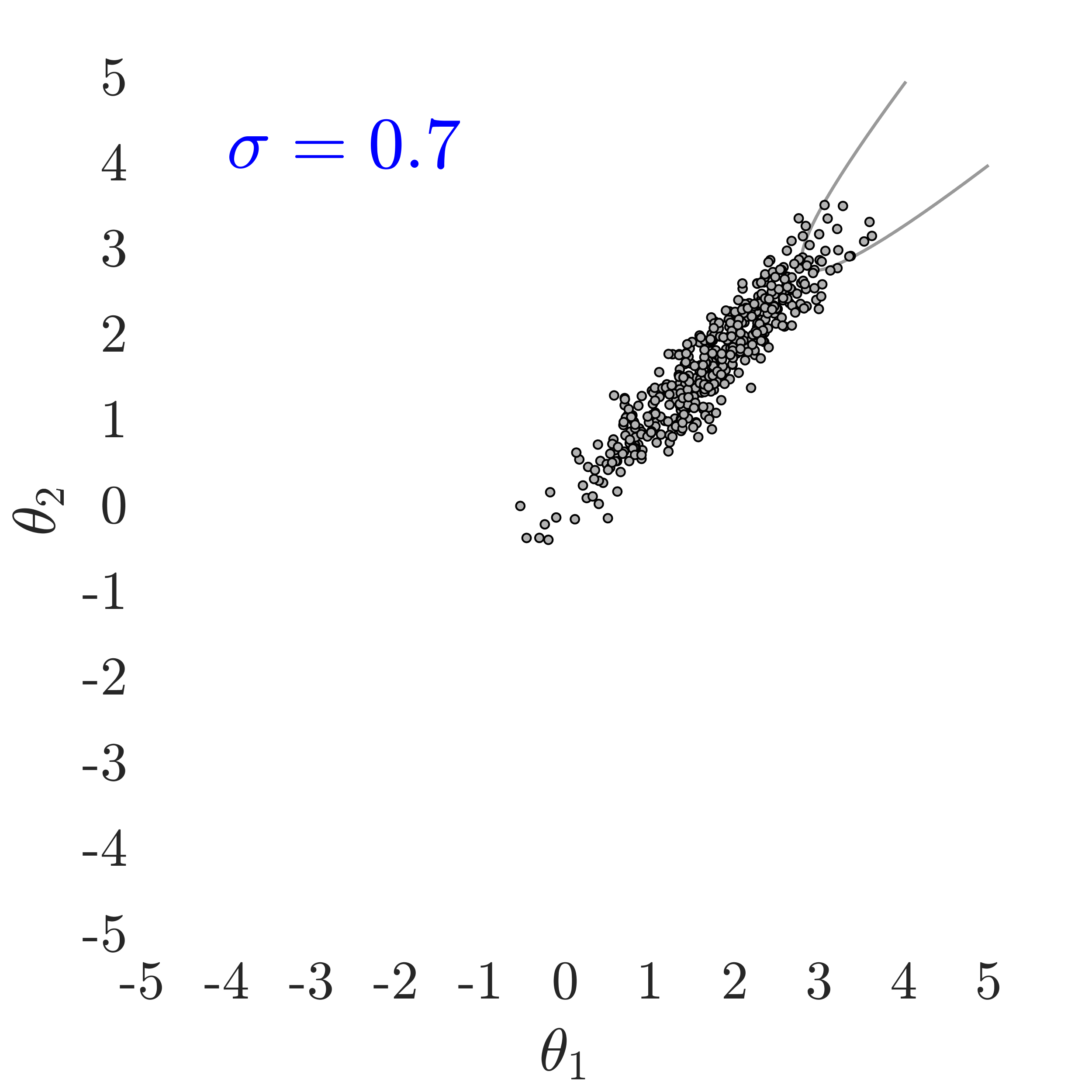}& \hspace*{-0.2in} \includegraphics[width=.245\textwidth,keepaspectratio]{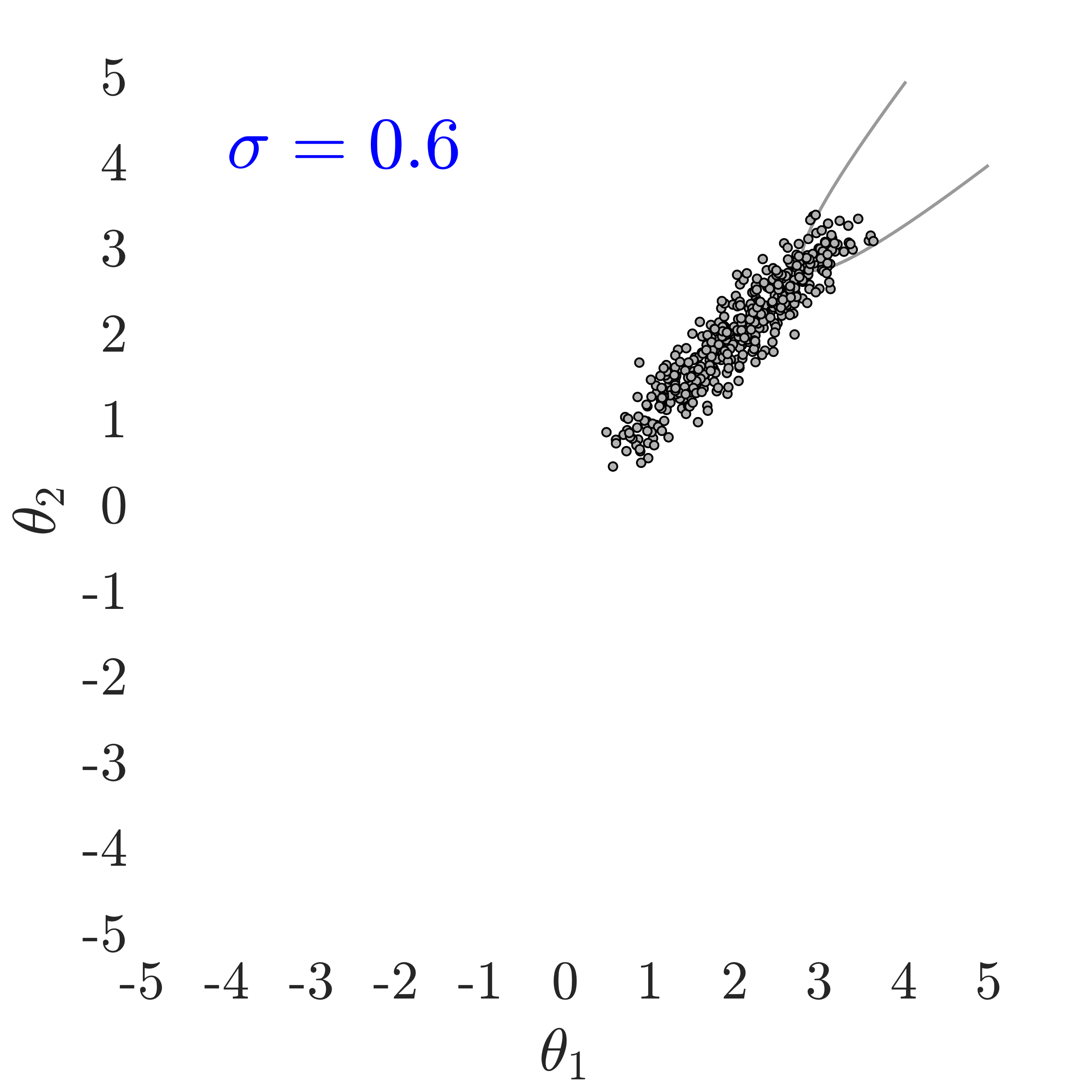}& \hspace*{-0.2in}
   \includegraphics[width=.245\textwidth,keepaspectratio]{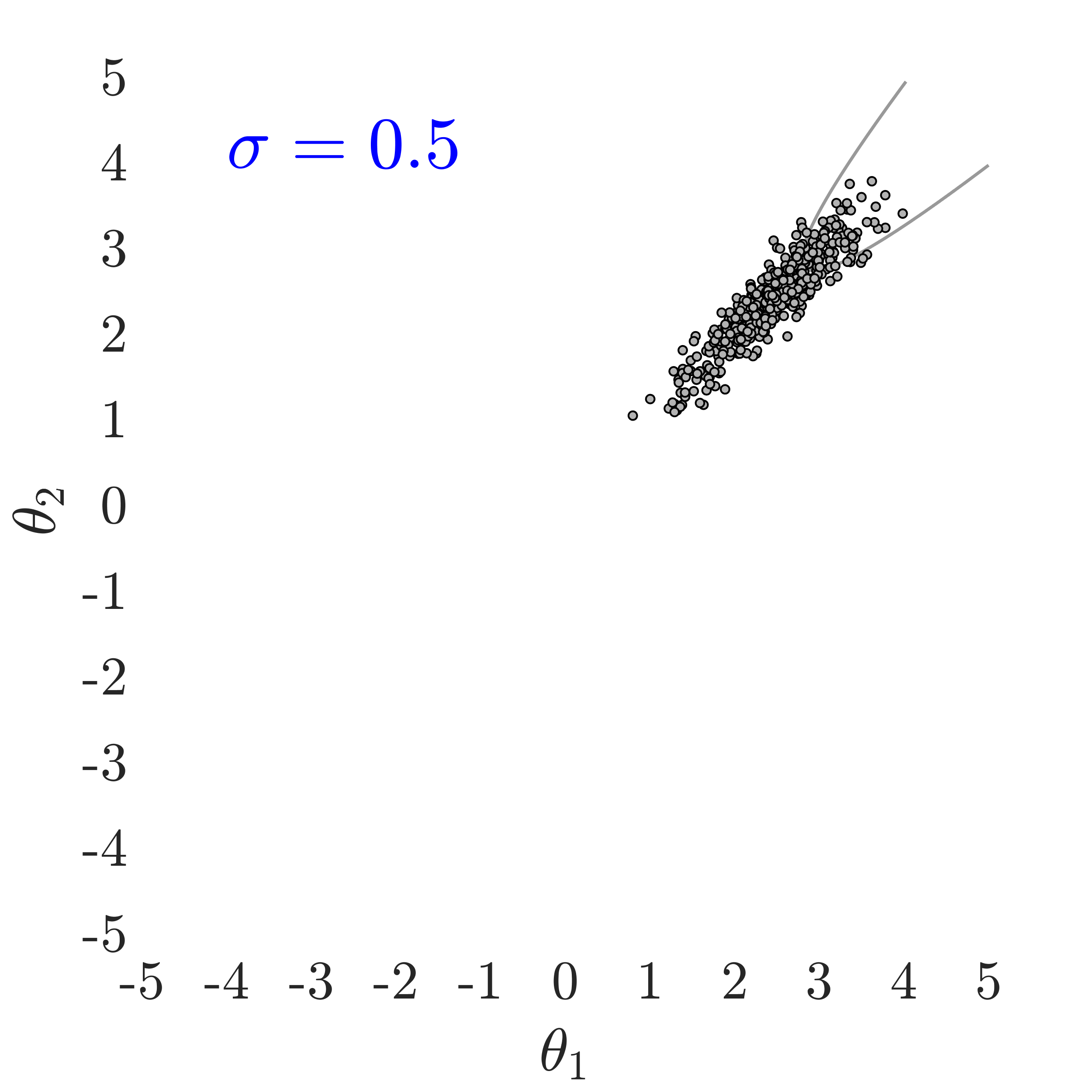}& \hspace*{-0.2in}
   \includegraphics[width=.245\textwidth,keepaspectratio]{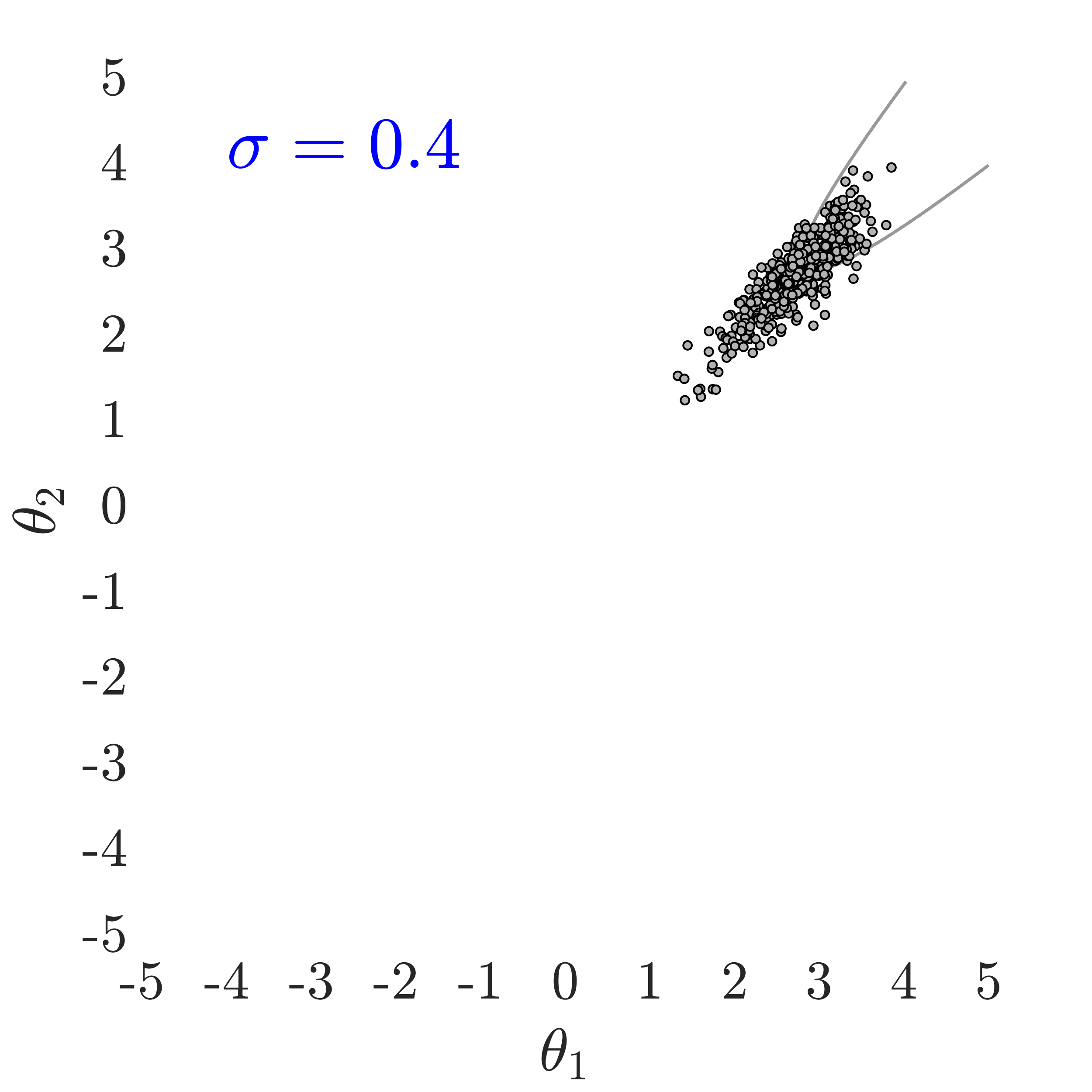}& \hspace*{-0.2in}
  \\
  $\mathop{\mathbb{E}}[\hat{P}_{F}] = \text{4.70E-6}$&
 $\mathop{\mathbb{E}}[\hat{P}_{F}]= \text{4.74E-6}$ &
   $\mathop{\mathbb{E}}[\hat{P}_{F}] = \text{4.71E-6}$ &
 $\mathop{\mathbb{E}}[\hat{P}_{F}]= \text{ 4.72E-6}$ 
  \vspace*{0.02in}
    \\
    
 C.o.V $ = \text{0.41}$&
   C.o.V  $= \text{0.29}$ &
   C.o.V  $= \text{0.20}$ &
  C.o.V  $= \text{0.15}$ 
    \vspace*{-0.15in}
  \\

\smallskip\\\hline\noalign{\smallskip}
 \end{tabular}
  \captionsetup{labelfont={color=Black}}
 \caption{Effect of the prescribed $\sigma$ on sampling from the target distribution; $\mu_{g}$ is calculated based on $p10$; (Ref. $P_{F}$ $\sim$ 4.73E-6).}
 \label{fig:footprint}
\end{figure}

\vspace{1 in}
For the mean parameter, $\mu_{g}$, we are interested in generally locating it into the failure region, $g(\boldsymbol\theta)<0$, to enhance the sampling efficiency. As such, we are describing $\mu_{g}$ through a percentile, $p$, of the logistic CDF and its quantile function, $\Upsilon_{g}$:
\begin{equation}
\Upsilon_{g} (p;\mu_{g},\sigma) = \mu_{g} + (\frac{\sqrt{3}}{\pi} \sigma) \ln \bigg (\frac{p}{1-p} \bigg )
\end{equation}
Placing a chosen percentile $p$ of the logistic CDF on the limit-state surface $g(\boldsymbol\theta) = 0$, results in $\Upsilon_{g} (p;\mu_{g},\sigma) = 0$ and the $\mu_{g}$ can be then given as:

\begin{equation}
\mu_{g} =  -(\frac{\sqrt{3}}{\pi} \sigma) \ln \bigg (\frac{p}{1-p} \bigg )  \label{eqmean}
\end{equation}
\noindent\ignorespacesafterend\parindent0pt \endtrivlist Therefore, as also shown in \cref{fig:4}, by reducing $p$, the $\mu_{g}$ of the likelihood function further moves into the failure domain and consequently more samples can infiltrate into this region and can contribute to the failure probability estimation. In \cref{fig:4}, $p50$, for example, denotes the \textit{50th} percentile ($p = 0.5$). In all examples presented in this paper, the $p10$ percentile, $p=0.1$, is used to define the value $\mu_{g}$, which is found to yield good efficiency. \cref{eqmean} and parameter $\sigma$ can now be used in \cref{finalTar} to fully define the non-normalized target distribution $\tilde{h}(\boldsymbol{\theta})$. 

\cref{fig:footprint} displays the effect of $\sigma$ on sampling the approximate target distribution, showcasing how smaller values of $\sigma$ push the samples further inside the failure regions. It also compares the $\mathop{\mathbb{E}}[\hat{P}_{F}]$ and Coefficient of Variation (C.o.V), computed according to \cref{PFF,eqqIS5}, based on 500 independent simulations with a number of model calls $\sim 900$ in all cases. Smaller values of $\sigma$ result in a more accurate estimate of $\mathop{\mathbb{E}}[\hat{P}_{F}]$ and a lower C.o.V, due to having the presence of more samples in the failure region.

\subsection{Inverse Importance Sampling}\label{IIS}
An original post-sampling step is devised at this stage, termed \textit{inverse importance sampling} (IIS), that can succesfully employ the already acquired samples and the normalized target distribution $h(\boldsymbol{\theta})$ as the importance sampling density function. Hence, the probability of failure can be now computed as follows:
\begin{align}
&\textit{P\textsubscript{F}} = \int_{\boldsymbol{\theta}} I_{F} (\boldsymbol{\theta}) \pi_{\Theta}(\boldsymbol{\theta}) d\boldsymbol{\theta}
 = \int_{\boldsymbol{\theta}} I_{F} (\boldsymbol{\theta})  \dfrac{\pi_{\Theta}(\boldsymbol{\theta})}{h(\boldsymbol{\theta})} h(\boldsymbol{\theta}) d\boldsymbol{\theta}\\
&\text{where}\, \, \,  h(\boldsymbol{\theta}) = \frac{\tilde{h}(\boldsymbol{\theta})}{C_{h}} \, \,, \, \, \, \tilde{h}(\boldsymbol{\theta}) = \ell_{g_{\boldsymbol\theta}}(\boldsymbol{\theta})  \pi_{\Theta}(\boldsymbol{\theta}),
\end{align} 
$\pi_{\Theta}(\boldsymbol{\theta})$ is the original distribution, 
$\tilde{h}(\boldsymbol{\theta})$ denotes the non-normalized sampling target distribution, and $C_{h}$ is its normalizing constant. After some simplifications we get:
\begin{align}    
&\textit{P\textsubscript{F}} = C_{h} \int_{\boldsymbol{\theta}} I_{F}(\boldsymbol{\theta}) \dfrac{1}{\ell_{g_{\boldsymbol\theta}}(\boldsymbol{\theta})} \, h(\boldsymbol{\theta})\,d\boldsymbol{\theta} = \bigg (\frac{1}{N} \sum_{i=1}^{N}\dfrac{I_{F}(\boldsymbol{\theta}_{i})}{\ell_{g_{\boldsymbol\theta}}(\boldsymbol{\theta}_{i})}\bigg ) \, C_{h} \label{PFF}
\end{align}
with $N$ the number of already acquired samples. To now calculate the normalizing constant $C_{h}$, once again we resort to an importance sampling scheme in the following manner:
\begin{align}
C_{h} = \int_{\boldsymbol{\theta}} \tilde{h}(\boldsymbol{\theta}) d\boldsymbol{\theta} = \int_{\boldsymbol{\theta}} \frac{\tilde{h}(\boldsymbol{\theta})}{Q(\boldsymbol{\theta})} Q(\boldsymbol{\theta})d\boldsymbol{\theta} = \frac{1}{M} \sum_{i=1}^{M} \dfrac{\tilde{h}(\boldsymbol{\theta^\prime}_{i})}{Q(\boldsymbol{\theta^\prime}_{i})}
\label{chh}
\end{align}
where $Q(.)$ can be a computed Gaussian Mixture Model (GMM), based again on the already available samples and the generic Expectation-Maximization (EM) algorithm \citep{mclachlan2000finite}, as indicatively seen in the right plot of \cref{fig2}. A typical GMM expression can be then given by:
\begin{align}
Q(\boldsymbol{\theta})= \sum_{k=1}^{\bm{K}} w_{k} \phi(\boldsymbol{\theta}\, ;\,\boldsymbol{\mu}_{k},\boldsymbol{\Sigma}_{k})
\end{align}       
where $\phi(.)$ is the PDF, $w_{k}$ is the weight, $\boldsymbol{\mu}_{k}$ is the mean vector and $\boldsymbol{\Sigma}_{k}$ is the covariance matrix of the $kth$ Gaussian component, that can all be estimated, for all components, by the EM algorithm \citep{mclachlan2000finite} based on the samples. When a GMM can be accurately fitted, the original HMCMC samples can be used directly to compute the normalizing constant $C_{h}$, i.e., $\boldsymbol{\theta^\prime}_{i}\equiv\boldsymbol{\theta}_{i}$, and $M=N$ in \cref{PFF} and \cref{chh}. A precisely fitted GMM, with a sufficiently large number of components, adequately acts as a representative distribution for the original samples. However, given that the accuracy of GMMs may often deteriorate, particularly in high-dimensional and challenging multi-modal cases, due to the large number of parameters that need to be identified \citep{papaioannou2019improved}, similar to many other mixture models, additional  $M$  samples from the computed $Q(.)$ can be required, just in order to accurately evaluate the normalizing constant $C_{h}$. In this latter, more general case, even a very approximately fitted GMM with diagonal covariance matrices, particularly appropriate for high-dimensional cases, works effectively, since it is solely used  for computing the constant $C_{h}$, while the already obtained HMCMC samples are instead used in \cref{PFF}. Drawing each of the i.i.d. $Q(.)$ samples requires only one model call and is called IIS sample in the following sections for the sake of clarity and simplicity. As a general guide, using IIS samples around $20\%$ of the total number of samples results in a good estimate for $C_{h}$ and eventually \cref{PFF}.

\subsection{Adaptation during the burn-in phase}\label{burnin}
A burn-in sampling phase is required in the ASTPA framework, regardless of the used MCMC sampling scheme. For the presented HMCMC methods in this work, the burn-in samples are usually adequate to be around $5\%-10\%$ of the total number of samples. To improve the overall sampling efficiency, we are taking advantage of this burn-in phase, that guides the chain to the appropriate target distribution, by also performing a series of adaptations and relevant parameter tuning steps. In particular, the stepsize $\varepsilon$ and the trajectory length $\tau$ are established in this burn-in stage, as explained in \cref{sec3.2}, and then used throughout the rest of the analysis, while the precondtioned mass matrix $\textbf{M}$, used in our QNp-HMCMC algorithm, is also computed in the burn-in phase and then utilized for the remainder of the sampling, as discussed in \cref{section3}.\par 
In addition to these sampling related parameters, we are also adjusting the target distribution in this sampling stage, through the parameters $\mu_{g}$ and $\sigma$. Following the discussion in \cref{sec4.2}, these two parameters are automatically evolving toward their constant values in this adaptive phase, to assist in guiding the samples to the final target distribution. The likelihood dispersion factor initiates with $\sigma_{0} = 1$ and then follows an exponential decay until its prescribed constant value $\sigma$, as: 
\begin{align}
\sigma_{iter} = a_{1} e^{(\dfrac{-iter}{a_{2}})}, \, \, \, \, \,
\text{where}  \, \,   a_{2} = \dfrac{(N_{BI} - 1)}{\ln (\frac{\sigma_{0}}{\sigma})} , \, \, a_{1} = \dfrac{\sigma_{0}}{e^{(\dfrac{- 1}{a_{2}})}}  
\end{align}
where $N_{BI}$ is the assigned number of total burn-in samples and $iter>=1$ denotes the iteration counter based on a fully completed leapfrog step. A similar exponential growth scheme is also used for $\mu_{g}$, starting based on $p50$ in \cref{eqmean}, i.e. $\mu_{p50}=0=g(\boldsymbol{\theta})$, and reaching the constant value $\mu_{g}=\mu_{p10}$ given by the prescribed $p10$ percentile in \cref{eqmean}, as:

\begin{align}
\mu_{iter} = b_{1} e^{ \big (\dfrac{-iter}{b_{2}} \big )} + \mu_{p50}, \, \, \, \, \,
\text{where}  \, \,   b_{2} = \dfrac{(N_{BI} - 1)}{\ln (\frac{0.0001}{\mu_{t}})} , \, \, b_{1} = \dfrac{0.0001}{e^{(\dfrac{- 1}{b_{2}})}}, \, \, \mu_{t} = \mu_{p10}  + \mu_{p50}    
\end{align}  
\vspace{-0.3in}

\begin{table}[t!]
\caption{Summary of ASTPA parameters and generic values beyond tuning/optimization}
\centering

\small
\begin{tabular}{cccccccc}
  \toprule[1.5pt]
  Dispersion factor ($\sigma$) & $\mu_{g(\boldsymbol\theta)}$   &Trajectory length ($\tau$)  & Step size ($\varepsilon$) & Total model calls &Burn-in& IIS samples\\
  \hline\noalign{\smallskip}
  [0.1 0.8] & $p10$ &0.7  &\textit{dual averaging} \citep{hoffman2014no} &5,000-10,000 & 10$\%$& 20$\%$\\

 \bottomrule[1.5pt]
\end{tabular}\label{tabelparam}
\end{table}

\subsection{Statistical properties of the estimator \texorpdfstring{$P_{F}$}{PF}}\label{analCOV}
Based on \cref{PFF} and given samples $[\boldsymbol{\theta}_{1},..., \boldsymbol{\theta}_{N}]$, the sample estimator $\hat{P}_{F}$ is expressed:

\begin{align}
\hat{P}_{F} =  \bigg (\frac{1}{N} \sum_{i=1}^{N}\dfrac{I_{F}(\boldsymbol{\theta}_{i})}{\ell_{g_{\boldsymbol\theta}}(\boldsymbol{\theta}_{i})}\bigg )
 \bigg( \frac{1}{M} \sum_{j=1}^{M} \dfrac{\tilde{h}(\boldsymbol{\theta^\prime}_{j})}{Q(\boldsymbol{\theta^\prime}_{j})}\bigg) = \tilde{P}_{F} \, \hat{C}_h
\label{eqISCoV}
\end{align}

where 
\begin{align}
\tilde{P}_{F}=\bigg (\frac{1}{N} \sum_{i=1}^{N}\dfrac{I_{F}(\boldsymbol{\theta}_{i})}{\ell_{g_{\boldsymbol\theta}}(\boldsymbol{\theta}_{i})}\bigg ) \,\, \text{and} \,\,  \hat{C}_h=\bigg( \frac{1}{M} \sum_{j=1}^{M} \dfrac{\tilde{h}(\boldsymbol{\theta^\prime}_{j})}{Q(\boldsymbol{\theta^\prime}_{j})}\bigg)
\label{eqISCoV1}
\end{align}

with variances:

\begin{align}
\sigma^{2}(\tilde{P}_{F}) = \frac{1}{N_{s}(N_{s}-1)} \sum_{i=1}^{N_{s}} \Bigg (\dfrac{I_{F}(\boldsymbol{\theta}_{i})}{\ell_{g_{\boldsymbol\theta}}(\boldsymbol{\theta}_{i})} - \tilde{P}_{F} \Bigg )^{2}
\label{eqIS3}
\end{align}

\begin{align}
\sigma^{2}(\hat{C}_{h}) = \frac{1}{M(M-1)} \sum_{j=1}^{M} \Bigg ( \dfrac{\tilde{h}(\boldsymbol{\theta^\prime}_{j})}{Q(\boldsymbol{\theta^\prime}_{j})} -\hat{C}_{h} \Bigg )^{2}
\label{eqIS4}
\end{align}
Assuming $\tilde{P}_{F}$ and $\hat{C}_h$ are independent random variables, the variance of $\hat{P}_{F}$ can be given as:
\vspace{0.15in}
\begin{align}
\sigma^{2}(\hat{P}_{F}) =\sigma^{2}(\tilde{P}_{F})\, \,\sigma^{2}(\hat{C}_{h})+
\sigma^{2}(\tilde{P}_{F})\,\hat{C}_{h}^2+\tilde{P}_{F}^2\,\sigma^{2}(\hat{C}_{h})
\end{align}

The Coefficient of Variation (C.o.V) can then be provided as: 

\begin{align}
\textrm{C.o.V} \approx \dfrac{\sigma(\hat{P}_{F})}{\hat{P}_{F}}  \label{eqqIS5}
\end{align}

where $N_{s}$ denotes the used Markov chain samples, taking into account the fact that the samples are not independent and identically distributed (i.i.d) in this case. HMCMC samplers typically showcase low autocorrelation and thus thinning the sample size from $N$ to $N_{s}$ to enhance independence is often not required. In any case, $N_{s}$ can be easily determined by examining the sample autocorrelation and choosing an appropriate thinning lag, if needed. In this work, for the C.o.V calculation and $N_{s}$ in \cref{eqIS3}, we used every $3^{rd}$ sample for all examples. The same thinning process can also be used for $\tilde{P}_{F}$ in \cref{eqISCoV1}, if wanted, although we have not done this in this work and all acquired $N$ samples have been used for the probability of failure estimation.

\subsection{Summary of the ASTPA parameters}\label{SumPar}

The required input parameters for the presented methodology are summarized in \cref{tabelparam}, together with some suggested generic values for reliability estimation problems. The constant likelihood dispersion factor $\sigma$ follows the suggestions in \cref{sec4.2} and can generally be in the range of $[0.1\, \, 0.8]$. The mean, $\mu_{g}$, is chosen to be provided by the $p10$ percentile in \cref{eqmean}, and the trajectory length $\tau$ can be based on the ESJD metric, as explained in \cref{sec3.2}, with a generic value being $0.7$. The dual averaging algorithm of \citep{hoffman2014no} is adopted here to automatically provide the used constant step size $\varepsilon$ in the non-adaptive sampling phase, and the required minimum number of model calls for the QNp-HMCMC method is roughly suggested to be 5,000-10,000 for high-dimensional problems and target probabilities lower than $10^{-4}$. Naturally, the required number of model calls and samples is case dependent and convergence of the estimator can be checked through \cref{PFF}, \cref{eqIS3}, and/or \cref{eqIS4}. For the burn-in phase less than $10\%$ of the total number of samples are often required, so a $10\%$ value can be generally suggested. Finally, to compute the normalizing constant $C_{h}$ of the approximate target, we can generally use IIS samples around $20\%$ of the total number of samples.
                
\section{Numerical Results}\label{section5}
Several numerical examples are studied in this section to examine the performance and efficiency of the proposed methods. In all examples, input parameters follow the provided guidelines in \Cref{SumPar}. To compute the normalizing constant $C_{h}$, IIS samples around $20\%$ of the total number of samples have been drawn from a computed GMM with diagonal covariance matrices and, generally, $10$ and $1$ Gaussian components for low and high dimensional problems, respectively. Results are compared with the Component-Wise Metropolis-Hastings based Subset Simulation (CWMH-SuS) \cite{au2001estimation}, with two proposal distributions, a uniform one of width 2 and a standard normal one. The adaptive Conditional Sampling (aCS) SuS variant introduced in \citep{papaioannou2015mcmc} is also used in all examples, implemented based on the online provided code by the authors \citep{PapaMatlab}. In all examples, the SuS parameters are chosen as $n_{s}$ = 1,000 and 2,000 for low- and high-dimensional problems, when needed, respectively, with $n_{s}$ the number of samples in each subset level, and $p_{0} = 0.1$, where $p_{0}$ is the samples percentile for determining the appropriate subsets. Comparisons are provided in terms of accuracy and computational cost, reporting the \textit{P\textsubscript{F}} estimation and the mean number of limit-state function calls. Analytical gradients are provided in all examples, hence one limit-state/model call can provide both the relevant function values and gradients. The number of limit-state function evaluations for the HMCMC-based methods has been determined based on reported C.o.V values $ \in [0.1,\, 0.40]$, as estimated by  500 independent simulations. C.o.V values estimated by the analytical expression in \cref{eqqIS5} are also reported in parenthesis, for both HMCMC and QNp-HMCMC methods.  The reference failure probabilities are obtained based on the mean estimation of 100 independent simulations using $10^{8}$ crude Monte Carlo samples, where applicable for larger reference probabilities, and SuS method with $n_{s}$ = 100,000 for smaller reference probabilities. The problem dimensions are denoted by $d$ and all ASTPA parameters are carefully chosen for all examples but are not optimized for any one. Comparative and perhaps improved alternate performance might thus be achieved with a different set of parameters.    

\subsection{Example 1: Nonlinear convex limit-state function}
The first example consists of a nonlinear convex limit-state function characterized by two independent standard normal random variables and a low failure probability ($P_{F}$ $\sim$ 4.73E-6): 
\begin{align}
g(\boldsymbol{\theta}) = 4 - \frac{1}{\sqrt{2}}\ (\theta_{1}+\theta_{2}) + 2.5\ (\theta_{1}-\theta_{2})^{2}\label{eq23}
\end{align}  
\cref{tabel1} compares the number of model calls, C.o.V and $\mathop{\mathbb{E}}[\hat{P}_{F}]$ obtained by SuS and the two HMCMC algorithms. For the HMCMC algorithms, the trajectory length $\tau$ is chosen equal to the default value of 0.7 and the likelihood dispersion factor, $\sigma$, is 0.4. The burn-in sample size is set to 150 samples. As shown, the suggested approaches perform noticeably better than SuS, and aCS-SuS results outperform the original SuS approach with two different proposal distributions. The QNp-HMCMC method exhibits an excellent performance with only 844 model calls in this problem, on average. The simulated QNp-HMCMC samples and the analytical target distribution based on the limit-state function of \cref{eq23} can be seen in the two figures on the right hand side in \cref{fig:footprint}.\par

\newcommand{\head}[1]{\textnormal{\textbf{#1}}}
\begin{table}[t!]
\caption{Performance of various methods for the nonlinear convex limit-state function in Example 1 ($d=2$)}
\centering

\footnotesize
\setlength\tabcolsep{4pt}
\begin{tabular}{p{1cm}p{4cm}ccccccc}
  \toprule[1.5pt]
  \multirow{8}{*}{\shortstack[l]{$\sigma=0.4$ \\ $\tau = 0.7$}} & 
  \multirow{2}{4cm}{\textbf{500 Independent Simulations}} & \multicolumn{2}{c}{\textbf{CWMH-SuS}} & \multicolumn{1}{c}{\head{aCS-SuS}}& \multicolumn{1}{c}{\head{HMCMC}} & \multicolumn{1}{c}{\head{QNp-HMCMC}} \\ 
  \cline{3-4}
  
  & & $U(-1,1)$ & $N(0,1)$\\ 

 \cmidrule(lr){2-7}
 &Number of total model calls & 5,462 & 5,552 &5,453 & 1,873 & 836\\
 &C.o.V &1.06 & 1.55 &0.94 & 0.14$\color{ForestGreen}($0.15$\color{ForestGreen})$ & 0.15$\color{ForestGreen}($0.15$\color{ForestGreen})$\\
 &$\mathop{\mathbb{E}}[\hat{P}_{F}]$ \ \ \ (Ref. $P_{F}$ $\sim$ 4.73E-6) & 4.88E-6 & 5.11E-8 &4.52E-6 &4.73E-6& 4.72E-6\\
 \bottomrule[1.5pt]

\end{tabular}\label{tabel1}
\end{table}

\subsection{Example 2: Parabolic/Concave limit-state function}
\begin{figure}[t!]

\centerline{\subfigure[]{\includegraphics[trim=0cm 3.5cm 0cm 3cm,width=0.39\textwidth]{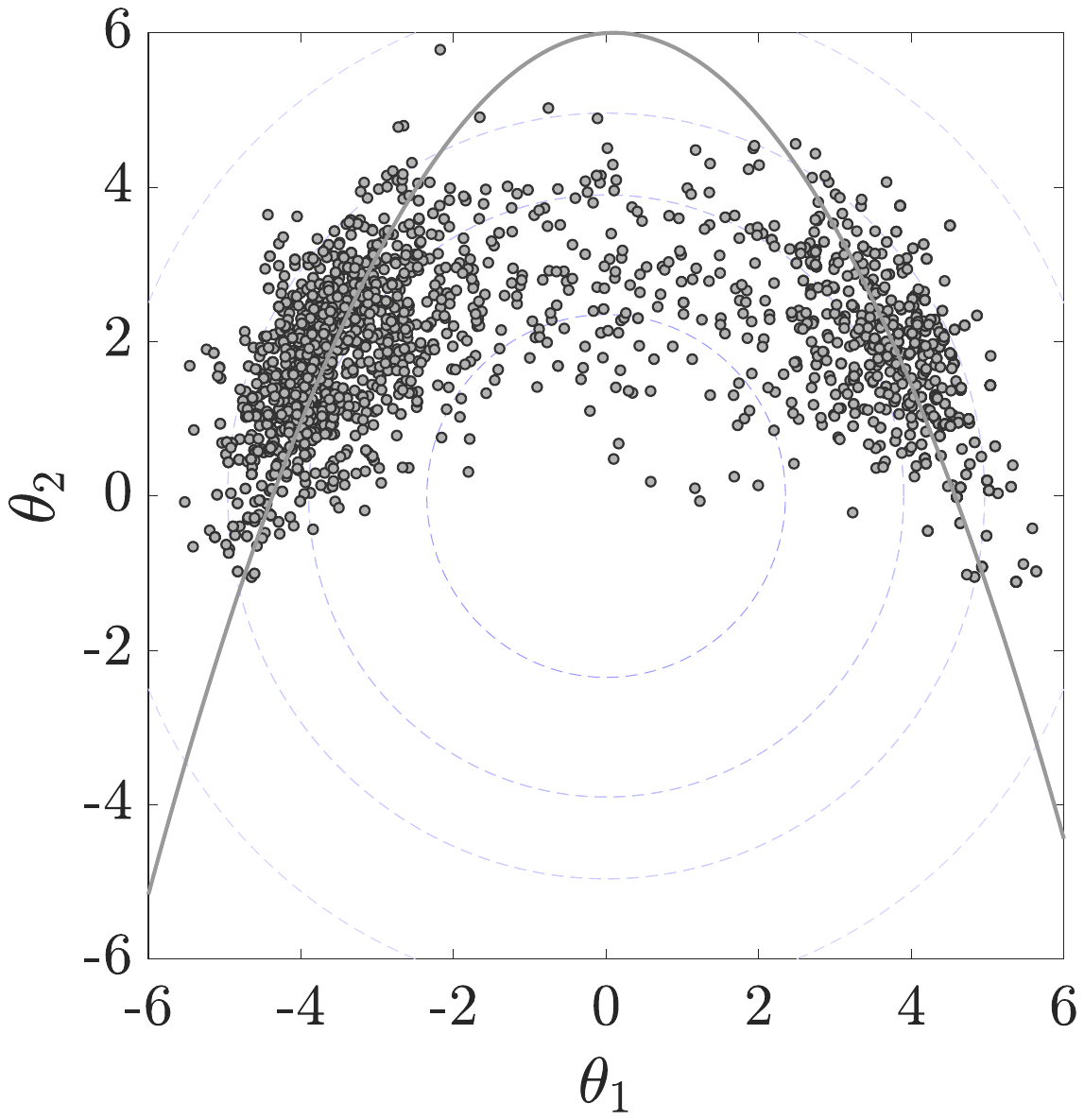}}
\quad
\subfigure[]{\includegraphics[trim=0cm 3.5cm 0cm 3cm,width=0.39\textwidth]{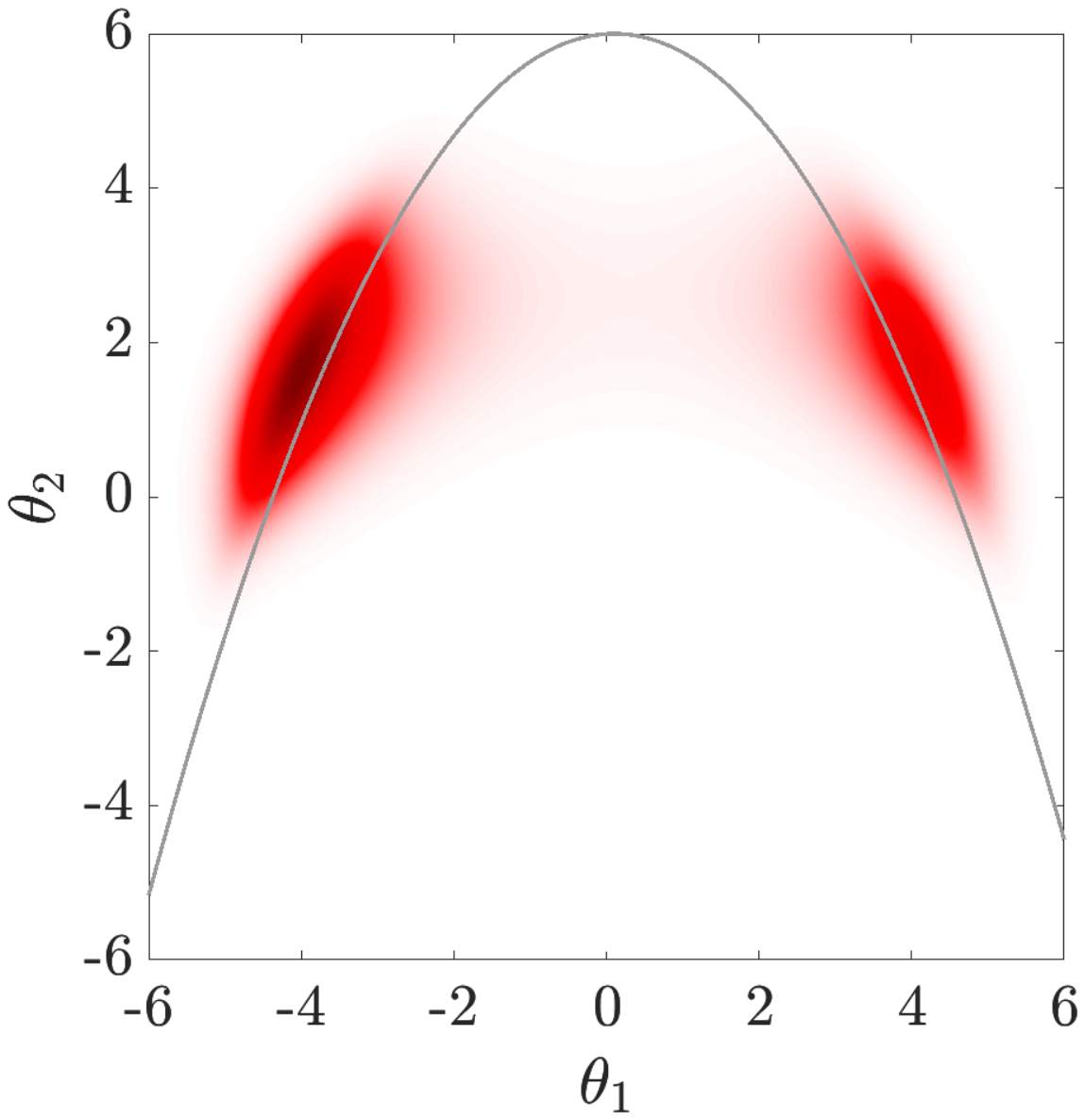}}}
  \captionsetup{labelfont={color=Black}}
\caption{Example 2: (a) Simulated samples from the target distribution, (b) Analytical target distribution. \vspace{-0.1in}}\label{fig6}
\end{figure}

This example is based on the following limit state function with two standard
normal random variables \citep{der1998multiple}:
\begin{align}
g(\boldsymbol{\theta}) = r - \theta_{2} - \kappa\ (\theta_{1}-e)^{2}
\end{align}   
where $r$, $\kappa$ and $e$ are deterministic parameters chosen as $r = 6$, $\kappa=0.3$ and $e = 0.1$. The probability of failure is 3.95E-5 and the limit-state function describes two main failure modes, as also seen in \cref{fig6}. For the HMCMC-based algorithms, the likelihood dispersion factor, $\sigma$, is 0.7, the burn-in sample size is 200 samples, and the trajectory length is set to $\tau=1$.\par
\cref{tabel2} compares the number of model calls, the C.o.V and the $\mathop{\mathbb{E}}[\hat{P}_{F}]$ obtained by all methods. The HMCMC-based approaches provide significantly smaller C.o.V values than SuS, with fewer model calls, and the HMCMC in particular outperforms all other methods in this two dimensional problem. The QNp-HMCMC samples are also shown in \cref{fig6} and can accurately represent the two important failure regions. The circular dash lines in \cref{fig6}(a) represent the probability contour of the standard normal space $\boldsymbol{\Theta}$, for probability levels of $10^{-2}$, $10^{-4}$, $10^{-6}$ and $10^{-10}$, similarly provided for \Cref{fig7,figHimmelblau2} as well.

\begin{table}[t!]
\caption{Performance of various methods for the parabolic/concave limit-state function in Example 2 ($d=2$)}
\centering

\footnotesize
\setlength\tabcolsep{4pt}
\begin{tabular}{p{1cm}p{4cm}ccccccc}
  \toprule[1.5pt]
  \multirow{8}{*}{\shortstack[l]{$\sigma=0.7$ \\$\tau = 1$}} & 
  \multirow{2}{4cm}{\textbf{500 Independent Simulations}} & \multicolumn{2}{c}{\textbf{CWMH-SuS}} & \multicolumn{1}{c}{\head{aCS-SuS}}& \multicolumn{1}{c}{\head{HMCMC}} & \multicolumn{1}{c}{\head{QNp-HMCMC}}\\ 

  \cline{3-4}

  & & $U(-1,1)$ & $N(0,1)$\\ 

 \cmidrule(lr){2-7}
 &Number of total model calls & 4,559 & 4,565 &4,562& 3,306&  3,306\\
 &C.o.V &0.62 & 0.65 & 0.63& 0.09$\color{ForestGreen}($0.06$\color{ForestGreen})$&0.09$\color{ForestGreen}($0.06$\color{ForestGreen})$\\
 &$\mathop{\mathbb{E}}[\hat{P}_{F}]$ \ \ \ \ (Ref. $P_{F}$ $\sim$ 3.95E-5) & 4.19E-5 & 4.14E-5 &4.09E-5 & 3.93E-5&3.88E-5\\
 \bottomrule[1.5pt]
\end{tabular}\label{tabel2}
\vspace{-0.1in}
\end{table}


\subsection{Example 3: Quartic bimodal limit-state function}
\begin{figure}[t!]

\centerline{\subfigure[]{\includegraphics[trim=0cm 3.5cm 0cm 3cm,width=0.39\textwidth]{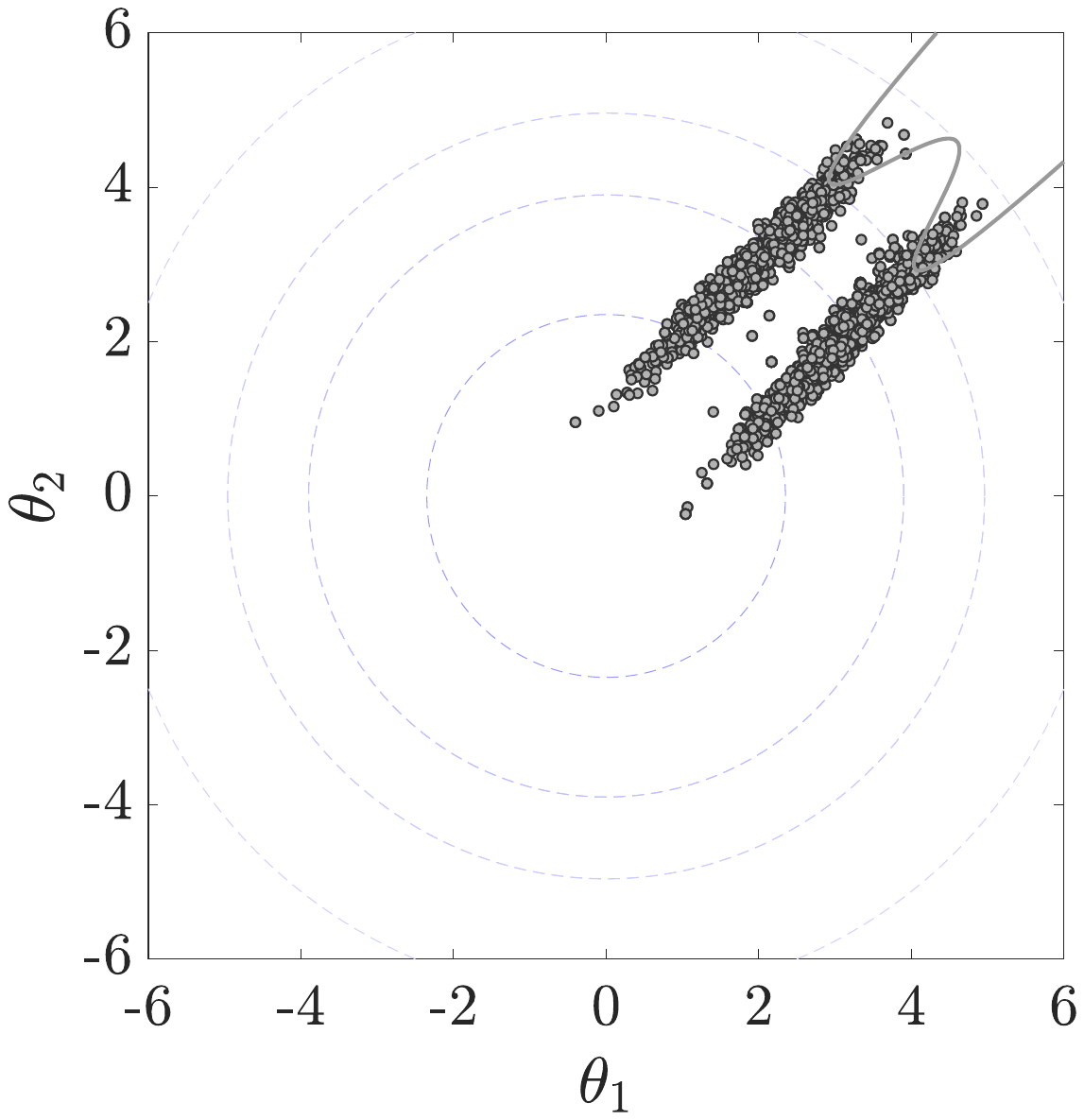}}
\quad
\subfigure[]{\includegraphics[trim=0cm 3.5cm 0cm 3cm,width=0.39\textwidth]{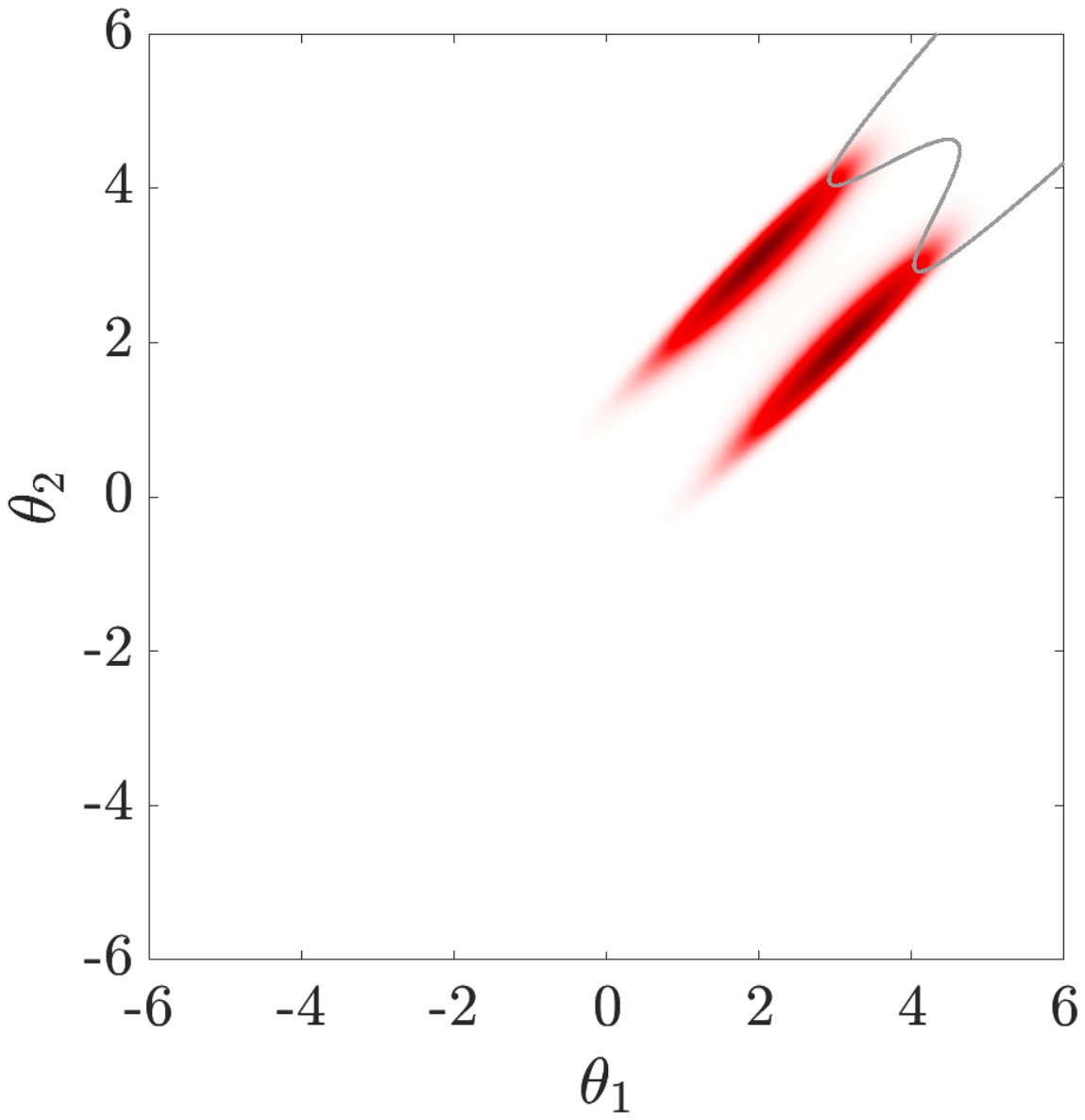}}}
  \captionsetup{labelfont={color=Black}}
\caption{Example 3: (a) Simulated samples from the target distribution, (b) Analytical target distribution.}\label{fig7}
\end{figure}

The third example is a quartic bimodal limit-state function with very low probability of failure ($P_{F}$ $\sim$ 5.90E-8), defined by the following limit-state function in the standard normal space: 
\begin{align}
g(\boldsymbol{\theta}) = 6.5 - \frac{1}{\sqrt{2}}\ (\theta_{1}+\theta_{2}) - 2.5\ (\theta_{1}-\theta_{2})^{2}+\ (\theta_{1}-\theta_{2})^{4}\label{eq25}
\end{align} 
\cref{tabel3} compares the performance of the HMCMC methods with the ones by SuS. 
The trajectory length is chosen $\tau=0.7$, and the likelihood dispersion factor, $\sigma$, is 0.5. The burn-in sample size is set to 200 samples. HMCMC-based methods, particularly QNp-HMCMC, once more provide significantly lower C.o.V values with fewer model calls than SuS approaches, that perform rather poorly in this example, with aCS-SuS performing the best among them. \cref{fig7} displays related target distribution based on the limit-state function of \cref{eq25}, as well as the samples generated by the QNp-HMCMC method.

\begin{table}[t!]
\caption{Performance of various methods for the quartic bimodal limit-state function in Example 3 ($d=2$)}
\centering

\footnotesize
\setlength\tabcolsep{4pt}
\begin{tabular}{p{1cm}p{4cm}ccccccc}
  \toprule[1.5pt]
  \multirow{8}{*}{\shortstack[l]{$\sigma=0.5$ \\$\tau = 0.7$}} & 
  \multirow{2}{4cm}{\textbf{500 Independent Simulations}} & \multicolumn{2}{c}{\textbf{CWMH-SuS}} & \multicolumn{1}{c}{\head{aCS-SuS}}& \multicolumn{1}{c}{\head{HMCMC}} & \multicolumn{1}{c}{\head{QNp-HMCMC}}\\ 

  \cline{3-4}

  & & $U(-1,1)$ & $N(0,1)$\\ 

 \cmidrule(lr){2-7}
 &Number of total model calls & 7,327 & 7,536 & 7,380 & 6,277 &2,696\\
 &C.o.V &1.64 & 2.45 &1.59& 0.28$\color{ForestGreen}($0.27$\color{ForestGreen})$&0.27$\color{ForestGreen}($0.27$\color{ForestGreen})$\\
 &$\mathop{\mathbb{E}}[\hat{P}_{F}]$ \ \ \ \ (Ref. $P_{F}$ $\sim$ 5.90E-8) & 6.13E-8 & 5.86E-8&6.12E-8 & 5.90E-8&5.89E-8\\
 \bottomrule[1.5pt]
\end{tabular}\label{tabel3}
\end{table}

\subsection{Example 4: The Himmelblau Function}
In nonlinear optimization, a commonly used fourth order polynomial test function is the so-called Himmelblau \citep{himmelblau1972applied} function. Here we adopt and modify this function, as:
	\begin{equation}
	\begin{aligned}
	g(\theta_{1},\theta_{2}) =&  \big (\dfrac{(0.75\theta_{1} - 0.5)^{2}}{1.81} + \dfrac{(0.75\theta_{2} - 0.5)}{1.81} - 11 \big )^{2} + \big (\dfrac{(0.75\theta_{1} - 1)}{1.81} + \dfrac{(0.75\theta_{2} - 0.5)^{2}}{1.81} - 7 \big )^{2} -\beta               \label{exHimmelblau}
	\end{aligned}
	\end{equation}
	
which is particularly suitable for reliability examples with multiple separated failure domains. $\theta_{1}$ and $\theta_{2}$ are assumed to be independent standard normal random variables and the constant $\beta$ is used to define different levels of the failure probability. \cref{tabelHimmelblau} compares the number of model calls, coefficient of variation and $\mathop{\mathbb{E}}[\hat{P}_{F}]$ obtained by SuS and the family of HMCMC algorithms. For the HMCMC algorithms, the trajectory length is chosen as $\tau=1$ and the burn-in is set to 200 samples. The $g(\textbf{0})$ is beyond the upper bound, $g(\textbf{0})> 7$, and as discussed in \cref{section4}, we perform the scaling with $g_{c}=g(\textbf{0})$$/$$4$. The likelihood dispersion factor $\sigma$ used in this example is mentioned in \cref{tabelHimmelblau}, and has been chosen according to the guidelines in \Cref{sec4.2}, considering both multimodality and the different levels of the failure probabilities. The Subset Simulation results are based on $n_{s}=\,$1,000.

\begin{figure}[t!]
		\centerline{\subfigure[]{\includegraphics[trim=0cm 3.5cm 0cm 3cm,width=0.39\textwidth]{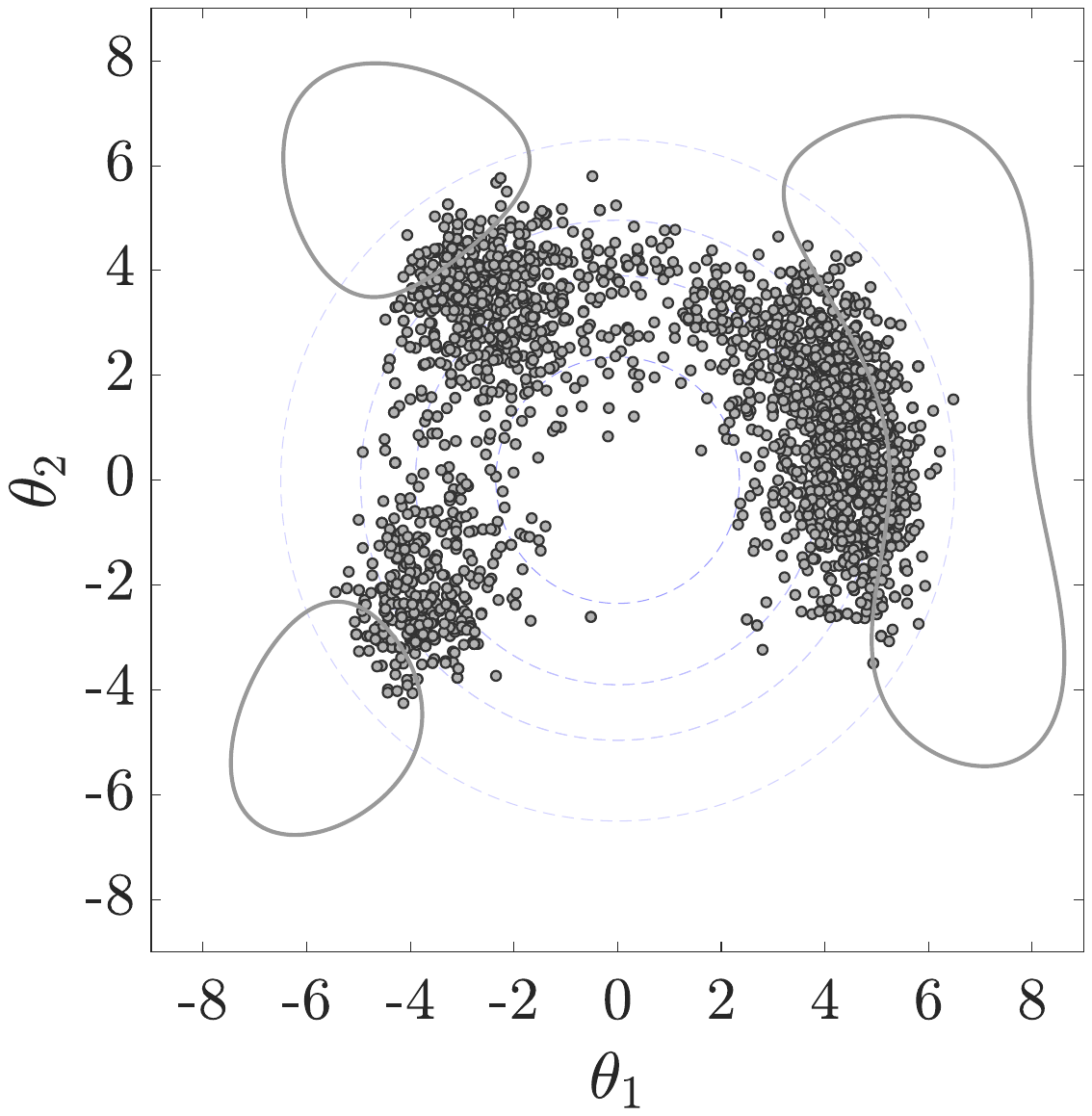}}
			\quad
			\subfigure[]{\includegraphics[trim=0cm 3.5cm 0cm 3cm,width=0.39\textwidth]{Himtarget.pdf}}}
			  \captionsetup{labelfont={color=Black}}
		\caption{Example 4: (a) Simulated samples from the target distribution, (b) Analytical target distribution.}\label{figHimmelblau2}
	\end{figure}

 It is again shown here that the HMCMC approach gives significantly smaller C.o.V than SuS and also outperforms it in terms of the $\mathop{\mathbb{E}}[\hat{P}_{F}]$ estimation. \cref{figHimmelblau2} also demonstrates that the HMCMC samples accurately describe all the three important failure regions.

\begin{table}[t!]
\caption{Performance of various methods for the Himmelblau limit-state function in Example 4 ($d=2$)}
\centering

\footnotesize

\setlength\tabcolsep{4pt}
 \begin{tabular}{p{1.2cm}p{4cm}ccccccc}
   \toprule[1.5pt]

   \multirow{7}{*}{\shortstack[l]{\vspace{0.15in}\\$\beta=95$\\$\sigma=0.5$ \\$\tau = 1$}} & \multirow{2}{4cm}{\textbf{500 Independent Simulations}} & \multicolumn{2}{c}{\textbf{CWMH-SuS}} & \multicolumn{1}{c}{\head{aCS-SuS}}& \multicolumn{1}{c}{\head{HMCMC}} & \multicolumn{1}{c}{\head{QNp-HMCMC}}\\
 
   \cline{3-4}

   & & $U(-1,1)$ & $N(0,1)$\\ 

  \cmidrule(lr){2-7}
   &Number of total model calls &3,833  &3,833 &3,821  &3,100 &3,100 \\
   &C.o.V &0.42 &0.50 &0.35 &0.10$\color{ForestGreen}($0.06$\color{ForestGreen})$  &0.13$\color{ForestGreen}($0.06$\color{ForestGreen})$ \\
   &$\mathop{\mathbb{E}}[\hat{P}_{F}]$ \ \ \  (Ref. $P_{F}$ $\sim$ 1.65E-4 ) &1.69E-4  &1.78E-4 &1.68E-4  &1.64E-4 &1.62E-4 \\
  \bottomrule[1.5pt]
     \multirow{3.5}{*}{\shortstack[l]{$\beta=50$\\$\sigma=0.4$ \\$\tau = 1$ }}\rule{0pt}{2.5ex} 
       &Number of total model calls & 6,471 &6,528  &6,463 &3,600 &3,600 \\
       &C.o.V &0.87 &1.67 &0.54&0.16$\color{ForestGreen}($0.10$\color{ForestGreen})$&0.16$\color{ForestGreen}($0.10$\color{ForestGreen})$\\
       &$\mathop{\mathbb{E}}[\hat{P}_{F}]$ \ \ \  (Ref. $P_{F}$ $\sim$ 2.77E-7) &3.14E-7 &3.33E-7 &2.89E-7&2.77E-7&2.77E-7\\
      \bottomrule[1.5pt]  
 \end{tabular}\label{tabelHimmelblau}
\end{table}

\subsection{Example 5: Cantilever beam}
\begin{figure}[t]
 \centering
 \includegraphics[width=0.6\textwidth]{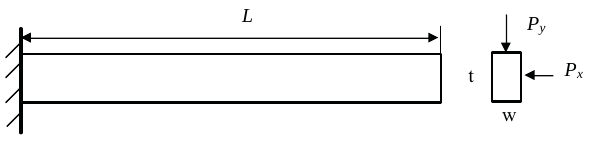}\vspace*{-0.2in}
 \caption{A cantilever beam case study.}
 \label{fig:13}
 \end{figure}

In this last two-dimensional example, a cantilever beam problem is studied \citep{du2005first}. The beam is illustrated in \cref{fig:13}, with cross-section width $w$, height $t$, beam length $L$, and transverse loads $P_{x}$ and $P_{y}$. The beam failure mode in this case is the maximum deflection exceeding the allowable value, $Y_{0}$, given by the limit-state function: 

 \begin{align}
  g(P_{x},P_{y}) = Y_{0} - \dfrac{4L^{3}}{Ewt} \sqrt{(\frac{P_{y}}{t^{2}})^{2} + (\frac{P_{x}}{w^{2}})^{2}} 
 \end{align}  
where $Y_{0} = 4.2\, in$ and $4.5\, in$, $E = 30\times10^{6} \, psi$ is the elastic modulus, $L = 100 \, in$, $w = 2 \, in$ and $t = 4 \, in$. $P_{x}$ and $P_{y}$ follow independent normal distributions $P_{x} \sim N(500,100) lb$ and $P_{y} \sim N(1000,100) lb$.
The normally distributed variables $P_{x}$ and $P_{y}$ are transformed into the standard normal space and the limit-state function in the $\boldsymbol{\Theta}$ space becomes:
\begin{align}
 g(\theta_{x},\theta_{y}) = Y_{0} - \dfrac{4L^{3}}{Ewt} \sqrt{(\frac{\mu_{y} + \theta_{y} \sigma_{y}}{t^{2}})^{2} + (\frac{\mu_{x} + \theta_{x} \sigma_{x}}{w^{2}})^{2}}\label{EqBeam} 
\end{align}  

\begin{table}[t!]
\caption{Performance of various methods for the cantilever beam in Example 5 ($d=2$)}
\centering

\footnotesize

\setlength\tabcolsep{4pt}
 \begin{tabular}{p{1.2cm}p{4cm}ccccccc}
   \toprule[1.5pt]
 
   \multirow{7}{*}{\shortstack[l]{\vspace{0.15in}\\$\sigma = 0.2$\\$Y_{0} = 4.2$ \\$\tau = 0.7$ }} & \multirow{2}{4cm}{\textbf{500 Independent Simulations}} & \multicolumn{2}{c}{\textbf{CWMH-SuS}} & \multicolumn{1}{c}{\head{aCS-SuS}}& \multicolumn{1}{c}{\head{HMCMC}} & \multicolumn{1}{c}{\head{QNp-HMCMC}}\\

   \cline{3-4}
 
   & & $U(-1,1)$ & $N(0,1)$\\ 

  \cmidrule(lr){2-7}
   &Number of total model calls &6,056  &6,069 &6,062  &1,900 &1,900 \\
   &C.o.V &0.80 &0.93 &0.50  &0.15$\color{ForestGreen}($0.17$\color{ForestGreen})$ &0.14$\color{ForestGreen}($0.17$\color{ForestGreen})$ \\
   &$\mathop{\mathbb{E}}[\hat{P}_{F}]$ \ \ \  (Ref. $P_{F}$ $\sim$ 1.01E-6 ) &1.07E-6  &1.08E-6 &1.05E-6  &1.01E-6 &1.01E-6 \\
  \bottomrule[1.5pt]
     \multirow{3.5}{*}{\shortstack[l]{$\sigma = 0.2$\\$Y_{0} = 4.5$ \\$\tau = 0.7$ }}\rule{0pt}{2.5ex} 
       &Number of total model calls &7,561  &7,586 &7,569 &3,200  &3,200 \\
       &C.o.V &1.11  &1.41 &0.60 &0.19$\color{ForestGreen}($0.24$\color{ForestGreen})$  &0.19$\color{ForestGreen}($0.24$\color{ForestGreen})$\\
       &$\mathop{\mathbb{E}}[\hat{P}_{F}]$ \ \ \  (Ref. $P_{F}$ $\sim$ 1.97E-8) &2.14E-8  &2.22E-8 &2.03E-8 &1.97E-8  &1.98E-8\\
      \bottomrule[1.5pt]  
 \end{tabular}\label{table8}
\end{table}

\noindent For the HMCMC-based algorithms, the trajectory length is chosen $\tau=0.7$ and the likelihood dispersion factor, $\sigma$, is 0.2. The burn-in sample size is set to 200 samples. \cref{table8} compares the number of model calls, C.o.V and $\mathop{\mathbb{E}}[\hat{P}_{F}]$ for two different failure probability levels obtained by SuS and the HMCMC-based methods. As shown, the HMCMC-based approaches noticeably again outperform the SuS methods. 

\subsection{Example 6: High-dimensional reliability example with linear limit-state function}
In this first high-dimensional example, a linear limit-state function of independent standard normal random variables is considered: 
\begin{align}
g(\boldsymbol{\theta}) = \beta - \frac{1}{\sqrt{d}}\ \sum_{i=1}^{d} \theta_{i}\label{ex5}
\end{align}  
where $d$ denotes again the related problem dimensions, and the pertinent probability of failure is equal to $\Phi(-\beta)$, independent of $d$, with $\Phi(.)$ the cumulative distribution function of the standard normal distribution. To investigate the effect of different failure probability levels, in relation to the HMCMC-based methods and SuS performance, a sequence of $\beta$ values for $d=100$ and $500$ are studied. \cref{tabel5} summarizes the comparative results through the mean number of model calls, mean $\mathop{\mathbb{E}}[\hat{P}_{F}]$ and C.o.V of the estimated probabilities. For both HMCMC-based methods, the likelihood dispersion factor, $\sigma$, is equal to 0.3, the burn-in sample size is 300, and the trajectory length, $\tau$, for all failure probability levels is chosen equal to 0.7, the default value. SuS results are again based on $n_{s}=\,$1,000.

\begin{figure}[t!]
\centerline{\subfigure[]{\includegraphics[clip, trim=0cm 3cm 1cm 3cm, width=0.324\textwidth]{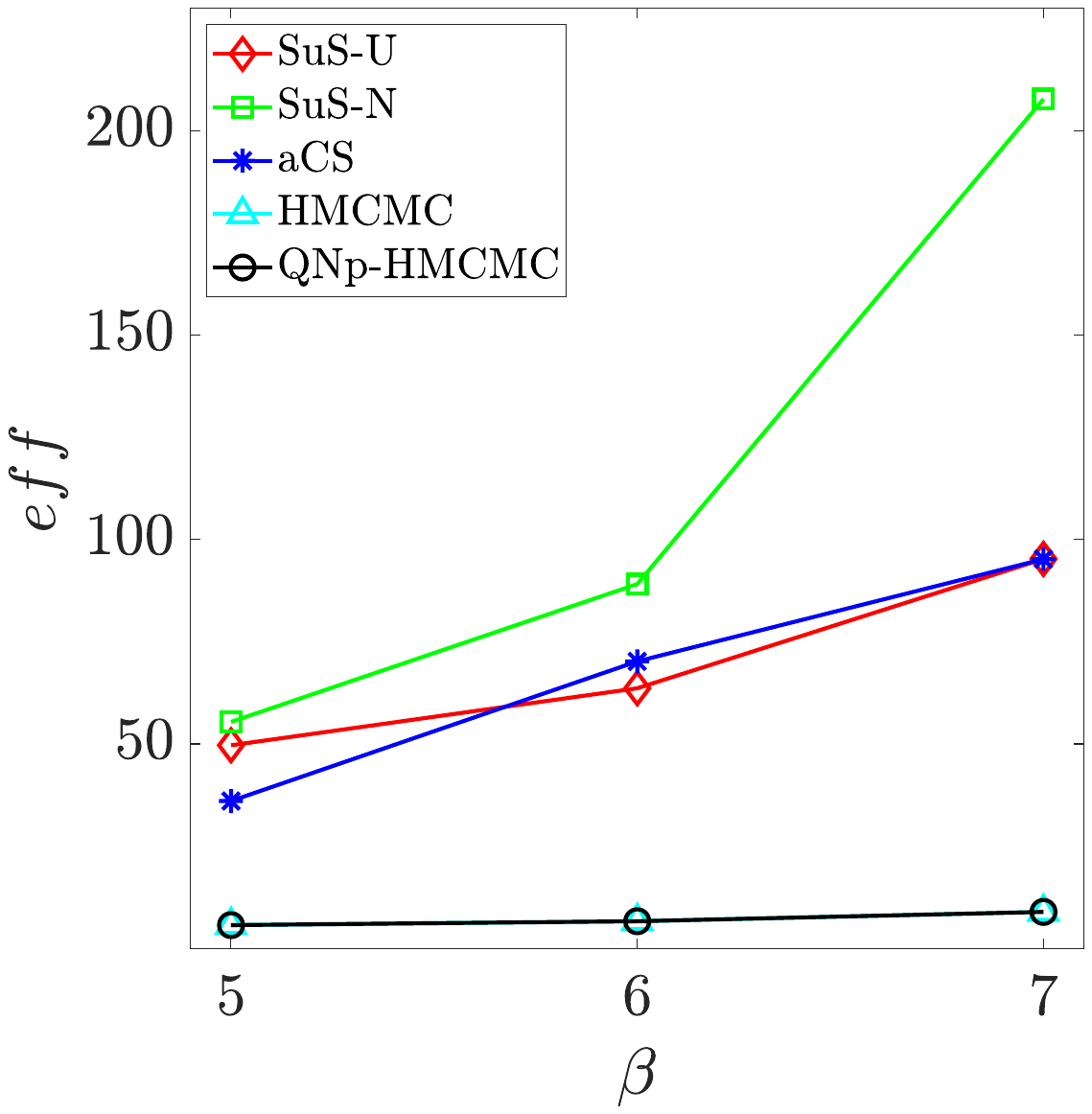}}
\quad
\subfigure[]{\includegraphics[clip, trim=0cm 3cm 1cm 3cm, width=0.324\textwidth]{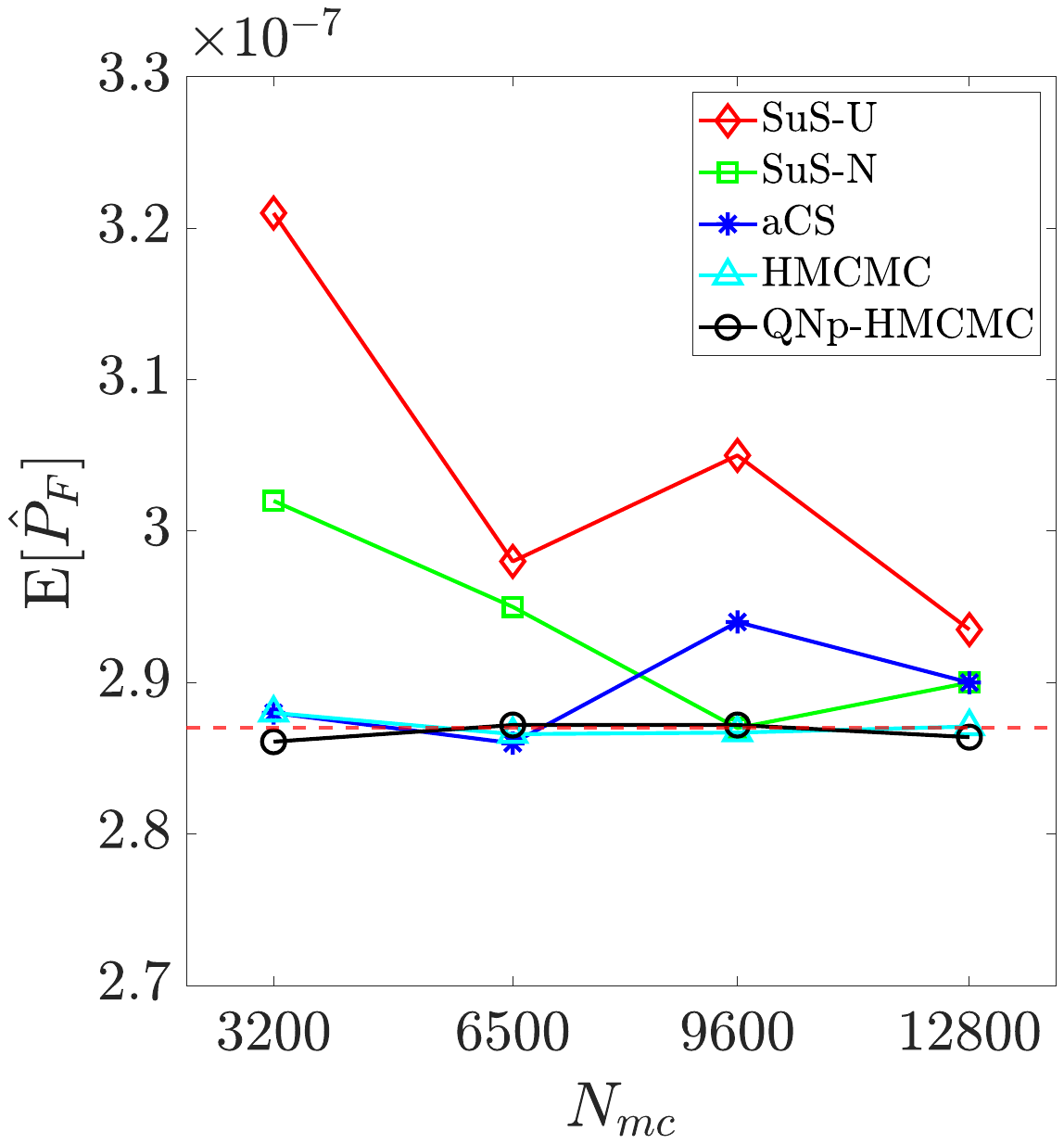}}
\quad
\subfigure[]{\includegraphics[clip, trim=0cm 3cm 1cm 3cm, width=0.324\textwidth]{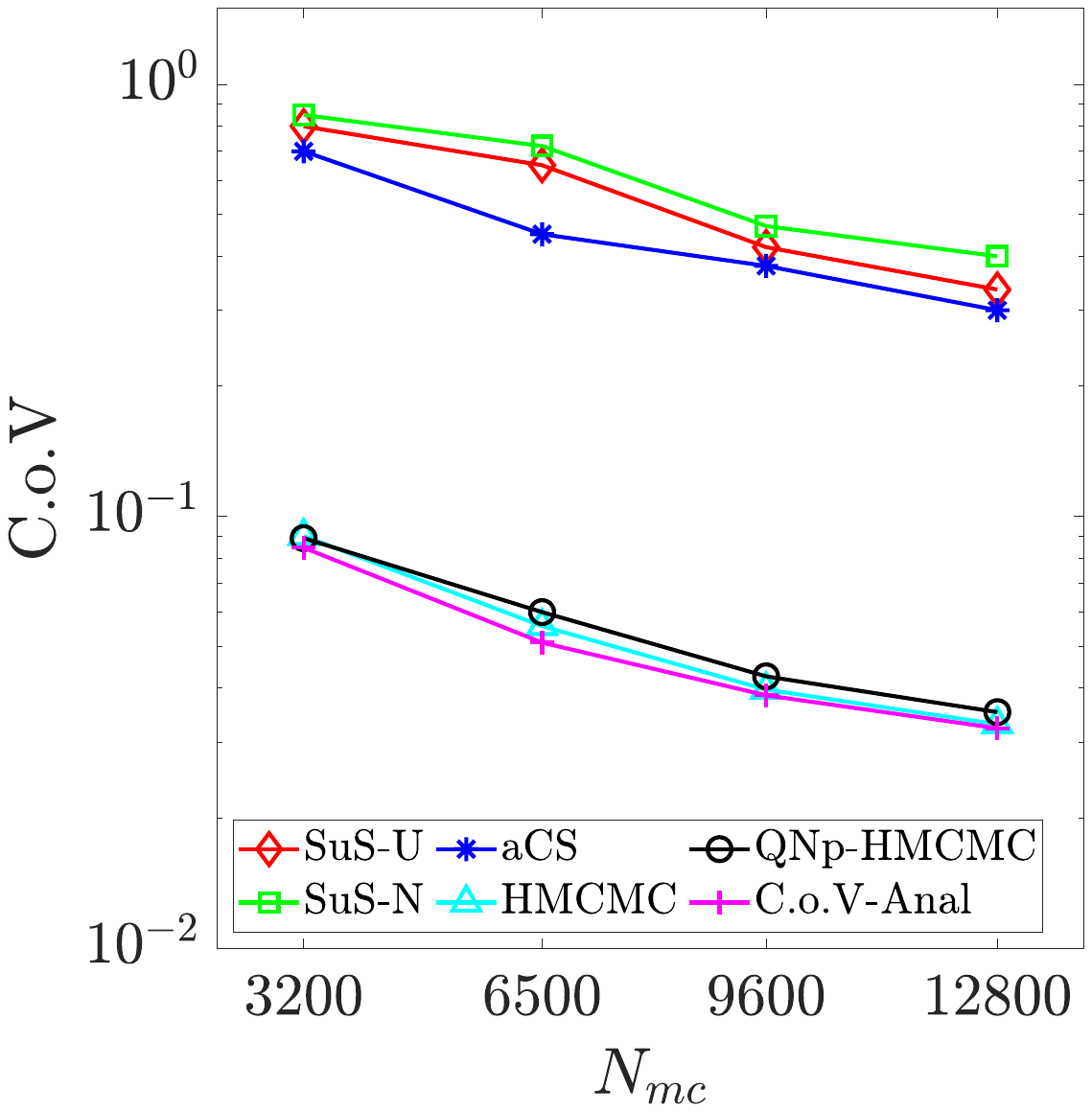}}}
  \captionsetup{labelfont={color=Black}}
\caption{Example 6: (a) \textit{eff} versus different $\beta$ values based on the results presented in \cref{tabel5} for $d=100$; a low \textit{eff} value indicates high efficiency. (b) Failure probability estimate against the number of model calls ($N_{mc}$) for $\beta = 5$ and $d=100$ (dash line is the reference failure probability). (c) Coefficient of variation  of the probability estimates plotted against the number of model calls for $\beta = 5$ and $d=100$ .\vspace{-0.2in}}\label{fig:9}
\end{figure}

As reported in \cref{tabel5}, the HMCMC-based approaches exhibit for all probability levels accurate and stable results in terms of C.o.V., outperforming all SuS results. Defining the "unit C.o.V" \textit{eff} as C.o.V =  \textit{eff}$/\sqrt{N_{mc}}$, with $N_{mc}$ the total number of model calls, an index can now be used that appropriately considers both accuracy and computational effort \citep{au2007application}, with a lower \textit{eff} value exhibiting higher efficiency, of course. \cref{fig:9}a displays the \textit{eff} variation in relation to the reliability index for $d=100$, confirming the HMCMC approaches efficiency for various failure probability levels. SuS-U and SuS-N in the figure stand for the SuS results with uniform and standard normal proposals, respectively. \cref{fig:9}b showcases the mean estimate for $\beta=5$ and $d=100$, based on the number of model calls. As shown, the HMCMC methods provide a consistent unbiased estimator after a certain, relatively small, number of model evaluations. Finally, in \cref{fig:9}c a similar plot is provided for the C.o.V results, with the proposed HMCMC-based framework exhibiting again excellent overall performance. The CoV-Anal curve in the figure represents the QNp-HMCMC C.o.V estimation based on the analytical expression in \cref{eqqIS5}.

\begin{table}[t!]
\caption{Performance of various methods for the high dimensional problem with linear limit-state function in Example 6}
\centering

\footnotesize

\setlength\tabcolsep{4pt}
\begin{tabular}{p{1cm}p{0.7cm}ccccccc}
  \toprule[1.5pt]
 
  \multirow{8}{*}{\shortstack[l]{\vspace{0in}\\$\beta=5$\\$\sigma = 0.3$\\ $\tau = 0.7$}} & \multirow{2}{*}{\hspace{0.07in}$d$} & \multirow{2}{4cm}{\textbf{500 Independent Simulations}} & \multicolumn{2}{c}{\textbf{CWMH-SuS}} & \multicolumn{1}{c}{\head{aCS-SuS}}& \multicolumn{1}{c}{\head{HMCMC}} & \multicolumn{1}{c}{\head{QNp-HMCMC}}\\  

  \cline{4-5}

  && & $U(-1,1)$ & $N(0,1)$\\ 

 \cmidrule(lr){2-8}

   &\multirow{3}{*}{\hspace{0.07in}100}&Number of total model calls & 6,418 & 6,443 &6,409 &2,225 &2,225\\
   &&C.o.V &0.62 & 0.69&0.45 &0.12$\color{ForestGreen}($0.12$\color{ForestGreen})$  &0.12$\color{ForestGreen}($0.11$\color{ForestGreen})$\\
   &&$\mathop{\mathbb{E}}[\hat{P}_{F}]$ \ \ \  (Ref. $P_{F}$ $\sim$ 2.87E-7) & 2.97E-7 & 2.94E-7 &2.86E-7&2.86E-7  &2.87E-7\\
  \bottomrule[1.5pt]  
  
  \multirow{4}{*}{\shortstack[l]{\vspace{-0.15in}\\$\beta=6$\\$\sigma = 0.3$\\ $\tau = 0.7$}}\rule{0pt}{2.5ex}
    &\multirow{3}{*}{\hspace{0.07in}100}&Number of total model calls & 8,711 & 8,798 &9,279&2,226  &2,228\\
    &&C.o.V &0.62 & 0.95&0.58 &0.14$\color{ForestGreen}($0.13$\color{ForestGreen})$  &0.14$\color{ForestGreen}($0.13$\color{ForestGreen})$\\
    &&$\mathop{\mathbb{E}}[\hat{P}_{F}]$ \ \ \  (Ref. $P_{F}$ $\sim$ 0.99E-9) & 1.05E-9 & 1.01E-9&1.03E-9 &0.98E-9  &0.99E-9\\
   \bottomrule[1.5pt]
   
    \multirow{4}{*}{\shortstack[l]{\vspace{-0.15in}\\$\beta=7$\\$\sigma = 0.3$\\ $\tau = 0.7$}}\rule{0pt}{2.5ex}
     &\multirow{3}{*}{\hspace{0.07in}100} &Number of total model calls & 11,458 & 11,473 &11,922&2,736 &2,735\\
     & &C.o.V &0.89 & 1.94&0.77 &0.17$\color{ForestGreen}($0.16$\color{ForestGreen})$   &0.17$\color{ForestGreen}($0.16$\color{ForestGreen})$\\
     & &$\mathop{\mathbb{E}}[\hat{P}_{F}]$ \ \ \  (Ref. $P_{F}$ $\sim$ 1.28E-12) & 1.36E-12 & 1.32E-12&1.31E-12 &1.28E-12  &1.28E-12\\
     \bottomrule[1.5pt]

         \multirow{4}{*}{\shortstack[l]{\vspace{-0.15in}\\$\beta=6$\\$\sigma = 0.3$\\ $\tau = 0.7$}}\rule{0pt}{2.5ex}
     &\multirow{3}{*}{\hspace{0.07in}500} &Number of total model calls & 8,760& 8,808&9,271&5,439&5,532\\
     & &C.o.V &0.67 & 1.05&0.60 &0.25$\color{ForestGreen}($0.19$\color{ForestGreen})$   &0.24$\color{ForestGreen}($0.19$\color{ForestGreen})$\\
     & &$\mathop{\mathbb{E}}[\hat{P}_{F}]$ \ \ \  (Ref. $P_{F}$ $\sim$ 0.99E-9) & 1.01E-9 & 1.01E-9&1.04E-9 &1.00E-9  &0.99E-9\\
     \bottomrule[1.5pt]

         \multirow{4}{*}{\shortstack[l]{\vspace{-0.15in}\\$\beta=7$\\$\sigma = 0.3$\\ $\tau = 0.7$}}\rule{0pt}{2.5ex}
     &\multirow{3}{*}{\hspace{0.07in}500} &Number of total model calls & 11,334 & 11,870 &11,908& 5,634 &5,583\\
     & &C.o.V &0.92 & 2.25&0.82 &0.27$\color{ForestGreen}($0.25$\color{ForestGreen})$   &0.30$\color{ForestGreen}($0.24$\color{ForestGreen})$\\
     & &$\mathop{\mathbb{E}}[\hat{P}_{F}]$ \ \ \  (Ref. $P_{F}$ $\sim$ 1.28E-12) & 1.45E-12 & 1.15E-12&1.39E-12 &1.16E-12  &1.16E-12\\
     \bottomrule[1.5pt]
\end{tabular}\label{tabel5}
\end{table}

\subsection{Example 7: High-dimensional problem with quadratic nonlinearity}
This example involves a quadratic limit-state function expressed in the standard normal space, as:
\begin{align}
g(\boldsymbol{\theta}) = \lambda - \frac{1}{\sqrt{d}}\ \sum_{i=1}^{d} \theta_{i} + 2.5\ \bigg (\theta_{1} - \sum_{j=2}^{\gamma} \theta_{j} \bigg )^{2}
\end{align}  
where $d$ is the problem dimension, $\lambda$ defines the level of the failure probability, and $\gamma$ affects the level of nonlinearity. To investigate the effect of dimensionality and nonlinearity, in relation to the HMCMC-based methods and SuS performance, different $\gamma$ values for $d=100$ and $200$ are studied. \cref{tabel6} presents the mean number of limit-state function evaluations, sample mean estimator, and C.o.V of the failure probabilities for various methods. For the HMCMC-based methods, the likelihood dispersion factor, $\sigma$, and the trajectory length, $\tau$, are shown in \cref{tabel6}, and the burn-in sample size is set to 500 samples. The Subset Simulation results for all approaches are based on $n_{s}=\,$2,000.\par

As seen in \cref{tabel6}, the QNp-HMCMC approach significantly outperforms all other methods, in all considered metrics, confirming its superiority, robustness, applicability and suitability in challenging, nonlinear, high-dimensional problems. By increasing the level of nonlinearity, through the parameter $\gamma$, HMCMC requires an excessive number of model calls in order to adequately explore the domain, while the QNp-HMCMC approach exhibits a largely invariant performance. \par

\begin{table}[t!]
\caption{Performance of various methods for the high dimensional problem with nonlinear limit-state function in Example 7}
\centering

\footnotesize

\setlength\tabcolsep{4pt}
\begin{tabular}{p{1.1cm}p{0.8cm}ccccccc}
  \toprule[1.5pt]
  \multirow{8}{*}{\shortstack[l]{\vspace{-0.05in}\\$\lambda=4.0$\\$\gamma=10$\\$\sigma = 0.5$\\ $\tau = 0.7$}}  & \multirow{2}{*}{\hspace{0.07in}$d$}& \multirow{2}{4cm}{\textbf{500 Independent Simulations}} & \multicolumn{2}{c}{\textbf{CWMH-SuS}} & \multicolumn{1}{c}{\head{aCS-SuS}}& \multicolumn{1}{c}{\head{HMCMC}} & \multicolumn{1}{c}{\head{QNp-HMCMC}}\\

  \cline{4-5}

  && & $U(-1,1)$ & $N(0,1)$\\ 

 \cmidrule(lr){2-8}

    &\multirow{3}{*}{\hspace{0.07in}100}   &Number of total model calls & 12,109 & 12,192&12,093 &21,198  &4,695 \\
     &  &C.o.V & 1.43 & 1.71&2.18 &0.16$\color{ForestGreen}($0.15$\color{ForestGreen})$  &0.16$\color{ForestGreen}($0.15$\color{ForestGreen})$\\
     &  &$\mathop{\mathbb{E}}[\hat{P}_{F}]$ \ \ \  (Ref. $P_{F}$ $\sim$ 1.15E-6) & 1.22E-6 & 1.31E-6&1.23E-6 &1.16E-6  &1.16E-6\\
      \bottomrule[1.5pt]

           \multirow{4}{*}{\shortstack[l]{\vspace{-0.14in}\\$\lambda=3.0$\\$\gamma=50$\\$\sigma = 0.5$\\ $\tau = 0.7$}}\rule{0pt}{2.5ex} 
    &\multirow{3}{*}{\hspace{0.07in}100}   &Number of total model calls & 13,758& 14,200&14,442 &53,664 &5,924 \\
     &  &C.o.V & 2.94& 4.04&5.78 &0.21$\color{ForestGreen}($0.21$\color{ForestGreen})$  &0.21$\color{ForestGreen}($0.21$\color{ForestGreen})$\\
     &  &$\mathop{\mathbb{E}}[\hat{P}_{F}]$ \ \ \  (Ref. $P_{F}$ $\sim$ 5.63E-7) & 4.29E-7 & 4.70E-7&5.86E-7 &5.69E-7  &5.63E-7\\
      \bottomrule[1.5pt]

           \multirow{4}{*}{\shortstack[l]{\vspace{-0.14in}\\$\lambda=0.7$\\$\gamma=100$\\$\sigma = 0.5$\\ $\tau = 0.7$}}\rule{0pt}{2.5ex} 
    &\multirow{3}{*}{\hspace{0.07in}100}   &Number of total model calls &14,697& 15,849&14,171 &154,440 &5,956 \\
     &  &C.o.V & 7.76 & 5.19&5.63&0.26$\color{ForestGreen}($0.26$\color{ForestGreen})$  &0.26$\color{ForestGreen}($0.26$\color{ForestGreen})$\\
     &  &$\mathop{\mathbb{E}}[\hat{P}_{F}]$ \ \ \  (Ref. $P_{F}$ $\sim$ 2.23E-6) & 2.40E-6 & 1.59E-6&2.83E-6 &2.26E-6  &2.24E-6\\
      \bottomrule[1.5pt]

           \multirow{4}{*}{\shortstack[l]{\vspace{-0.14in}\\$\lambda=2.5$\\$\gamma=100$\\$\sigma = 0.6$\\ $\tau = 0.7$}}\rule{0pt}{2.5ex} 
    &\multirow{3}{*}{\hspace{0.07in}200}   &Number of total model calls & 11,785 & 12,098&13,872&78,414&6,510\\
     &  &C.o.V & 3.57 & 3.67&3.43 &0.25$\color{ForestGreen}($0.25$\color{ForestGreen})$  &0.25$\color{ForestGreen}($0.24$\color{ForestGreen})$\\
     &  &$\mathop{\mathbb{E}}[\hat{P}_{F}]$ \ \ \  (Ref. $P_{F}$ $\sim$ 5.06E-6) & 5.89E-6 & 5.33E-6&4.62E-6 &5.08E-6  &5.05E-6\\
     
      \bottomrule[1.5pt]
                 \multirow{4}{*}{\shortstack[l]{\vspace{-0.14in}\\$\lambda=0.5$\\$\gamma=200$\\$\sigma = 0.6$\\ $\tau = 0.7$}}\rule{0pt}{2.5ex} 
    &\multirow{3}{*}{\hspace{0.07in}200}   &Number of total model calls & 17,271 & 19,370&17,805 &297,596  &8,575 \\
     &  &C.o.V & 8.09 & 11.85&6.64 &0.31$\color{ForestGreen}($0.31$\color{ForestGreen})$  &0.29$\color{ForestGreen}($0.29$\color{ForestGreen})$\\
     &  &$\mathop{\mathbb{E}}[\hat{P}_{F}]$ \ \ \  (Ref. $P_{F}$ $\sim$ 1.19E-6) & 1.36E-6 & 1.40E-6&1.08E-6 &1.18E-6  &1.17E-6\\
      \bottomrule[1.5pt]
\end{tabular}\label{tabel6}
\end{table}

\begin{figure}[t!]
\centerline{\subfigure[]{\includegraphics[clip, trim=0cm 3cm 1cm 3cm, width=0.324\textwidth]{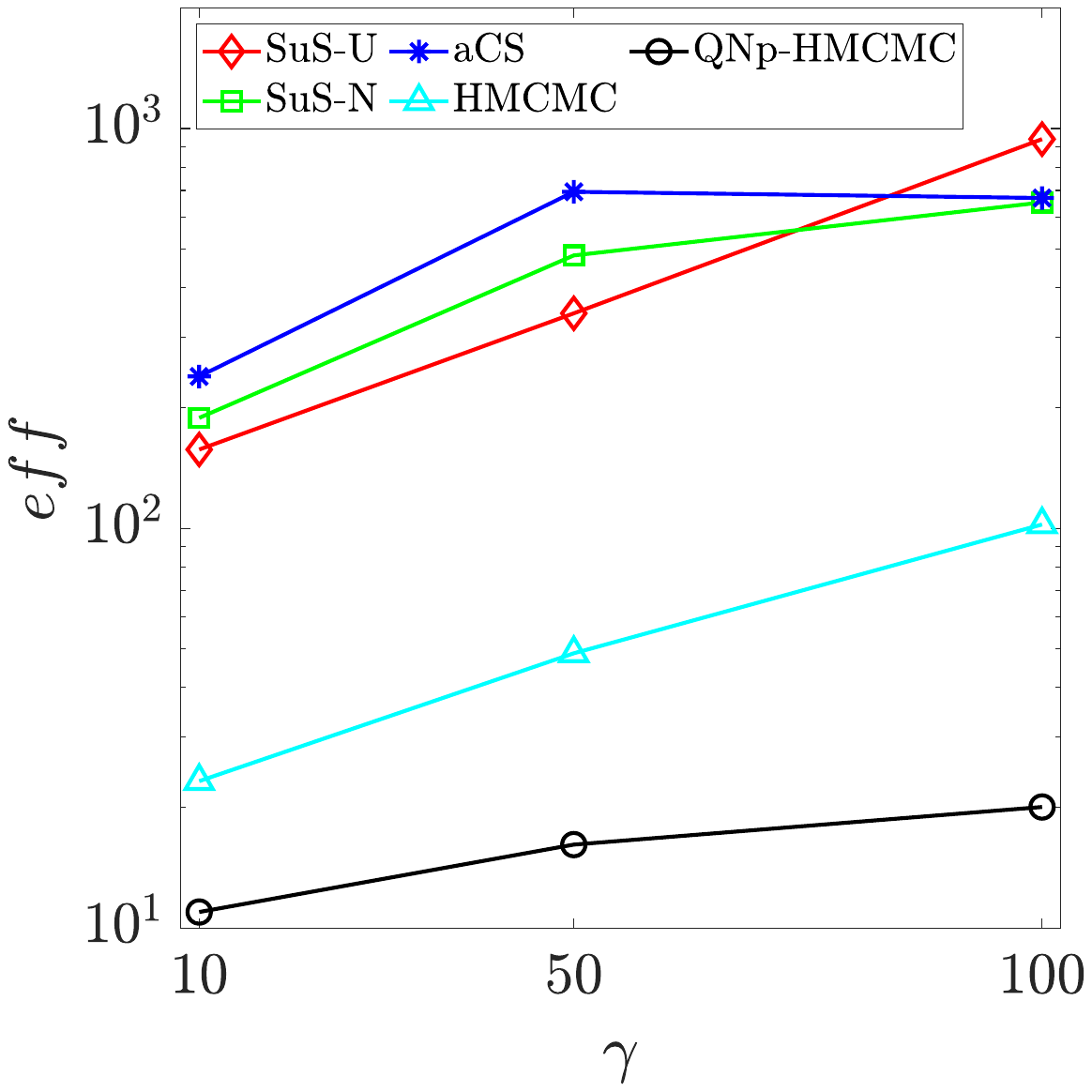}}
\quad
\subfigure[]{\includegraphics[clip, trim=0cm 3cm 1cm 3cm, width=0.324\textwidth]{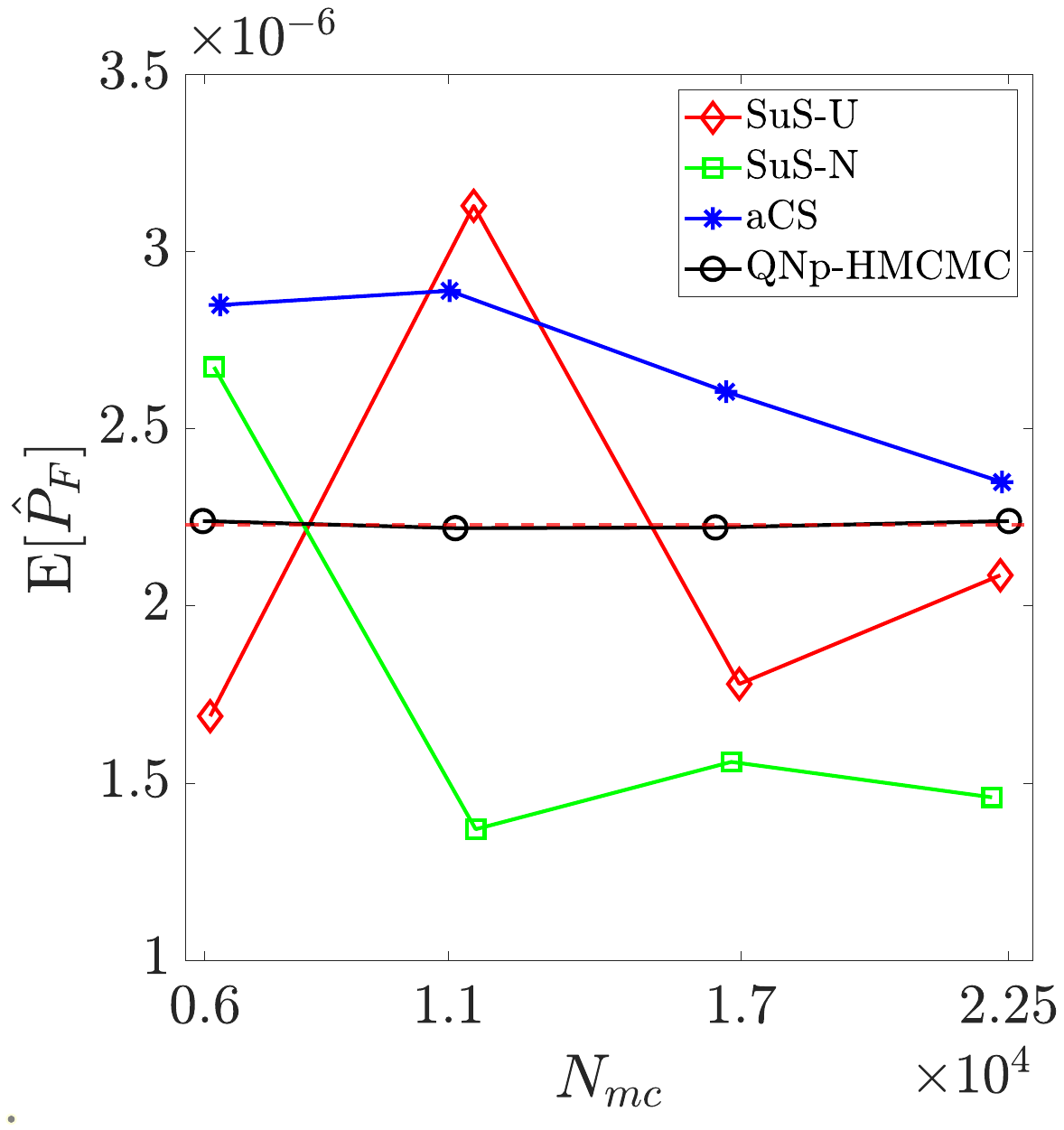}}
\quad
\subfigure[]{\includegraphics[clip, trim=0cm 3cm 1cm 3cm, width=0.324\textwidth]{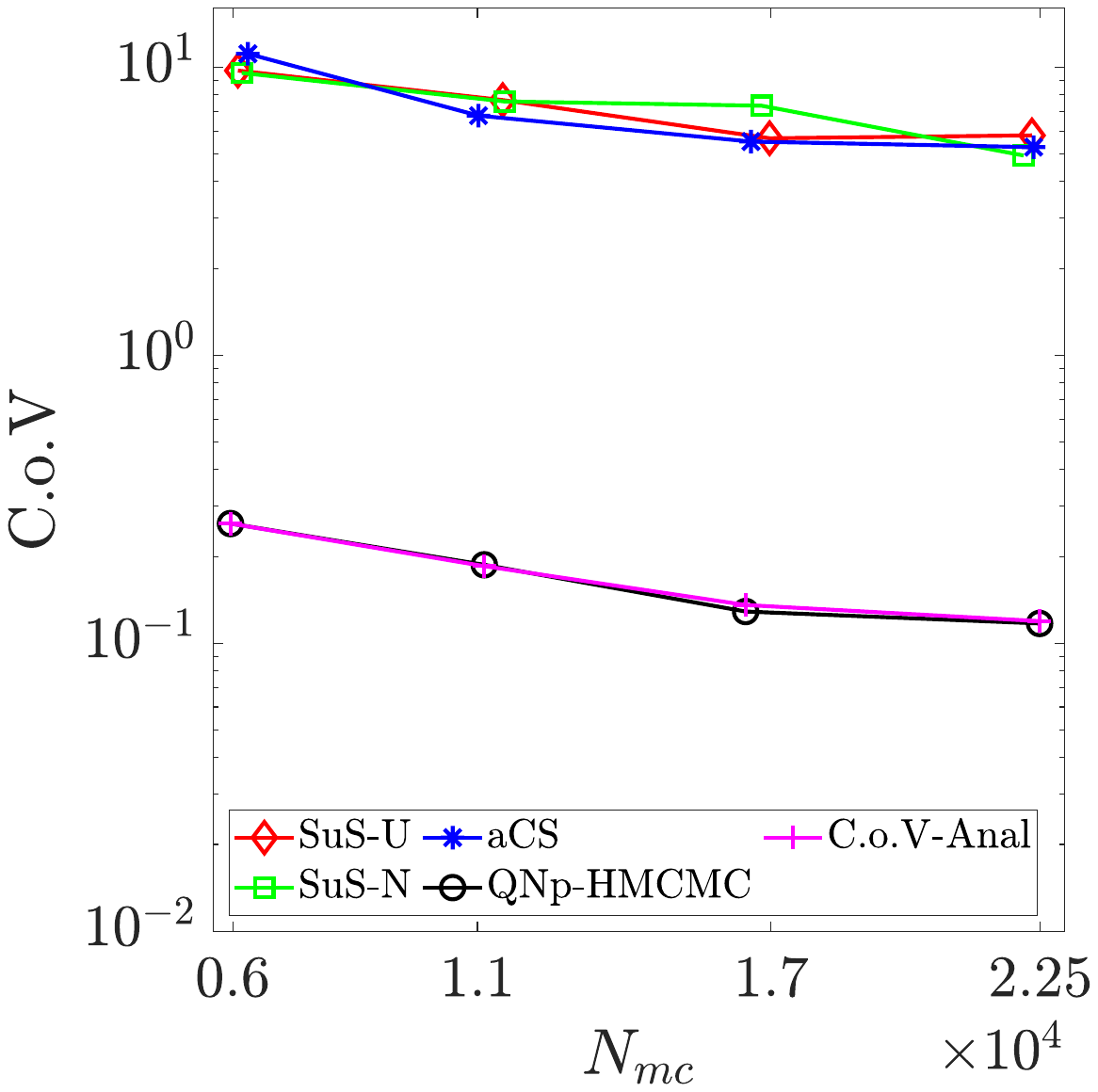}}}\vspace*{-0.1in}
  \captionsetup{labelfont={color=Black}}
\caption{Example 7: (a) \textit{eff} versus different $\gamma$ values based on the results presented in \cref{tabel6} for $d=100$; a low \textit{eff} value indicates high efficiency. (b) Failure probability estimate against the number of model calls ($N_{mc}$) for $d=100$ and $\gamma = 100$ (the dash line is the reference failure probability). (c) Coefficient of variation estimates against the number of model calls for $d=100$ and $\gamma = 100$.}\label{fig:10}
\vspace{-0.2in}
\end{figure}

In \cref{fig:10}a, the computed \textit{eff} values with respect to $\gamma$ for $d=100$ are reported, based on \cref{tabel6} results, confirming the QNp-HMCMC efficiency and robust behavior in all considered cases of nonlinearity. \cref{fig:10}b and \cref{fig:10}c study the effect of the number of model calls ($N_{mc}$) on the mean estimate and C.o.V values for the case of $d=100$ and $\gamma=100$. The QNp-HMCMC method exhibits the best results and significantly outperforms all other approaches, while its results also consistently improve with increased sample sizes. HMCMC results are not reported in these two figures because the needed number of model calls in order to get meaningful results is quite high for this high-dimensional nonlinear problem, deeming this approach non-competitive in this case, and further noting the superior QNp-HMCMC performance and suitability in challenging high-dimensional spaces. The CoV-Anal curve in \cref{fig:10}c again represents the QNp-HMCMC C.o.V estimation based on the analytical expression in \cref{eqqIS5}, showcasing excellent agreement with numerical results.

\begin{figure}[t!]
\centerline{\subfigure[]{\includegraphics[clip, trim=0cm 3cm 0cm 3cm, width=0.39\textwidth]{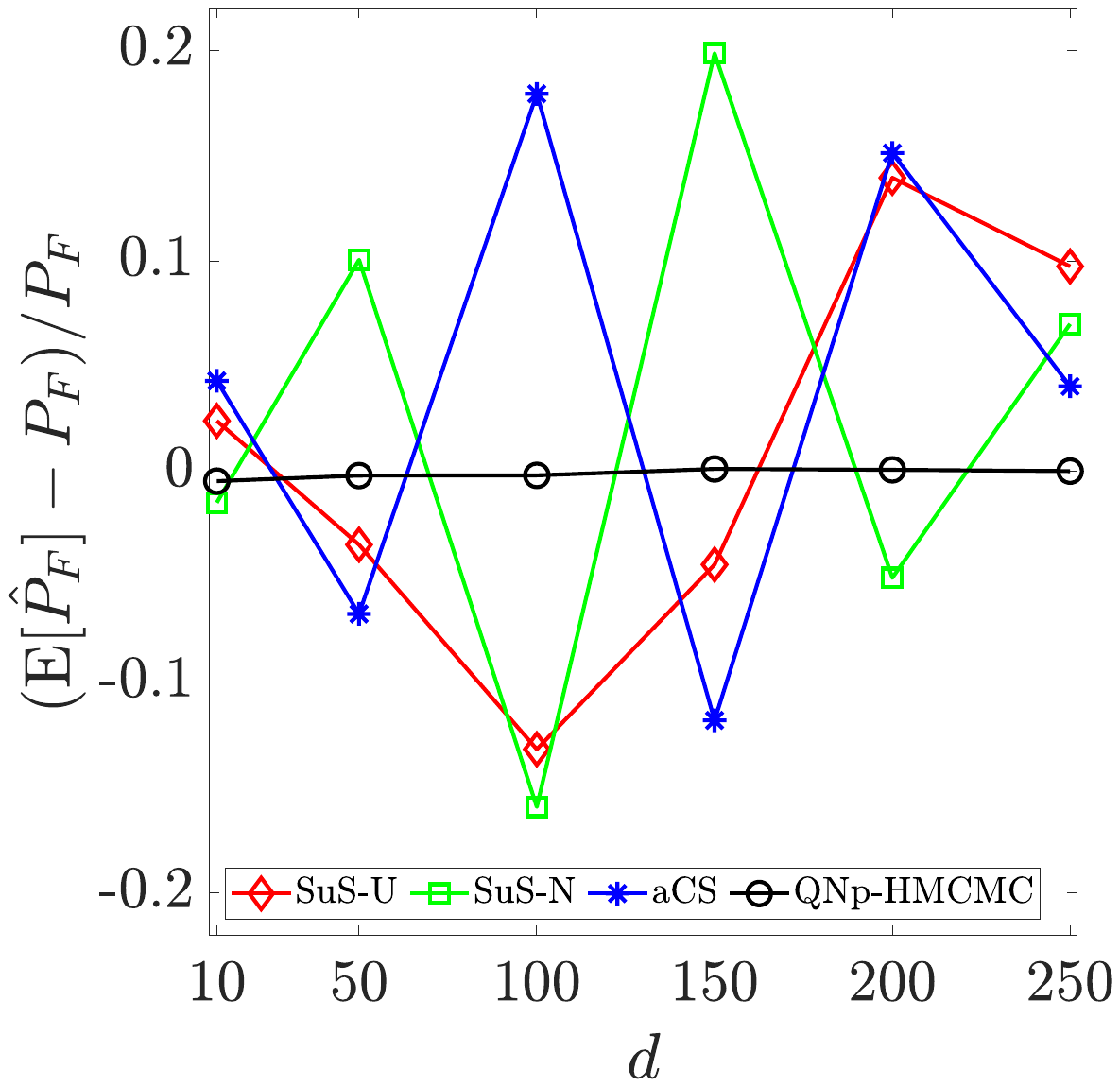}}
\quad
\subfigure[]{\includegraphics[clip, trim=0cm 3cm 0cm 3cm, width=0.39\textwidth]{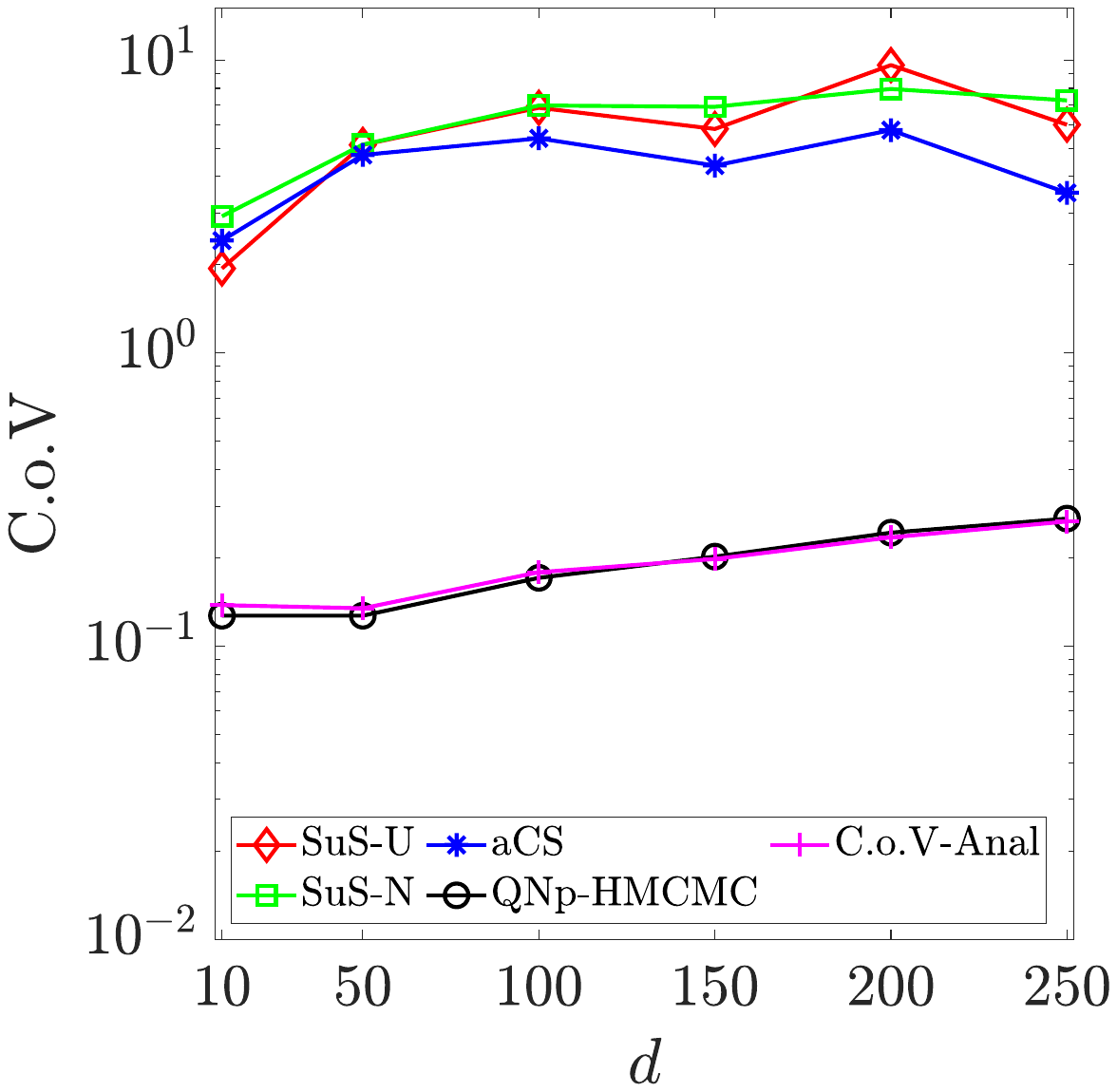}}}\vspace*{-0.1in}
  \captionsetup{labelfont={color=Black}}
\caption{(a) Relative bias of the $\mathop{\mathbb{E}}[\hat{P}_{F}]$ estimator for problems with different number of dimensions $d$ and the $\gamma = d$ cases in Example 7, based on the same number of model calls, approximately 11,000-12,000, for all methods. (b) Coefficient of variation of the estimates with respect to the number of dimensions $d$, for the $\gamma = d$ cases.}\label{fig:12}
\end{figure}

Finally, \cref{fig:12} investigates the effect of the number of dimensions on the relative bias and C.o.V of the estimates, for the $\gamma=d$ cases. The threshold $\lambda$ is adjusted to have a failure probability of around $10^{-6}$.  Again here, HMCMC results are not reported since they require a quite higher number of limit-state function evaluations, after $d>=50$ or so, than all other methods, to achieve meaningful and comparable results. The same number of model calls is used for all methods shown in \cref{fig:12}, approximately 11,000-12,000 total model calls, which differs from \cref{tabel6} settings and results for the similar cases that are studied and reported there,  and the results are based on 500 independent simulations. QNp-HMCMC results are shown to provide an essentially unbiased estimator and very low C.o.V. values for all analyzed cases, in contrast to SuS results, reporting very high C.o.V. values. The analytical C.o.V expression in \cref{eqqIS5} is also here in very good agreement with numerical results.
\par

\subsection{Example 8: High-dimensional highly nonlinear problem}
\begin{figure}[t!]
	\centerline{\subfigure[]{\includegraphics[clip, trim=0cm 3cm 0cm 3cm, width=0.39\textwidth]{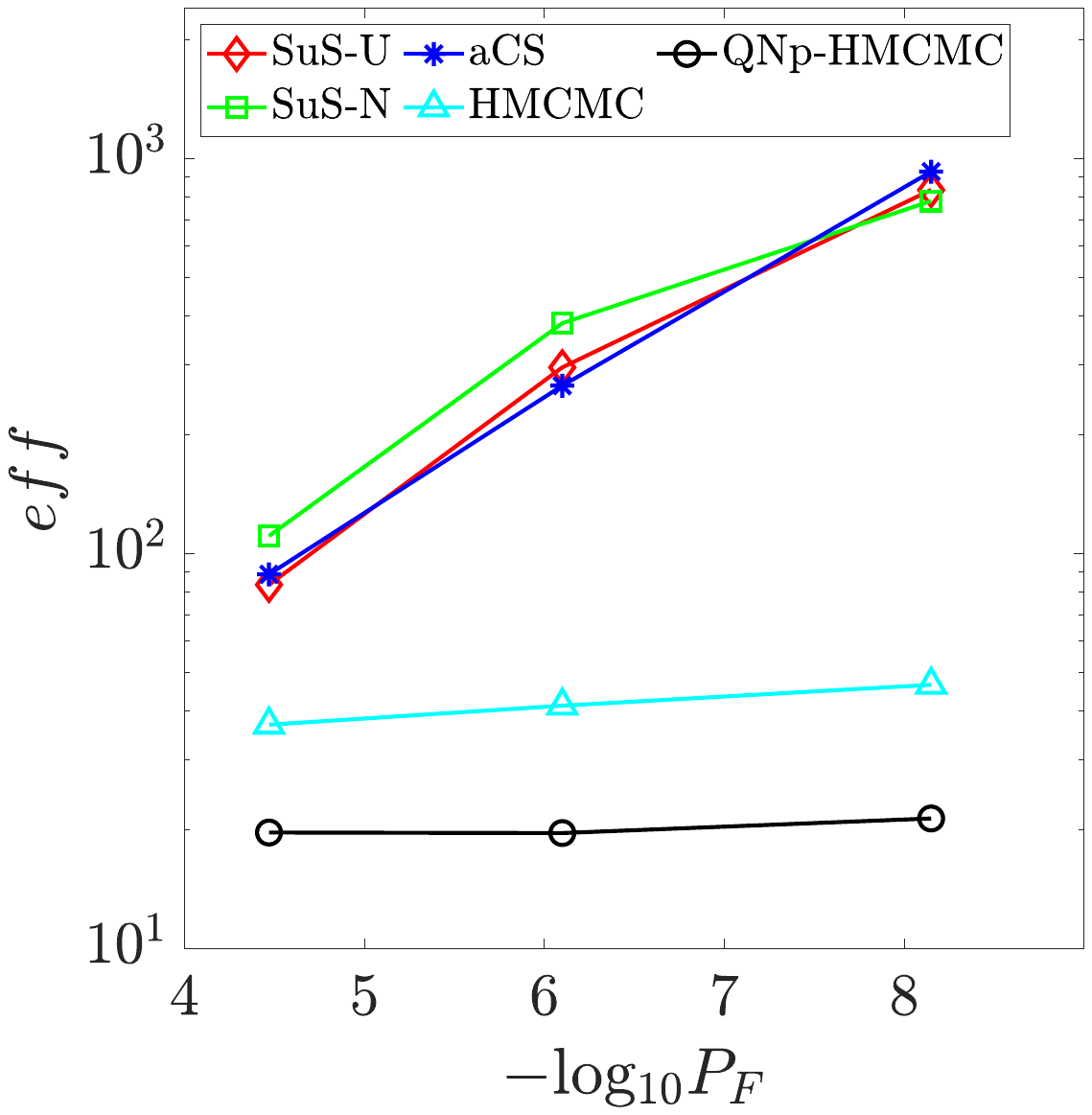}}
		\quad
		\subfigure[]{\includegraphics[clip, trim=0cm 3cm 0cm 3cm, width=0.39\textwidth]{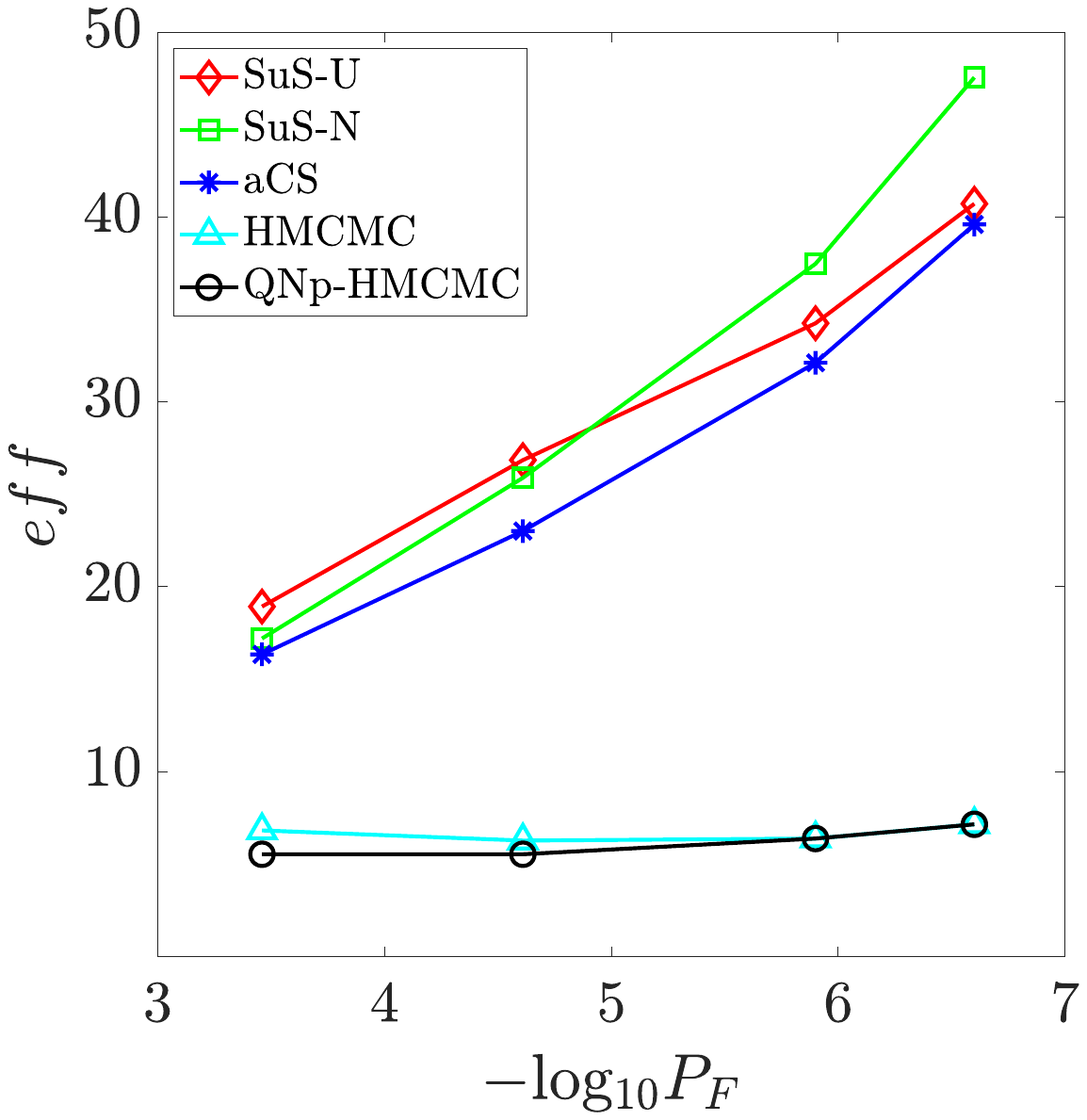}}}\vspace*{-0.1in}
		  \captionsetup{labelfont={color=Black}}
	\caption{Computed \textit{eff} values for various failure probabilities, for a) Example 8 with $d=100$, and b) Example 9 with $d=102$. A low \textit{eff} value exhibits high efficiency.}\label{fig:15}
\end{figure}

To further investigate the QNp-HMCMC performance in challenging high-dimensional nonlinear problems, the limit-state function in this example is expressed in the standard normal space, as: 
\begin{align}
g(\boldsymbol{\theta}) = Y_{0} - \frac{1}{\sqrt{d}}\ \sum_{i=1}^{d} \theta_{i} + 2.5\ \bigg (\theta_{1} - \sum_{j=2}^{10} \theta_{j} \bigg )^{2} + \bigg (\theta_{11} - \sum_{k=12}^{14} \theta_{k} \bigg )^{4} + \bigg (\theta_{15} - \sum_{l=16}^{17} \theta_{l} \bigg )^{8}\label{eq:31}
\end{align} 
where $d$ is the problem dimension, equal to 100. The parameters for the HMCMC approaches are chosen as $\sigma=0.6$, $\tau=0.7$ and 500 burn-in samples. \cref{tablenonlin} summarizes the computed results for various probability levels, acquired by adjusting the $Y_{0}$ threshold value, and as shown the QNp-HMCMC approach once more achieves excellent results, superior to all other methods. In \cref{fig:15}a the \textit{eff} metric is accordingly reported for the three different failure probability levels in \cref{tablenonlin}, with QNp-HMCMC exhibiting the best and stable performance, with very significant efficiency gains in relation to other methods, and particularly for decreasing failure probability levels.

\begin{table}[t!]
\caption{Performance of various methods for the limit-state function of \cref{eq:31} in Example 8 ($d=100$)}
\centering

\footnotesize

\setlength\tabcolsep{4pt}
\begin{tabular}{p{1.1cm}p{4cm}ccccccc}
  \toprule[1.5pt]
  \multirow{7}{*}{\shortstack[l]{\vspace{0.13in}\\$Y_{0}=2.5$\\$\sigma = 0.5$\\ $\tau = 0.7$ }} & \multirow{2}{4cm}{\textbf{500 Independent Simulations}} & \multicolumn{2}{c}{\textbf{CWMH-SuS}} & \multicolumn{1}{c}{\head{aCS-SuS}}& \multicolumn{1}{c}{\head{HMCMC}} & \multicolumn{1}{c}{\head{QNp-HMCMC}}\\

  \cline{3-4}

  & & $U(-1,1)$ & $N(0,1)$\\ 

 \cmidrule(lr){2-7}
  &Number of total model calls &9,380 &9,593  &10,593 &28,026 &7,295\\
  &C.o.V &0.86 &1.13&0.86 &0.22$\color{ForestGreen}($0.21$\color{ForestGreen})$ &0.23$\color{ForestGreen}($0.21$\color{ForestGreen})$ \\
  &$\mathop{\mathbb{E}}[\hat{P}_{F}]$ \ \ \  (Ref. $P_{F}$ $\sim$ 3.40E-5 ) &3.38E-5  &3.26E-5&3.62E-5 &3.37E-5 &3.41E-5 \\
 \bottomrule[1.5pt]
   \multirow{4}{*}{\shortstack[l]{\vspace{-0.14in}\\$Y_{0}=3.5$\\$\sigma = 0.5$\\ $\tau = 0.7$}}\rule{0pt}{2.5ex}
     &Number of total model calls &12,948  &13,704 &13,965 &32,009 &7,924\\
     &C.o.V & 2.60 &3.27 &2.25 &0.23$\color{ForestGreen}($0.22$\color{ForestGreen})$  &0.22$\color{ForestGreen}($0.22$\color{ForestGreen})$\\
     &$\mathop{\mathbb{E}}[\hat{P}_{F}]$ \ \ \  (Ref. $P_{F}$ $\sim$ 7.96E-7) & 8.26E-7 & 7.32E-7&7.99E-7 &7.97E-7  &7.96E-7\\
    \bottomrule[1.5pt] 
      \multirow{4}{*}{\shortstack[l]{\vspace{-0.14in}\\$Y_{0}=4.5$\\$\sigma = 0.5$\\ $\tau = 0.7$}}\rule{0pt}{2.5ex} 
        &Number of total model calls &17,304 &17,793 &17,974 &31,948 &7,889 \\
        &C.o.V &6.33 &5.85 &6.91&0.26$\color{ForestGreen}($0.25$\color{ForestGreen})$  &0.24$\color{ForestGreen}($0.24$\color{ForestGreen})$\\
        &$\mathop{\mathbb{E}}[\hat{P}_{F}]$ \ \ \  (Ref. $P_{F}$ $\sim$ 6.75E-9) & 8.84E-9 & 4.68E-9&7.32E-9 &6.93E-9  &6.86E-9\\
       \bottomrule[1.5pt]
\end{tabular}\label{tablenonlin}
\end{table}

\begin{figure}[t!]
 \centering
 \includegraphics[width=0.59\textwidth]{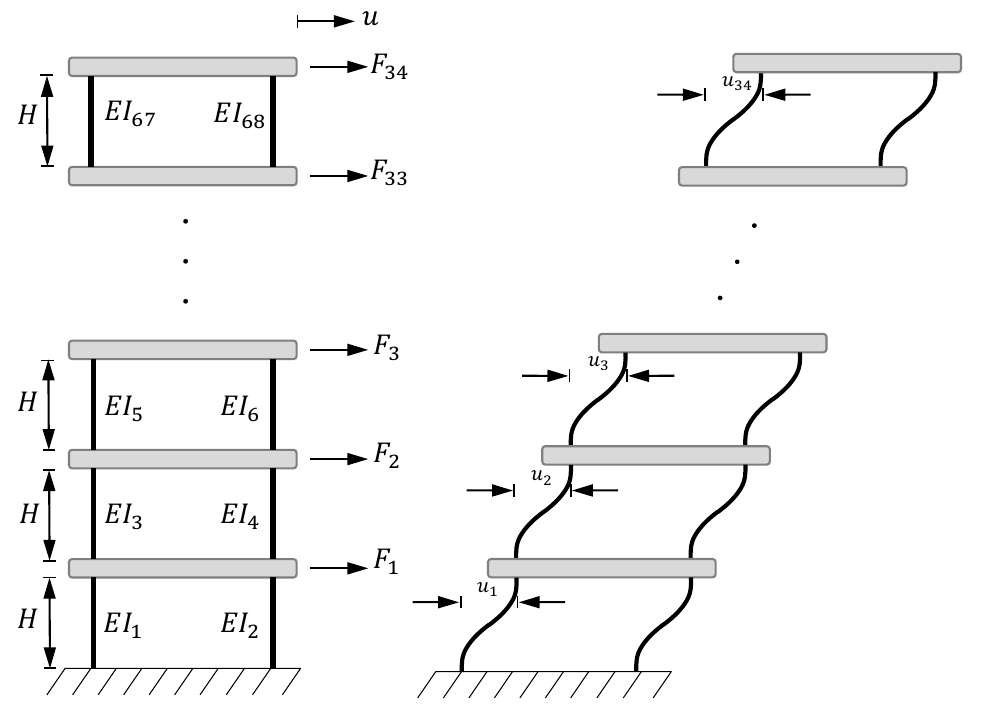}
 \caption{A thirty four-story structure under static loads.}
 \label{fig:14}
 \end{figure}

\newpage
\subsection{Example 9: A thirty four-story structural example}
A thirty four-story structure is analyzed in this modified example from \citep{bucher2009computational}, as represented in \cref{fig:14}, and is subjected to thirty four static loads $F_{i}, i = 1,2,...,34$. The floor slabs/beams are assumed to be rigid and all columns have identical length, $H$ = 4 $\mathrm{m}$, and different flexural stiffnesses $EI_{k}, k = 1,2,...,68$. Loads and stiffnesses are random variables and the total number of random variables is $d=102$ in this problem. The loads are assumed normally distributed, with a mean value of 2 $\mathrm{kN}$ and a C.o.V of 0.4, while stiffnesses are also normally distributed, with a mean value of 20 $\mathrm{MNm^{2}}$ and a C.o.V of 0.2. Based on linear elastic behavior and excluding gravity effects, the top story displacement, $u$, can be calculated by adding the interstory relative displacements, $u_{i}$, as:
 \begin{equation}
 \begin{aligned}
u_{34} = \dfrac{F_{34} H^{3}}{12(EI_{67}+EI_{68})}, \, \, u_{33} = \dfrac{(F_{33} + F_{34}) H^{3}}{12(EI_{65}+EI_{66})}&, \, \, ... \, \, , \, \, u_{2} = \dfrac{(\sum_{i=2}^{34} F_{i}) H^{3}}{12(EI_{3}+EI_{4})},\, \, u_{1} = \dfrac{(\sum_{i=1}^{34} F_{i}) H^{3}}{12(EI_{1}+EI_{2})} \\
u = &\sum_{i =1}^{34} u_{i}\\
g = & \, Y_{0} - u
\end{aligned}
 \end{equation}  
and $g$ is the used limit-state function indicating failure when \textit{u} exceeds a threshold value $Y_{0}$, chosen as 0.21, 0.22, 0.23 and 0.235 m. All random variables are transformed into the standard normal space $\boldsymbol{\Theta}$ for the analysis, and for the HMCMC-based methods the likelihood dispersion factor, $\sigma$, equals 0.3, and the burn-in samples are 400. In this example, 0<g(\textbf{0})< 2, and as discussed in \cref{section4}, we perform the scaling with $g_{c}=g(\textbf{0})$$/$$4$.  \par

 \begin{table}
 \caption{Performance of various methods for the thirty four-story structure in Example 9 ($d=102$)}
 \centering
\footnotesize
\setlength\tabcolsep{4pt}
 \begin{tabular}{p{1.3cm}p{4cm}ccccc}
   \toprule[1.5pt]
   \multirow{7}{*}{\shortstack[l]{\vspace{0.14in}\\$Y_{0}=0.21$\\$\sigma = 0.3$\\$\tau = 0.7$}} & \multirow{2}{4cm}{\textbf{500 Independent Simulations}} & \multicolumn{2}{c}{\textbf{CWMH-SuS}} & \multicolumn{1}{c}{\head{aCS-SuS}}& \multicolumn{1}{c}{\head{HMCMC}} & \multicolumn{1}{c}{\head{QNp-HMCMC}}\\

   \cline{3-4}

   & & $U(-1,1)$ & $N(0,1)$\\ 

  \cmidrule(lr){2-7}
   &Number of total model calls &7,400  &7,400 &7,400 &2,741 &2,520 \\
   &C.o.V &0.22 &0.20 &0.19 &0.13$\color{ForestGreen}($0.12$\color{ForestGreen})$ &0.11$\color{ForestGreen}($0.11$\color{ForestGreen})$ \\
   &$\mathop{\mathbb{E}}[\hat{P}_{F}]$ \ (Ref. $P_{F}$ $\sim$ 3.47E-4) &3.56E-4  &3.46E-4  &3.39E-4 &3.45E-4 &3.47E-4 \\
   \bottomrule[1.5pt]
  \multirow{3}{*}{\shortstack[l]{\vspace{0in}\\$Y_{0}=0.22$\\$\sigma = 0.3$\\$\tau = 0.7$}}\rule{0pt}{2.5ex}
   &Number of total model calls &9,200  &9,200 &9,200 &2,723 &2,519 \\
   &C.o.V &0.28 &0.27 &0.24 &0.12$\color{ForestGreen}($0.11$\color{ForestGreen})$ &0.11$\color{ForestGreen}($0.10$\color{ForestGreen})$ \\
   &$\mathop{\mathbb{E}}[\hat{P}_{F}]$ \ (Ref. $P_{F}$ $\sim$ 2.48E-5) &2.52E-5  &2.54E-5  &2.42E-5 &2.48E-5 &2.48E-5 \\
   \bottomrule[1.5pt]
  \multirow{3}{*}{\shortstack[l]{\vspace{0in}\\$Y_{0}=0.23$\\$\sigma = 0.3$\\$\tau = 0.7$}}\rule{0pt}{2.5ex}
   &Number of total model calls &11,468  &11,475 &11,470 &2,819 &2,819 \\
   &C.o.V &0.32 &0.35 &0.30 &0.12$\color{ForestGreen}($0.11$\color{ForestGreen})$ &0.12$\color{ForestGreen}($0.10$\color{ForestGreen})$ \\
   &$\mathop{\mathbb{E}}[\hat{P}_{F}]$ \ (Ref. $P_{F}$ $\sim$ 1.26E-6) &1.27E-6  &1.30E-6  &1.22E-6 &1.26E-6 &1.26E-6 \\
  \bottomrule[1.5pt]
  \multirow{3}{*}{\shortstack[l]{\vspace{0in}\\$Y_{0}=0.235$\\$\sigma = 0.3$\\$\tau = 0.7$}}\rule{0pt}{2.5ex}
  &Number of total model calls &12,804  &12,836 &12,815 &3,019&3,019 \\
  &C.o.V &0.36 &0.42 &0.35 &0.13$\color{ForestGreen}($0.11$\color{ForestGreen})$ &0.13$\color{ForestGreen}($0.11$\color{ForestGreen})$ \\
  &$\mathop{\mathbb{E}}[\hat{P}_{F}]$ \ (Ref. $P_{F}$ $\sim$ 2.56E-7) &2.62E-7  &2.53E-7  &2.41E-7 &2.52E-7 &2.50E-7 \\
   \bottomrule[1.5pt]

 \end{tabular}\label{tabel9}
 \end{table}

\cref{tabel9} summarizes all computed results, that once more validate the outstanding performance of the proposed HMCMC-based framework, particularly in high-dimensional, very low target probability problems. In \cref{fig:15}b, these findings are further supported by the reported \textit{eff} metric for the four considered failure probability levels in \cref{tabel9}, showcasing again important advantages in relation to other methods.

\section{Conclusions}\label{section6}
A novel approach for estimation of rare event and failure probabilities, termed \textit{Approximate Sampling Target with Post-processing Adjustment} (ASTPA), is presented in this paper,  suitable for low- and high-dimensional problems, very small probabilities, and multiple failure modes. ASTPA can provide an accurate, unbiased probability estimation with an efficient number of limit-state function evaluations. The basic idea of ASTPA is to construct a relevant target distribution by weighting the high-dimensional random variable space through a likelihood model, using the limit-state function. Although this framework is general, to sample from this target distribution we utilize gradient-based Hamiltonian MCMC schemes in this work, including our newly developed \textit{Quasi-Newton based mass preconditioned HMCMC algorithm} (QNp-HMCMC) that can sample very adeptly, particularly in difficult cases with high-dimensionality and very small rare event probabilities. Finally, an original post-sampling step is also devised, using the introduced \textit{inverse importance sampling} procedure, based on the samples. The performance of the proposed methodology is carefully examined and compared very successfully against Subset Simulation in a series of low- and high-dimensional problems. As a general guideline, QNp-HMCMC is recommended to be used for problems with more than 20 dimensions, where traditional HMCMC schemes may not perform that well. However, even in lower dimensions QNp-HMCMC performs extremely well and is still a very competitive algorithm. Since we are utilizing gradient-based sampling methods in this work, all of our analyses and results are based on the fact that analytical gradients can be computed. In cases where numerical schemes are needed for the gradient evaluations, then HMCMC methods will not be competitive in relation to Subset Simulation. It should also be pointed out that different feature combinations of the HMCMC and QNp-HMCMC algorithms can be possible, based on problem-specific characteristics. Some of the ongoing and future works are directed toward exploring various ASTPA variants, sampling directly from non-Gaussian spaces, without the need for Gaussian transformations, and estimating high-dimensional first-passage problems under various settings \citep{PapakonICASP2023, PapakonICOSSAR2022}.

\section*{Acknowledgements}
This material is based upon work partially supported by the U.S. National Science Foundation under CAREER Grant No. 1751941. The authors would also like to thank Prof. Dr. Daniel Straub and Dr. Iason Papaioannou at the Technical University of Munich, for very fruitful scientific discussions in relation to the presented formulation, and particularly for their helpful insights on the construction of the target distribution.

\bibliographystyle{ieeetr}
\bibliography{references}  

\end{document}